\documentclass[reprint,aps,prl,groupedaddress,10pt]{revtex4-2}

\usepackage{packages}
\usepackage{notations}

\usepackage{stackengine}

\begin{document}

\setlength{\abovedisplayskip}{4pt}
\setlength{\belowdisplayskip}{4pt}

\title{The crossover from the Macroscopic Fluctuation Theory to\\  the Kardar-Parisi-Zhang equation controls 
the large deviations\\ beyond Einstein's diffusion
}

\author{Alexandre Krajenbrink}
\email{alexandre.krajenbrink@cambridgequantum.com}
\affiliation{Quantinuum and Cambridge Quantum Computing, Cambridge, UK}
\author{Pierre Le Doussal}
\affiliation{Laboratoire de Physique de l'\'Ecole Normale Sup\'erieure, CNRS, ENS $\&$ PSL University, Sorbonne Universit\'e, Universit\'e de Paris, 75005 Paris, France}

\date{\today}

\begin{abstract}
We study the crossover from the macroscopic fluctuation theory (MFT) which describes 1D stochastic diffusive systems at {\it late times},
to the weak noise theory (WNT) which describes the Kardar-Parisi-Zhang (KPZ) equation at {\it early times}. We focus on the example 
of the diffusion in a time-dependent random field, observed in an atypical direction which induces an asymmetry. 
The crossover is described by a non-linear system which interpolates between the derivative and the standard  non-linear Schrodinger equations in imaginary time. We solve this system using the inverse scattering method for mixed-time boundary conditions
introduced by us to solve the WNT. We obtain the rate function which describes the large deviations of the
sample-to-sample fluctuations of the cumulative distribution of the tracer position. 
It exhibits a crossover as the asymmetry is varied, recovering both MFT and KPZ limits.
We sketch how it is consistent with extracting the asymptotics of a Fredholm determinant formula,
recently derived for sticky Brownian motions. The crossover mechanism studied here 
should generalize to a larger class of models described by the MFT. Our results apply to
study extremal diffusion beyond Einstein's theory. 
\end{abstract}

\maketitle

{\it Introduction} For one-dimensional stochastic systems with a diffusive scaling at large time, such as the symmetric exclusion process (SEP),
the macroscopic fluctuation theory (MFT) \cite{BertiniMFT2015} provides a powerful framework to describe 
the large deviations of the density and current \cite{DerridaMFTReview2007}.
Upon introduction of an asymmetry or driving, such as in the asymmetric 
exclusion process (ASEP) \cite{DerridaReviewASEP}, the diffusive scaling breaks down above some scale, 
and the large scale behavior of the model is usually described by the Kardar-Parisi-Zhang (KPZ) universality class
\cite{TWASEP2009}. 
A paradigmatic member of this class is the KPZ equation \cite{KPZ}, which can be obtained as the continuum limit of the ASEP with a weak
asymmetry \cite{bertini1997stochastic}. The large deviations for the KPZ equation at short time can be described 
using the so-called weak noise theory (WNT)
\cite{Korshunov,Baruch,smith2019time}. It is a close cousin of the MFT, both reduce the calculation 
of large-deviation rate functions to solving saddle point partial non-linear differential equations, not always an easy task. 
A natural question is to understand, in presence of a small but relevant asymmetry, the nature of the crossover 
from the MFT to the WNT. We can expect that is should be somewhat subtle since the MFT describes the large deviations at large time, while the WNT describes the large deviations for the KPZ equation at short time. 

Recently, exact solutions of the WNT equations were obtained by us \cite{UsWNT2021,UsWNTFlat2021}.
It required to extend the inverse scattering method of \cite{ZS,AblowitzKaup1974} to mixed-time boundary conditions
on the so-called $\{P,Q\}$ system, a close cousin of the non-linear Schrodinger equation (NLS). In this paper
we show on an example that the crossover from the MFT to the WNT can be realized as the crossover from the
derivative non-linear Schrodinger equation (DNLS) \cite{kaup1978exact} to the NLS equation. We focus 
on a model for the diffusion of a particle (also called a tracer) at position $y(\tau)$ convected by a centered Gaussian random field
$\eta(y,\tau)$ which is white noise in time and short-range correlated in space, described by a Langevin equation
\be 
\frac{\rmd y(\tau)}{\rmd t}= \sqrt{2} \eta(y(\tau),\tau)+ \chi(\tau) \, ,
\ee
where $\chi$ is a standard white noise in time. Equivalently, the
probability density function (PDF) for the particle position in a given realization of $\eta$, $q_\eta(y,\tau)=\langle \delta(y(\tau)-y) \rangle_\chi$,
obeys the Fokker-Planck equation 
\be 
\label{FP} 
 \partial_\tau q_\eta(y,\tau) = \partial_y^2 q_\eta(y,\tau) - \partial_y ( \sqrt{2} \eta(y,\tau) q_\eta(y,\tau) )\, .
\ee
This model, and its discrete random walk versions, has been revisited recently
\cite{BarraquandCorwinBeta,TTPLD,CorwinGu,TTPLDBeta,BarraquandSticky,GBPLDModerate}. 
The typical behavior is rather dull, and given by the random field average $\overline{q_{\eta}(y,\tau)}$
which yields standard diffusion $y \sim \sqrt{\tau}$. However, 
in the space-time directions which are {\it atypical} for the random walk, e.g. $y \sim v \, \tau$,
it exhibits interesting sample to sample fluctuations related to the KPZ class \cite{BarraquandCorwinBeta}.
In fact, in the small $v$ regime (more precisely for $y \sim \tau^{3/4}$), it maps to the KPZ equation itself
(as predicted in \cite{TTPLD,TTPLDBeta} and proved in \cite{GBPLDModerate}, see also \cite{CorwinGu}). 
These predictions found interesting applications in quantum models with noise, for observables 
dominated by atypical trajectories \cite{bernard2020entanglement}.
They also lead to interesting predictions for extremal diffusion
\cite{BarraquandCorwinBeta,BarraquandThesis,TTPLD,GBPLDModerate}, i.e. for the
position of the maximum of $N$ independent particles, see below.
It turns out that 
Eq.~\eqref{FP} also arises from a lattice gas model of heat transfer, the 
Kipnis-Marchioro-Presutti (KMP) model \cite{KMP}, to which the MFT has been applied
\cite{BertiniPRL2005,bertini2005large,DerridaGershenfeld,Lecomte,KrapivskyMeerson,Zarfaty,BodineauDerrida,BertiniMFT2015,Tailleur2007,Hurtado,Peletier,Shpielberg,grabsch2021closing,poncet2021generalized}.
Hence we anticipate a crossover from the MFT to the WNT when focusing on less and less
typical directions. It is an interesting and open question to understand 
how the large-time large deviations of this model match the short-time large 
deviations of the KPZ equation.

In this paper we show that this crossover is described by the so-called interpolating
system, see \eqref{interpolating} below. Using inverse scattering methods we provide the solution for this system
and obtain the large deviation function of a particular observable. At the end we sketch how the result agrees
with the asymptotic behavior of a Fredholm determinant formula for this observable obtained in \cite{BarraquandSticky}
for a related model of sticky Brownian motions.
We consider a particle which is at position $y=0$ at time $\tau=0$ and study the 
statistics of the probability $Z(Y,T)$ that at time $\tau=T$ it is found to the right of $y=Y$
\be \label{defZ} 
Z(Y,T) = {\mathbb{P}}(y(T) > Y|y(0)=0) 
\ee 
We also need to introduce $H(Y,T)$ the logarithm of this probability, 
our observable of interest here. It also equals 
\be \label{Z} 
Z(Y,T) =e^{H(Y,T)}= \int_{Y}^{+\infty} \rmd y \, q_\eta(y,T) 
\ee 
with $q_\eta(y,0)=\delta(y)$. Note that $H(Y,T) \in [-\infty,0]$
since $Z=Z(Y,T) \in [0,1]$. $Z(Y,T)$ is itself a random variable that fluctuates depending on the realization of $\eta(y,\tau)$, which
from now on is a standard white noise in space and time. 
We consider the diffusive scaling so that $Y,T$ are large, with $Y=\xi \sqrt{T}$, where $\xi=\mathcal{O}(1)$ 
is fixed and plays the role of the asymmetry parameter.
We are interested in the tails of the PDF of $Z=Z(Y,T)$, equivalently of $H=H(Y,T)$,
which take the large deviation forms for $T \gg 1$
\be \label{largePhi} 
{\cal P}(Z) \sim e^{- \sqrt{T} \hat \Phi(Z)} \; , \quad {\cal P}(H) \sim e^{- \sqrt{T} \Phi(H)} \; ,
\ee 
where $\Phi(H)=\hat \Phi(Z=e^H)$ is the rate function which we want to compute,
together with its (implicit) dependence in $\xi$. 

We perform a change of variable $y= x \sqrt{T}$, $\tau= t T$, and $\sqrt{T} q_\eta(y,\tau)=Q_{\tilde \eta}(x,t)$ so that \eqref{FP} becomes
\be \label{eqmoresc} 
 \partial_t Q_{\tilde \eta} = \partial_x^2 Q_{\tilde \eta}(x,t) - T^{-1/4} \partial_x (\sqrt{2} \tilde \eta(x,t) Q_{\tilde \eta}(x,t) )
\ee
where $\tilde \eta$ is a standard white noise and $Z(Y,T) = \int_{\xi}^{+\infty} \rmd x \, Q_{\tilde \eta}(x,1)$. To obtain 
$\Phi(H)$ in \eqref{largePhi} we will first calculate the rate function $\Psi(z)$ which is defined
from the generating function
\be \label{eq:gener} 
\overline{ \exp( - z \sqrt{T} Z(Y,T) )  } \sim \exp(- \sqrt{T} \Psi(z)) 
\ee 
Since $Z$ is a random variable taking values in $[0,1]$,
$\Psi(z)$ is defined for any real $z$, with $\Psi'(z) \in [0,1]$
\cite{SM}. Using \eqref{largePhi} one can compute the expectation value 
in the l.h.s of \eqref{eq:gener} for $T \gg 1$ via a saddle point method and obtain
the relation
\be \label{Legendre0} 
\Psi(z) = \min_{H \leq 0} [\Phi(H) + z e^H ] = \min_{Z \in [0,1]} [\hat \Phi(Z) + z Z ] 
\ee
which shows that $\Psi(z)$ and $\Phi(H)$ are Legendre transforms 
\cite{footnoteLT}. 
The minimum in \eqref{Legendre0} is attained at $H=H(z)$ which is
a solution of $\Phi'(H)=-z e^H$. Anticipating, a remarkable feature of the present problem
is that for $\xi>\sqrt{8}$ this equation has more than one solution, which leads 
to different possible branches for $\Psi(z)$. In that case, we will call
the "optimal" $\Psi(z)$ the function defined from \eqref{Legendre0} (and \eqref{eq:gener}),
i.e. as a global minimum, which will thus exhibit {\it a first-order transition},
with a jump in $\Psi'(z)$. Our strategy will be to compute all branches
of $\Psi(z)$, which allow to reconstruct $\Phi(H)$ and $\hat \Phi(Z)$. 

\paragraph{Interpolating system.} To do so we note that, as in \cite{UsWNT2021}, the l.h.s of \eqref{eq:gener} can be represented as a path
integral
\be \label{path} 
\iint \mathcal{D}Q \mathcal{D}P e^{-\sqrt{T} (S[P,Q] + z \int_{\xi}^{+\infty} \rmd x \, Q(x,1))} 
\ee 
where the associated dynamical action is
\be
S[P,Q] =  \int_0^1 \rmd t \int_\R \rmd x [ P (\partial_t - \partial_x^2) Q - Q^2 (\partial_x P)^2 ]
\ee 
and $P \sqrt{T}$ is the response field. In the large $T$ limit
the path integral in \eqref{path} is controled by its saddle point. Taking the functional derivatives w.r.t.
$\{P,Q\}$, introducing the field $R(x,t)=\partial_x P(x,t)$, and performing a Galilean transformation $x \to x - \xi t$ 
to bring back $\xi$ to zero (see details in \cite{SM}), we arrive at the system of coupled equations
\begin{equation} \label{interpolating}
\begin{split}
    \partial_t Q &= \partial_x^2 Q+2 \beta \partial_x (Q^2 R) + 2 g Q^2 R \\
  -  \partial_t R &= \partial_x^2 R- 2 \beta \partial_x(Q R^2) + 2 g Q R^2
    \end{split}
\end{equation}
with $\beta=-1$ \cite{footnote1} and $g=- \beta \xi/2$ and with the mixed-time boundary conditions
\be \label{init}
Q(x,t=0)= \delta(x) \quad , \quad R(x,t=1) = \Lambda \delta(x) 
\ee 
with $\beta \Lambda=z e^{ - \frac{\xi^2}{4}}$ \cite{footnote2}. 
Once this system is solved, the value of $\Psi(z)$ is obtained from
the saddle point via \cite{SM}
\be  \label{Psi} 
\Psi'(z) = \int_{0}^{+\infty} \rmd x \, Q(x,1) e^{- \frac{1}{2} x \xi - \frac{\xi^2}{4}} 
\ee 
and $\Psi(0)=0$,
where $Q(x,1)$ is the $z$-dependent solution of the above system.
This system interpolates between (i) the $\{P,Q\}$ system for $\beta=0$ (with $P$ called $R$ here), i.e. the cousin of the NLS equation \cite{ZS,AblowitzKaup1974} which
controls the WNT of the KPZ equation \cite{UsWNT2021,UsWNTFlat2021}, and (ii) the cousin of the DNLS equation \cite{kaup1978exact} for $g=0$, which controls the MFT
for this model for $\xi=0$ \cite{NaftaliDNLS}. Thus as $\xi=\frac{Y}{\sqrt{T}}$ is increased, $g$ increases and in the 
limit of large $\xi$, which corresponds to atypical directions, one recovers the large deviations
associated to the KPZ equation (see below). Remarkably, this interpolating system is again integrable \cite{wadati}. 
We will thus extend the inverse scattering analysis of our previous work \cite{UsWNT2021,UsWNTFlat2021}
on the $\{ P,Q\}$ system. Note in passing that the functions $Q(x,t)$ and $R(x,t)$ are not even for $\beta \neq 0$
but as in the $\{P,Q\}$ system they still enjoy the symmetry
\be 
\label{eq:SymmetrySolution}
R(x,t) = \Lambda  Q(-x,1-t) \, .
\ee 
\paragraph{Inverse scattering solution of the interpolating system.} It is a simple generalization of our previous
works  \cite{UsWNT2021,UsWNTFlat2021} so we will sketch it. The Lax pair of linear differential equation reads $\partial_x \vec v= U_1 \vec v$, $\partial_t \vec v= U_2 \vec v$
where $\Vec{v}=(v_1,v_2)^\intercal$ is a two component vector (depending on $x,t,k$) 
where
\begin{equation} \label{eq:LaxPairU} 
U_1=
\begin{pmatrix}
- \frac{\I k}{2}  & - (g + \I \beta k) R(x,t) \\ 
&\\
Q(x,t) &  \frac{\I k}{2} 
\end{pmatrix}
\, , \, U_2= 
\begin{pmatrix}
{\sf A} & {\sf B}\\
{\sf C} & -{\sf A}
\end{pmatrix}
\end{equation} 
with ${\sf A}=\frac{k ^2}{2} -(g+\I \beta k ) Q  R$, ${\sf B}=-(g+\I \beta k ) \left((\I k -\p_x) R+2 \beta  Q R^2\right)$,
        ${\sf C}=(\p_x+\I k) Q+2 \beta  Q^2 R$. One can check that
        the compatibility condition 
        $\partial_t U_1-\partial_x U_2 + [U_1,U_2]=0$ recovers \eqref{interpolating}.
        Let $\Vec{v}=e^{ k^2 t/2} {\phi}$ with $ {\phi}=(\phi_1,\phi_2)^\intercal$ and $\Vec{v}=e^{- k^2 t/2} {\bar{\phi}}$
be two independent solutions of the linear problem such that 
at $x \to -\infty$, $\phi \simeq (e^{-\I k x/2},0)^\intercal$ and $\bar \phi \simeq (0,-e^{\I k x/2})^\intercal$.
Assuming from now on that $\{Q,R\}$ vanish at infinity, the $x \to +\infty$ behavior of these solutions
defines scattering amplitudes
\be \label{eq:plusinfinity} 
\phi \underset{x \to +\infty}{\simeq}
\begin{pmatrix}
a(k,t)e^{-\frac{\I  kx}{2}}\\b(k,t)e^{\frac{\I  kx}{2}}
\end{pmatrix}  ~,~
\bar{\phi} \underset{x \to +\infty}{\simeq}
\begin{pmatrix}
\tilde{b}(k,t)e^{-\frac{\I  kx}{2}}\\ -\tilde{a}(k,t)e^{\frac{\I  kx}{2}}
\end{pmatrix}
\ee 
Plugging this form into the $\partial_t$ equation of the Lax pair at $x \to +\infty$,
one finds 
a very simple time dependence, 
$a(k,t)=a(k)$ and $b(k,t)=b(k)e^{-k^2 t}$, 
$\tilde a(k,t)=\tilde a(k)$ and $\tilde b(k,t)=\tilde b(k)e^{k^2 t}$.
Another normalization relation is obtained from the Wronskian of the two solutions 
, $a(k) \tilde{a}(k)+b(k)\tilde{b}(k)=1$.

Integrating the $\partial_x$ equation of the Lax pair successively for $\bar{\phi}$ and $\phi$ at $t=0$
and at $t=1$, using \eqref{init}, allows to obtain (see \cite{SM}) 
\be \label{bbt} 
\tilde b(k)= (g + \I \beta k) \Lambda e^{-k^2} \quad , \quad b(k)=1
\ee 
and 
\bea \label{aat} 
&& a(k) = 1- (g + \I \beta k) \Lambda Q_-(k) \\
&& \tilde a(k) = 1- (g + \I \beta k) \Lambda Q_+(k) \nonumber
\eea 
where we have defined the half-Fourier transforms
\begin{equation}
    Q_\pm(k)=\int_{\mathbb{R}^\pm} \rmd x \, Q(x,1) e^{-\I k x}
\end{equation}
From the normalization relation one also obtains 
\be  \label{conservation} 
a(k) \tilde a(k) = 1 - b(k) \tilde b(k) = 1- (g+\I \beta k)\Lambda e^{-k^2} 
\ee
which $Q_\pm(k)$ must satisfy. For $\beta=0$ this equation was first obtained 
by us in Ref.~\cite{UsWNT2021} and used recently in 
\cite{mallick2022exact}. As noted there, it is akin to the Fourier transform
of the Wiener-Hopf formulae obtained in \cite[Eqs.~(S65)--(S66)]{grabsch2021closing}.
Our Eq.~\eqref{conservation} is thus the natural extension to arbitrary $g,\beta$. 

Taking these relations in the large $k$ limit
we obtain that $Q(x,1)$ has a jump at $x=0$, with some relation between the
right and left values $Q(0^\pm,1)$. We now follow similar manipulations
as in the recent work \cite{NaftaliDNLS}, the details are given in \cite{SM}.
As $k \to \infty$ one has
\bea \label{largek} 
&& Q_\pm(k) \simeq \pm \frac{1}{\I k} Q(0^\pm,1) \\
&&     a(k) \simeq 1+\beta \Lambda Q(0^-,1) \\
&&     \tilde{a}(k) \simeq 1-\beta \Lambda Q(0^+,1)  
\eea
Equation~\eqref{conservation} at $k \to \infty$ thus implies a first relation
\begin{equation} \label{relationQ} 
    (1-\beta \Lambda Q(0^+,1))( 1+\beta \Lambda Q(0^-,1))=1
\end{equation}
The complete solution of \eqref{conservation} is given by \cite{SM}
\begin{equation}
\begin{split} 
    a(k)& = (1+\beta \Lambda Q(0^-,1) ) e^{\Phi_+(k)} \\
    \tilde{a}(k)&= (1-\beta \Lambda Q(0^+,1) ) e^{\Phi_-(k)} \label{soluatilde}
    \end{split}
\end{equation}
where 
\be \label{Mpm} 
\begin{split}
    \Phi_{\pm}(k)=& \pm \int_\R \frac{\rmd q}{2\I \pi}\frac{\log(1-(g+\I \beta q)\Lambda e^{-q^2} )}{q-k\mp \I 0^+} \\
     =& \pm \dashint_\R \frac{\rmd q}{2\I \pi}\frac{\log(1-(g+\I \beta q)\Lambda e^{-q^2} )}{q-k} 
   \\
   & + \frac{1}{2} \log(1-(g+\I \beta k)\Lambda e^{-k^2}) 
   \end{split}
\ee
The first expression for $\Phi_{\pm}(k)$ is valid 
for $k$ in the complex upper/lower half plane
including the real line, while the second is valid for real $k$ only. 
In the limit $\beta \to 0$ one has $Q(0^+,1)=Q(0^-,1)$ and one
recovers the same formula as first obtained in \cite{UsWNT2021}.
In the limit $g \to 0$ one recovers the recent result in \cite{NaftaliDNLS}. 

We still need to determine the two unknown
constants $Q(0^\pm,1)$ which are related by \eqref{relationQ}.
Combining \eqref{aat} and \eqref{soluatilde} we obtain the relation
\be \label{rel2} 
(g + \I \beta k) \Lambda Q_\mp(k) = 1 - (1 \pm \beta \Lambda Q(0^\mp,1)) e^{\Phi_\pm(k)}
\ee 
which is valid for $\Im(k) \in \mathbb{R}^{\pm}$. Taken at $k=\frac{\I g}{\beta}= - \I \frac{\xi}{2}$
one obtains
\be \label{rel3} 
1 \pm \beta \Lambda Q(0^\mp,1) = e^{- \Phi_\pm(\I g / \beta)}
\ee 
where, for $g/\beta \neq 0$
\be  \label{opposite} 
\Phi_{\pm}(\I g / \beta)= \pm \beta \int_\R \frac{\rmd q}{2 \pi}\frac{\log(1-(g+\I \beta q)\Lambda e^{-q^2} )}{g+ \I \beta q} 
\ee 
are opposite real numbers, so that \eqref{rel3} is compatible with \eqref{relationQ}. 
As discussed below and in \cite{SM}, Eqs.~\eqref{Mpm} and \eqref{opposite} are valid only for $\Lambda g<1$. 

\paragraph{Specialization to the MFT problem.} We now compute $\Psi(z)$ from Eq.~\eqref{Psi} and replace $\beta=-1$, $g=-\beta \frac{\xi}{2}$
and $\beta \Lambda = z e^{- \xi^2/4}$. We note that the r.h.s. of \eqref{Psi} is equal to 
\be \label{psi2} 
\Psi'(z)= Q_+(k=- \I \frac{\xi}{2}) e^{ - \frac{\xi^2}{4}} = 1 - Q_-(k=- \I \frac{\xi}{2})e^{ - \frac{\xi^2}{4}}
\ee
where the second equality comes from the conservation of probability (see \eqref{cons1}).
These quantities can be obtained taking derivatives. 
Taking a derivative w.r.t. $k$ of \eqref{rel2} at $k=\frac{\I g}{\beta}= - \I \frac{\xi}{2}$ and
using \eqref{rel3} one obtains (see details in \cite{SM})
\be \label{zPsiFinal0}
 z \Psi'(z) = 
\dashint_\R \frac{\rmd q}{2 \pi}\frac{\log(1- z (\I q - \frac{\xi}{2}) e^{-q^2 - \frac{\xi^2}{4}} )}{(\I q - \frac{\xi}{2})^2} + z \Theta(-\xi)
\ee
where here and below we use the convention that $\Theta(0)=1/2$ and the principal part is needed only for $\xi=0$.
Integrating over $z$ one obtains
\be \label{PsiFinal0} 
\begin{split}
\Psi(z) &= -
\dashint_\R \frac{\rmd q}{2 \pi}\frac{\mathrm{Li}_2( z (\I q - \frac{\xi}{2}) e^{-q^2 - \frac{\xi^2}{4}} )}{(\I q - \frac{\xi}{2})^2} +z\Theta(-\xi)\\
\end{split}
\ee 
where the last term guarantees analyticity of $\Psi(z)$ in $\xi$. Denoting here $\Psi_\xi(z)$
to indicate the dependence in $\xi$ and performing the change $q \to -q$ in the integrand we see that it obeys the symmetry
\be 
\label{eq:symmetryPsi}
\Psi_{-\xi}(z)= \Psi_{\xi}(-z) + z 
\ee 
which is expected from the definition \eqref{eq:gener}, since upon the symmetry $y \to -y$ in \eqref{FP} and \eqref{Z},
the PDF of $Z(Y,T)$ must be the same as the PDF of $1-Z(-Y,T)$.
For $\xi=0$ one can check (see \cite{SM}) that \eqref{PsiFinal0} 
is consistent with the result in \cite{NaftaliDNLS}. 

Expanding \eqref{PsiFinal0} in series of $z$ one predicts the cumulants of the probability $Z=Z(Y,T)$ in \eqref{defZ}. The first one
is the typical value $\overline{Z}= Z_{\rm typ}(\xi)=e^{H_{\rm typ}(\xi)}$ (i.e. in a typical random field $\eta$)
\bea \label{Ztyp} 
 Z_{\rm typ}(\xi) &&= \Psi'(0) =  - 
\dashint_\R \frac{\rmd q}{2 \pi} \frac{e^{-q^2 - \frac{\xi^2}{4}} }{\I q - \frac{\xi}{2}} +\Theta(-\xi) \\
&& = \frac{1}{2} {\rm Erfc}\left(\frac{\xi}{2}\right) = \int_\xi^{+\infty} \frac{\rmd x}{\sqrt{4 \pi} } e^{-\frac{x^2}{4}} \nn
\eea 
as expected since the mean (and typical) behavior is standard diffusion. The second cumulant
is predicted as $\overline{Z(Y,T)^2}^c \simeq - T^{-1/2} \Psi''(0) = \frac{1}{4 \sqrt{2 \pi T }} e^{- \frac{\xi^2}{2}}$,
as confirmed by a direct weak-noise expansion, see Section~\ref{sec:cum} in \cite{SM}.

\paragraph{Branch cuts, branches of $\Psi(z)$ and the rate function $\Phi(H)$.} We will determine in this section the rate function $\Phi(H)$ for $\xi\geq 0$  (for $\xi<0$ we rely on the symmetry \eqref{eq:symmetryPsi}).
From our expression for $\Psi(z)$ a priori one can now determine the 
rate function for the PDF's in \eqref{largePhi} by inverting the Legendre transform \eqref{Legendre0},
which gives the parametric representation
\be 
\Phi(H)= \Psi(z)- z e^H  , \quad \Psi'(z) = e^H 
\label{eq:ParametricRepresentation}
\ee 
and in terms of $Z$, 
\be 
\hat \Phi(Z)= \Psi(z)- z Z , \quad Z = \Psi'(z)\, .
\label{eq:ParametricRepresentationZ}
\ee 
As mentioned in the introduction, the 
parametric representation \eqref{eq:ParametricRepresentationZ} can lead to different different branches, i.e. a multi-valuation of $\Psi(z)$.
The "optimal" $\Psi(z)$, i.e. solution of the Legendre transform \eqref{Legendre0}, is defined as the minimum over the different branches.

\begin{table*}[t!]
    \centering
    \begin{tabular}{c | c | c | c }
         $\xi$ & $ 0\leq \xi \leq \xi_1$ & \makecell{$\xi_1\leq \xi \leq \xi_2$\\ $z_{c1}<z_{c2}<z_c$} & \makecell{$ \xi_2\leq \xi$\\$z_{c1}<z_c<z_{c2}$} \\[1ex]
        \hline 
        \hline &&&\\[-1ex]
        $\Delta(z)=$  & 
        $\begin{cases}
        0, &z_c<z \\
        \Delta_1(z),  &z<z_c 
        \end{cases}
        $
        &
       $\begin{cases}
        0, &z_c<z \\
        \Delta_1(z), &z_{c1}<z<z_c\\
        \Delta_2(z), \,  &z_{c1}<z<z_{c2}\\
        \Delta_3(z), \, & z<z_{c2}
        \end{cases}$ 
        & 
        $\begin{cases}
        0, &z_c<z \\
        \Delta_1(z), &z_{c1}<z<z_c\\
        \Delta_2(z), \,  &z_{c1}<z<z_{c}\\
        \Delta_2(z)-\Delta_1(z), \,  &z_{c}<z<z_{c2}\\
        \Delta_3(z)-\Delta_1(z), \,  &z_{c}<z<z_{c2}\\
        \Delta_3(z), \, & z<z_{c}
        \end{cases}$ 
    \end{tabular}
    \caption{Determination of the jump function $\Delta(z)$ in the different phases in the case $\xi\geq 0$. 
    One has $z_c= - \frac{2}{\xi} e^{\xi^2/4} \leq 0$ and the points $z=z_{c1}$ and $z=z_{c2}$ are turning points
    which depend on $\xi$. In the interval $z\in [z_{c1},z_{c2}]$, the function $\Delta(z)$ is multi-valued (i.e. it has several branches) 
    due to these turning points. The definition of $\Delta_\ell$ is given in \eqref{eq:JumpPsi}.}
    \label{tab:MainTextTableJump}
\end{table*}

The origin of these different branches can be traced to
the ambiguity which remains for $\Psi(z)$ since 
we have not specified the determination of the logarithm in Eq.~\eqref{zPsiFinal0}.
In practice, the functions $\log(1-x)$, and ${\rm Li}_2(x)$ in \eqref{PsiFinal0},
admit a branch cut for $x>1$. There are thus branch cuts in the complex plane for $q$,
and for some values of $\{z, \xi \}$ one of these branch cuts may cross the integration axis,
see Figures in \cite{SM}. 
These branch cuts originate from the values of $q$ such that the argument of the logarithm in Eq.~\eqref{zPsiFinal0} vanishes. 
Parameterizing the integration variable as $q=\I p$, we then have to find the solutions (i.e. the zeroes) of the following equation  
\begin{equation}
\label{eq:BranchCutEquation}
 e^{- p^2 + \frac{\xi^2}{4}} + z (p + \frac{\xi}{2})=0   \, .
\end{equation}
For $z > z_c$ 
where $z_c=- \frac{2}{\xi} e^{\xi^2/4}\leq 0$, there is never a branch cut crossing the real axis, see \cite{SM}, hence Eqs.~\eqref{zPsiFinal0} and \eqref{PsiFinal0} are valid in this regime and determine what we call the {\it main branch}. 

For $z< z_c$ all real solutions of Eq.~\eqref{eq:BranchCutEquation} for $p$ are negative
and as consequence one branch cut crosses the real axis \cite{SM}. It is then necessary to obtain the analytical
continuation of Eqs.~\eqref{zPsiFinal0} and \eqref{PsiFinal0} to any $z$ by deforming the contour of integration for $q$ to avoid this branch cut. 
In the easiest case this is possible in the complex plane, and in other cases one needs to consider the Riemann sheets, which
leads to more branches and multi-valuation. The analysis is involved
and detailed in \cite{SM}. Here we summarize the main results. The general formula for
$\Psi(z)$ takes the form
\be \label{continuation1} 
\Psi(z) = \Psi_0(z) + \Delta(z) 
\ee 
where $\Psi_0(z)$ is the same integral as in \eqref{PsiFinal0} \cite{footnotePsi0},
and $\Delta(z)$ is the jump contribution from the branch cut, which is
discussed below. The convention $\Delta(z)=0$ defines the main branch of $\Psi(z)$.
The other branches and the form of $\Delta(z)$ as a function of $\xi$ and $z$ are
shown in Table~\ref{tab:MainTextTableJump}. 

To understand Table~\ref{tab:MainTextTableJump} one needs to first discuss the behavior
of the real zeroes of \eqref{eq:BranchCutEquation} which are the relevant one to
determine $\Psi(z)$.  For $z_c\leq z\leq 0$, there is always one positive zero to \eqref{eq:BranchCutEquation} denoted $p_1=p_{1}(z,\xi)$. For $z < z_c$, the zeroes of \eqref{eq:BranchCutEquation} are all negative  and their number is:
\begin{enumerate}
    \item for $0<\xi<\xi_1=\sqrt{8}$, there is one zero $p_{1}(z,\xi)$;
    \item for $\xi_1<\xi$ and $z \in ]z_{c1},z_{c2}[$ there are three zeroes $p_{1}(z,\xi)>p_{2}(z,\xi)>p_{3}(z,\xi)$. 
    The zeroes degenerate, i.e. $p_{1}=p_{2}$ for $z=z_{c1}$ and $p_{2}=p_{3}$ for $z=z_{c2}$ which define $z_{c1},z_{c2}$.
    For $z>z_{c2}$, there is only one zero $p_{1}(z,\xi)$. For $z < z_{c1}$, there is only one zero $p_{3}(z,\xi)$. 
\end{enumerate}
Note that $z_{c1}<z_{c2}<0$, with $z_{c1}=z_{c2}$ at $\xi=\xi_1$, 
and their explicit expression and dependence on $\xi$ is given in \cite[Eq.~\eqref{eq:FormulaZcP}]{SM}.

\begin{figure*}
    \centering
    \stackinset{l}{1.1cm}{b}{1.6cm}{\includegraphics[scale=0.2]{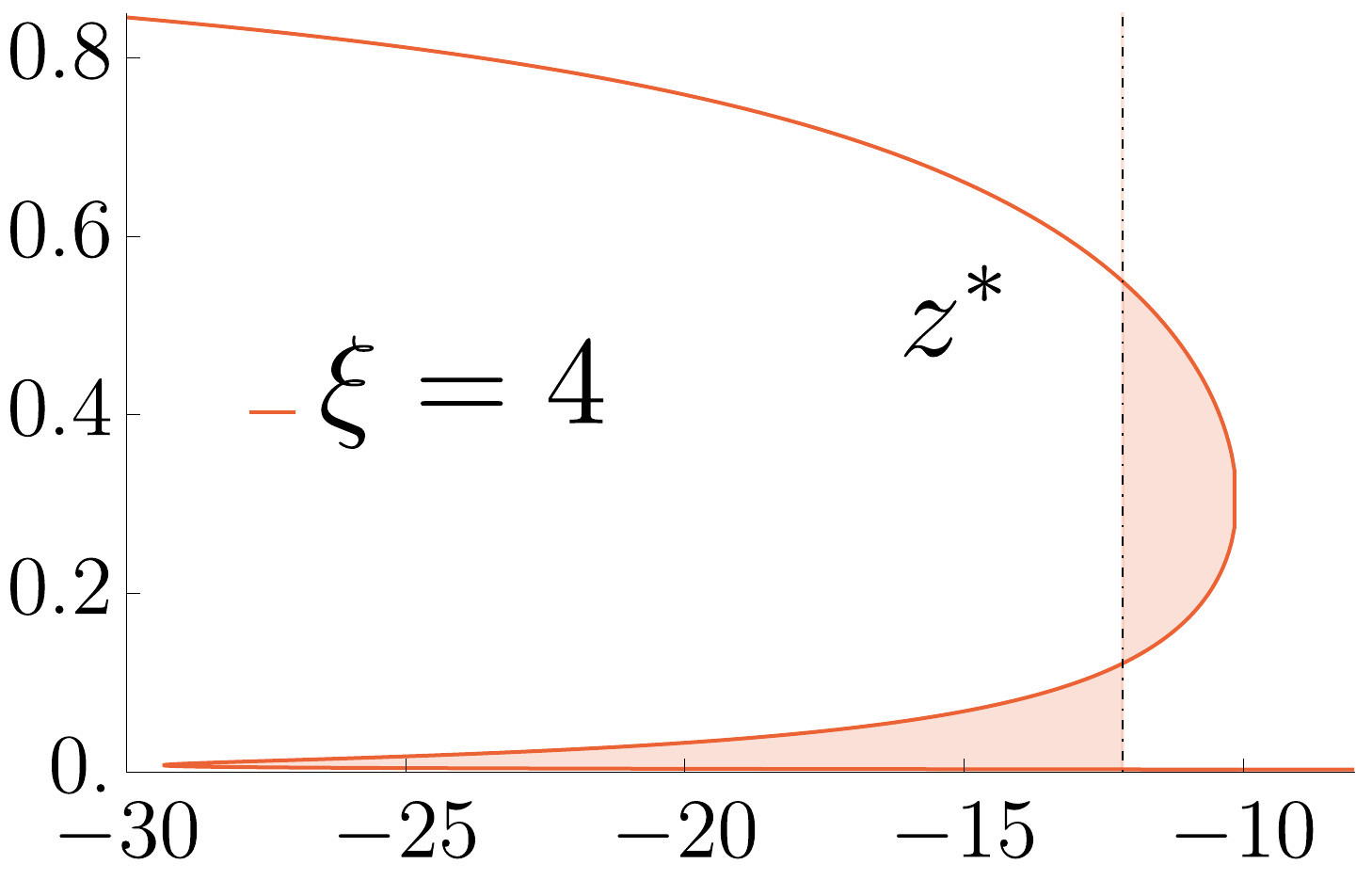}}{\includegraphics[scale=0.55]{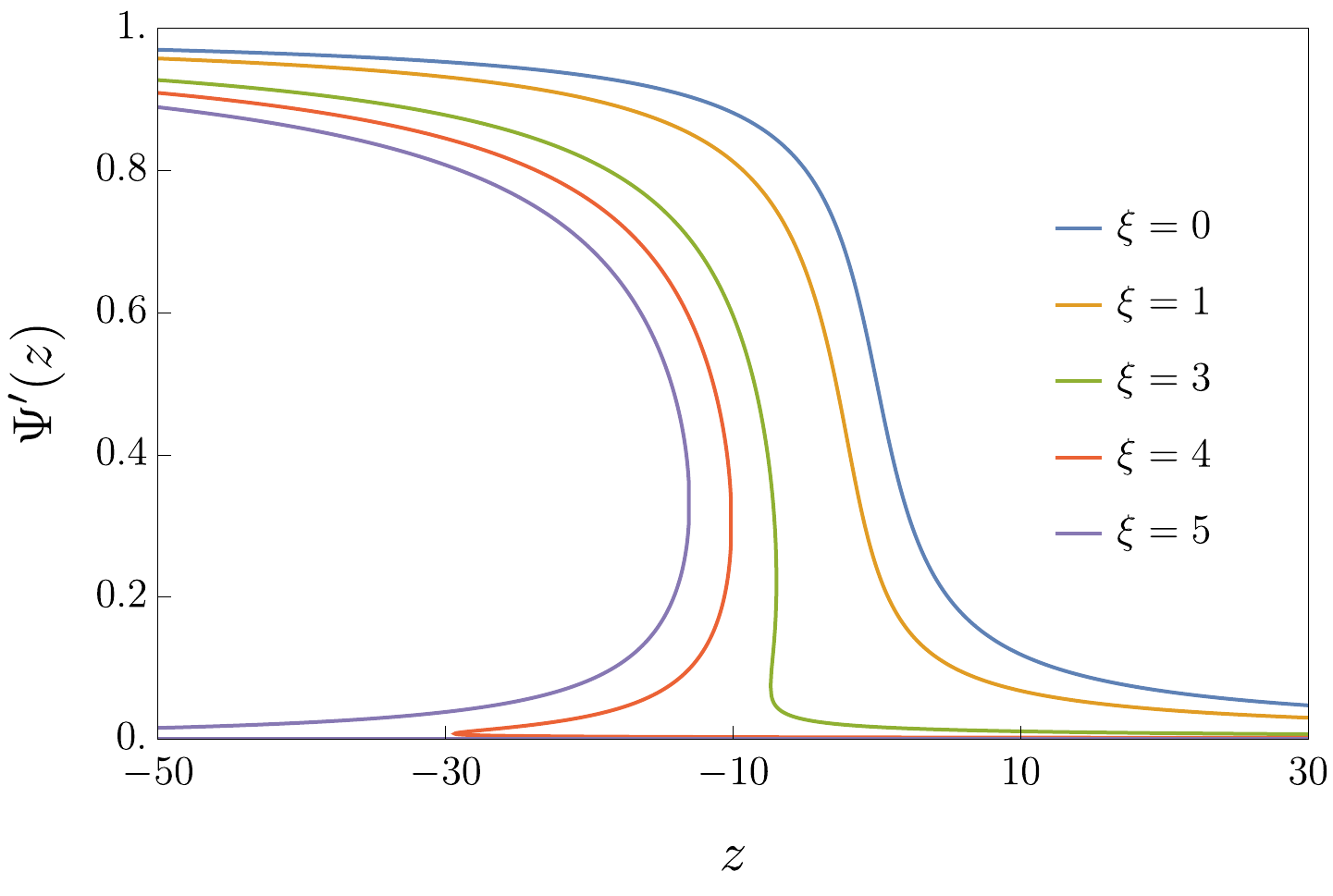}}
    \includegraphics[scale=0.55]{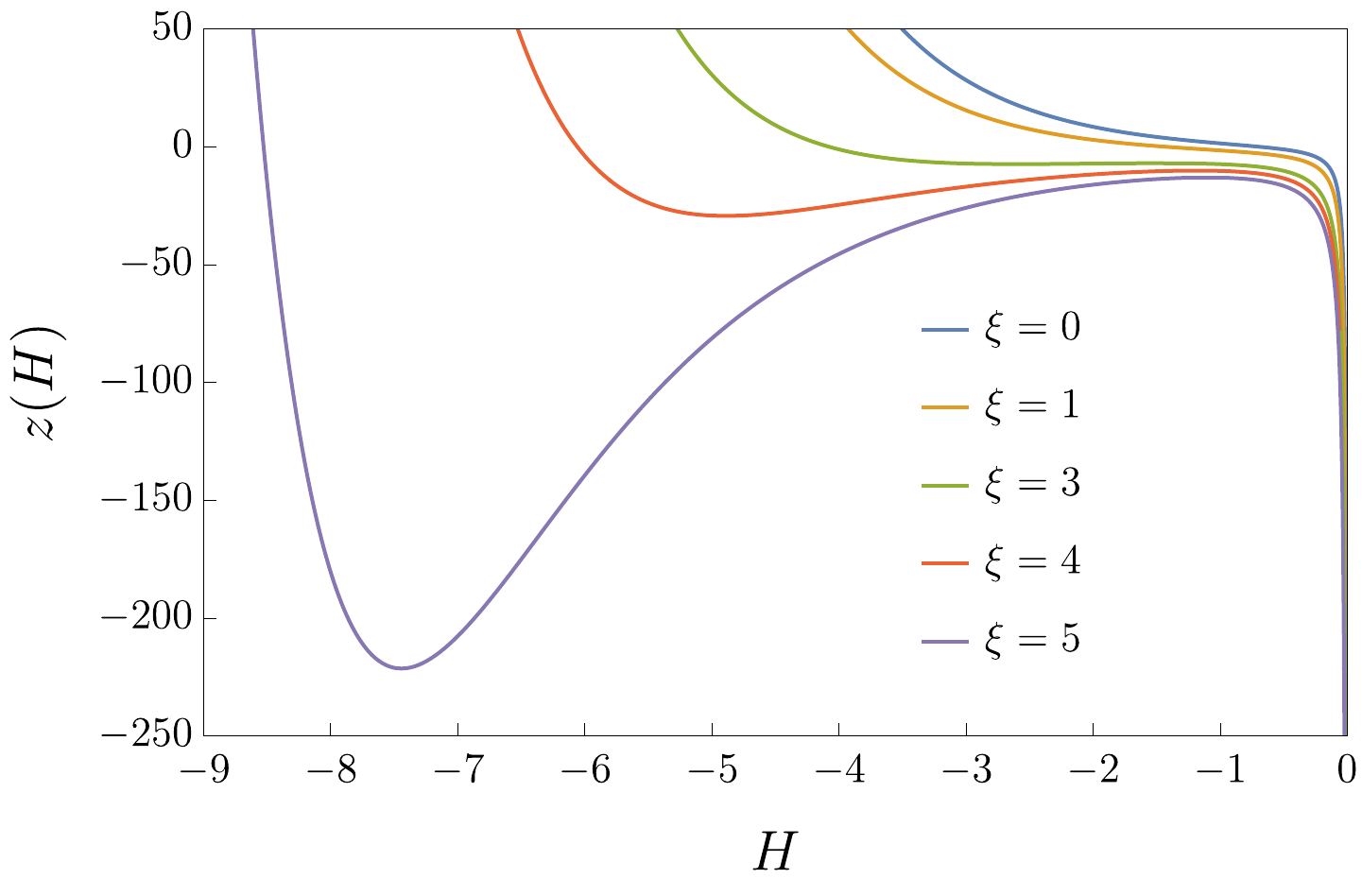}\\
    \includegraphics[scale=0.55]{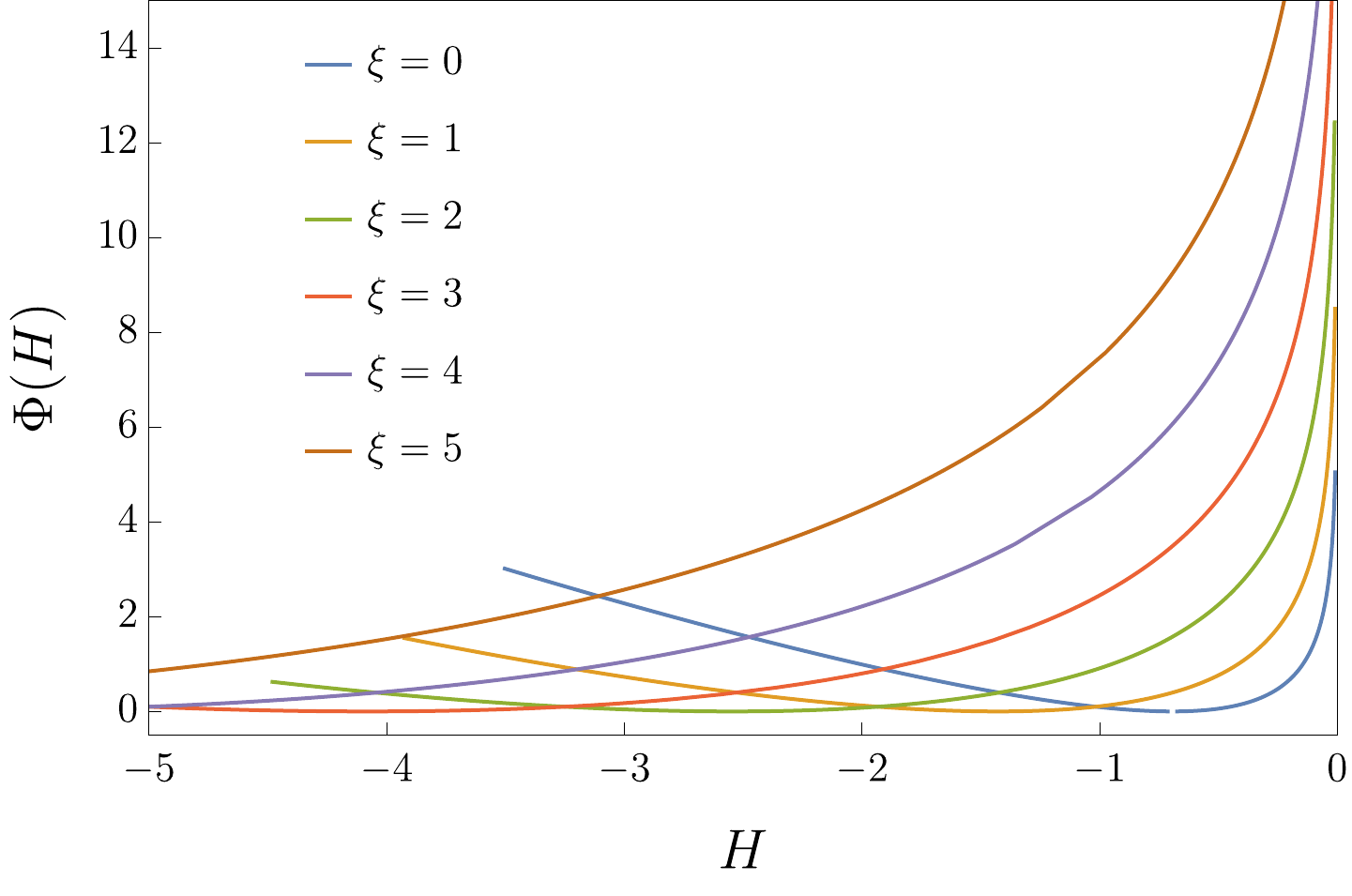}
    \includegraphics[scale=0.55]{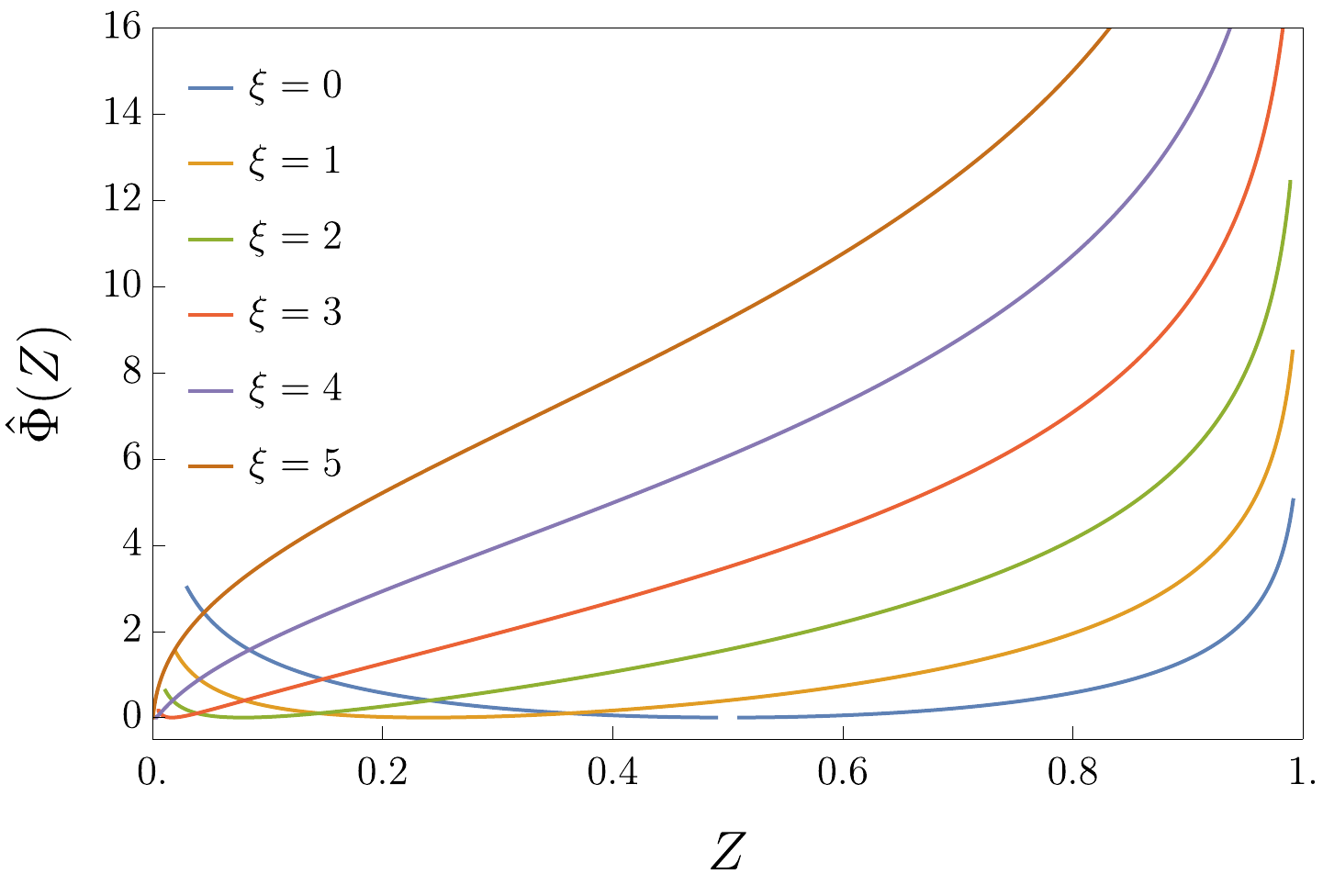}
    \caption{For $\xi=(0,1,2,3,4,5)$ we plot the following. \textbf{(Top Left)} The derivative rate function $\Psi'(z)$ from Table~\ref{tab:MainTextTableJump} as a function of $z$,
    with $\Psi'(+\infty)=0$ and $\Psi'(-\infty)=1$ (all the branches are shown).
    For $\xi > \xi_1$ and $z \in [z_{c1},z_{c2}]$ the function is multi-valued (see text).
    ({\bf Inset}) First order transition: at $z=z^*$ such that the areas of the two shaded regions become equal
    the value of (the optimal) $\Psi'(z)$ jumps from one branch to the other, shown for $\xi=4$.
\textbf{(Top Right)} The function $z=z(H)$ from the Legendre transform \eqref{eq:ParametricRepresentation}. The reciprocal
function $H(z)$ is multi-valued for $\xi > \xi_1$ and $z \in [z_{c1},z_{c2}]$. 
\textbf{(Bottom Left)} The large deviation rate function $\Phi(H)$ versus $H$, obtained using the parametric representation \eqref{eq:ParametricRepresentation}
and Table~\ref{tab:MainTextTableJump}. As $\xi$ increases, the location $H_{\rm typ}$
    of the minimum at $\Phi(H_{\rm typ})=0$ is shifted towards negative values. 
    \textbf{(Bottom Right)} The rate function $\hat \Phi(Z)$ versus $Z$. For $\xi=0$, it is symmetric around $Z=0.5$ and
    one recovers the result of \cite{NaftaliDNLS} (in general the symmetry is $Z(\xi)\leftrightarrow 1-Z(-\xi)$). 
    For large values $\xi>\xi_1$, $\hat \Phi(Z)$ develops a concave part which is responsible for
    the first-order phase transition.}
    \label{fig:MainTextFigs}
\end{figure*}

To come back to $\Psi(z)$ we now define a jump function for $\ell=\{1,2,3 \}$ as
\be 
\label{eq:JumpPsi}
\Delta_\ell(z) =    -\int_z^{z_c}\frac{\rmd z'}{z'}\frac{4p_{\ell}(z',\xi)}{\xi(2p_{\ell}(z',\xi)+\xi)}
\ee
see \cite[Eqs.~\eqref{De1}~\eqref{eq:FinalFormulaDelta}]{SM} for more explicit formula. Our result, as we now discuss,
is that the jump $\Delta(z)$ in Eq.~\eqref{continuation1} is always 
a linear combination of the $\Delta_\ell(z)$. 

Remarkably, the behavior of $\Psi(z)$ exhibits three "phases" depending on the value of $\xi$ with respect to the two critical values $\xi_1= \sqrt{8}$ and $\xi_2\simeq 3.13$  \cite[Eq.~\eqref{eq:CriticalXi}]{SM}, see Table~\ref{tab:MainTextTableJump}.
The function $\Delta(z)$ is multi-valued (i.e. it has several branches) 
for $\xi > \xi_1$ and $z \in [z_{c1},z_{c2}]$. Using the corresponding expressions for $\Psi(z)=\Psi_0(z)+\Delta(z)$ 
one can compute $\Psi'(z)$ for each branch, which is shown in Fig.~\ref{fig:MainTextFigs} (top left).
Using the parametric system \eqref{eq:ParametricRepresentation} one 
obtains the relation between $z$ and $H$, which reads $Z=e^H = \Psi'(z)$
and is shown in Fig.~\ref{fig:MainTextFigs} (top right). Note that $z(H)$ is single-valued
but $H(z)$ may not be. One also obtains the rate function $\Phi(H)$, plotted in Fig.~\ref{fig:MainTextFigs} 
(bottom left)
and $\hat \Phi(Z) $, plotted in Fig.~\ref{fig:MainTextFigs} (bottom right).
We now comment on these plots.

We start with $\xi < \xi_1$. In that case, see Fig.~\ref{fig:MainTextFigs}, the function $\Psi'(z)$
is nicely decreasing from $\Psi'(-\infty)=1$ to $\Psi'(+\infty)=0$ and it leads to a function 
$H(z)$ which is single-valued and monotonous. In Table~\ref{tab:MainTextTableJump}, the appearance of $\Delta_1(z)$ below $z_c$ is due to the fact that the zero $p_1$ becomes negative and the branch cut of the logarithm in \eqref{zPsiFinal0} crosses the real axis.

For $\xi>\xi_1$ the function $\Psi'(z)$ is multi-valued in the interval $z \in [z_{c1},z_{c2}]$,
as can be seen in Fig.~\ref{fig:MainTextFigs}. Outside of $z \in [z_{c1},z_{c2}]$, $\Psi'(z)$ is monotonously decreasing and it still has the correct limits $\Psi'(-\infty)=1$ to $\Psi'(+\infty)=0$. As a consequence of the multivaluation of $\Psi'(z)$, 
the corresponding function $z(H)$, shown in Fig.~\ref{fig:MainTextFigs}, is not monotonous
anymore, which implies that $H$ as a function of $z \in [z_{c1},z_{c2}]$ has three branches, $H_j(z)$, $j=1,2,3$.
These correspond to the three extrema of $\Phi(H)+ z e^H$, and among these extrema only one is the absolute minimum. In Table~\ref{tab:MainTextTableJump}, the appearance of $\Delta_2(z)$ and $\Delta_3(z)$ arise from the fact that (i) at the turning point $z=z_{c1}$ we stop following the first zero $p_1$ and start following $p_2$ instead, (ii) at the turning point $z=z_{c2}$ we stop following the second zero $p_2$ and start following $p_3$ instead. The turning points are located where the consecutive zeroes $p_i(z,\xi)$ coalesce.

For $\xi\geq \xi_2$, the ordering between $z_c$ and $z_{c2}$ changes. Hence to follow the second zero $p_2$ until its coalescence with $p_3$, one needs to cross $z=z_c$ where the branch cut of the logarithm in \eqref{zPsiFinal0} crosses again the real axis, requiring to take into account the jump $\Delta_1(z)$ again.

\paragraph{Multi-valuation and first-order transition.}
To interpret the $S$-shape form of $\Psi'(z)$ shown with all its branches
in Figure~\ref{fig:MainTextFigs} (top left), we recall that the optimal 
$\Psi'(z) = \langle Z \rangle_z$ is the expectation value of the random
variable $Z$ under the $z$-dependent tilted measure
\be \label{tiltedm} 
{\cal P}(Z) e^{- \sqrt{T}  z Z} \sim e^{- \sqrt{T} (\hat \Phi(Z) + z Z) } 
\ee 
The key point is that for $\xi>\xi_1$ the function $\hat \Phi(Z)$ has a concave part, see Fig.~\ref{fig:MainTextFigs} (bottom right).
As a consequence, for $z \in [z_{c1},z_{c2}]$ the tilted measure \eqref{tiltedm} develops three extrema at $Z_j(z)=e^{H_j(z)}$,
solutions of $\hat \Phi'(Z)=-z$. They lead to the
three branches of $\Psi'(z)=Z_j(z)$. Equivalently, there are 
three extremal values $H_j(z)$ in \eqref{Legendre0} solutions of \eqref{eq:ParametricRepresentation}.
The "optimal" $\Psi(z)$ is determined by the minimum in
\eqref{eq:ParametricRepresentation}, hence it is given by 
\be 
\Psi(z) = \min_{j=1,2,3} [\hat \Phi(Z_j) + z Z_j ]
\ee 
and the optimal $j$ switches from $j=1$ to $j=3$ at $z=z^*(\xi)$ where $z^*$ is the solution of \cite{SM}
\be 
\Delta_1(z^*)=\Delta_3(z^*) \, .
\ee 
It is also the point given by an equal area law on the curve $\Psi'(z)$, as in standard
magnetization versus field curve for a first-order phase transition, 
see Fig.~\ref{fig:MainTextFigs} (top left, inset).
The points $Z=\{Z_1, Z_3 \}$ are "stable" whereas $Z=Z_2$ is "unstable". The optimal rate function $\Psi(z)$ thus exhibits a first-order transition. This type
of transition occurs in other large deviation problems
\cite{TouchetteReview2018}.

\paragraph{Solitons.} Let us discuss the significance of the multiple branches 
in terms of the nature of the solutions of the interpolating system \eqref{interpolating}. 
For any value of $\xi$, the logarithm in the integrand of $\Phi_{\pm}$ in \eqref{Mpm} has branch cuts for $q$ in the complex plane.
Equivalently, the product $a(k) \tilde a(k)$ in \eqref{conservation} 
vanishes for some complex $k=k_s= \I p_c$ where $p_c$ are generic complex solutions 
of \eqref{eq:BranchCutEquation}, indicating the spontaneous generation of a soliton \cite{ZS}. This means that additional solutions with
a solitonic component are possible, as was the case for the WNT of the KPZ equation \cite{UsWNT2021,UsWNTFlat2021}. 
For that problem, by obtaining the exact solution of the $\{P,Q\}$ system for any space-time point, we were
able to show that the multi-valuation of $\Psi(z)$ was equivalent to the coexistence of two solutions
(in that case with and without a soliton) for the same mixed-time boundary conditions. Here, for $\xi > \sqrt{8}$ the multi-valuation
of $\Psi(z)$ similarly indicates the coexistence of three solutions for $\Lambda g \in [z_{c2}/z_c,z_{c1}/z_c]$ 
(a $\xi$-dependent interval), at least two of them being solitonic. Each of these solutions give rise to a different value $Z_j(z)$, i.e. of the value of the right hand side of Eq.~\eqref{Psi}. The precise nature and interactions of these solitons will be investigated in 
a subsequent work \cite{UsInterpolatingNext}.

\paragraph{Large $\xi$ limit and convergence to KPZ.} We now consider the limit where the tracer particle is located extremely far, i.e. $\xi \to +\infty$. In that limit we can approximate in \eqref{PsiFinal0}
$\I q - \frac{\xi}{2} \simeq  - \frac{\xi}{2}$ and define  $\tilde z = z \frac{\xi}{2} e^{- \frac{\xi^2}{4}}=-z/z_c$
to  obtain
\be \label{KPZ1} 
\Psi_0(z) \simeq - 
\dashint_\R \frac{\rmd q}{2 \pi}\frac{\mathrm{Li}_2( -z   \frac{\xi}{2} e^{-q^2 - \frac{\xi^2}{4}} )}{(\frac{\xi}{2})^2}= \frac{4}{\xi^2} \Psi_{\rm KPZ}(\tilde z) 
\ee 
where 
\be \label{KPZ2}
\Psi_{\rm KPZ}(\tilde z)  = -\frac{ 1}{\sqrt{4 \pi}} \mathrm{Li}_{5/2}(-\tilde z )
\ee 
is the main branch of the large-deviation rate function for the height field $h_{\rm KPZ}(0, T_{\rm KPZ})$ of the KPZ equation
with droplet initial condition. 
This rate function was
obtained in \cite{le2016exact} from a Fredholm determinant formula and
in \cite{UsWNT2021} from the exact solution of the WNT, i.e. of the $\{P,Q\}$ system. 
It admits a second branch denoted $\Psi_{\rm KPZ}(\tilde z)+ \Delta_{\rm KPZ}(\tilde z)$,
which is also recovered, see below. 

Hence, at the level of the large deviations, the MFT in the regime $Y \sim \sqrt{T}$ recovers,
in the large $\xi=\frac{Y}{\sqrt{T}}$ limit, the result of the WNT for the KPZ equation
valid for small KPZ time $T_{\rm KPZ} \ll 1$. Comparing \cite{le2016exact,UsWNT2021} and
the present result \eqref{KPZ1} shows  that the correspondence between the 
MFT time $T$ and the KPZ time $T_{\rm KPZ}$ reads \cite{SM}
\be  \label{correspondence} 
T_{\rm KPZ}= \frac{Y^4}{16 T^3} = \frac{\xi^4}{16 T}
\ee 
This can be compared with \cite{GBPLDModerate} where it was shown, in the different scaling 
regime $Y \sim T^{3/4}$, i.e. $T_{\rm KPZ}=\mathcal{O}(1)$, that in law $Z(Y,T) \simeq \frac{Y}{2 T}  e^{- \frac{Y^2}{4 T} } 
e^{ h_{\rm KPZ}(0, T_{\rm KPZ}) }$, with the same $T_{\rm KPZ}$ as in \eqref{correspondence}
(see \cite{SM} for details). Since $\tilde z = z \frac{\xi}{2} e^{- \frac{\xi^2}{4}}$,
the two results match perfectly, showing that no intermediate regime exists between
the diffusive scaling $Y \sim \sqrt{T}$ and the of the finite-time KPZ equation scaling $Y \sim T^{3/4}$ 
(note that the large-time Tracy-Widom KPZ class universality is seen only for $Y \gg T^{3/4}$). 

Finally, as detailed in \cite{SM}, we obtain the convergence at large $\xi$ of the rate function 
for the logarithm $H=\log Z$, to the rate function of the reduced KPZ height $H_{\rm KPZ}$
\be \label{PhiPhiKPZ} 
\Phi(H) \simeq \frac{4}{\xi^2} \Phi_{\rm KPZ}(H_{\rm KPZ})
\ee 
with the correspondence $H=- \frac{\xi^2}{4} - \log(\frac{\xi}{2}) + H_{\rm KPZ}$ and 
$\Phi_{\rm KPZ}$ is the rate function for the KPZ equation, see
details and definitions in \cite{SM}. 

We now discuss what happens to the other branches of $\Psi(z)$ at large $\xi$. We show 
how the second branch of the KPZ rate function and the value of its jump, $\Delta_{\rm KPZ}$, 
is recovered in the limit. Recovering this second branch, which
exists for $-1\leq  \tilde z<0$, is necessary 
for \eqref{PhiPhiKPZ} to hold for all $H_{\rm KPZ} \in \mathbb{R}$. 
To this aim, we first define the rescaled critical values of $z$ as 
\begin{equation}
    \tilde{z}_c=-\frac{z_c}{z_c}, \; \tilde{z}_{c1}=-\frac{z_{c1}}{z_c}, \; \tilde{z}_{c2}=-\frac{z_{c2}}{z_c} \, .
\end{equation}
and take their large $\xi$ limit which read
\begin{equation}
    \tilde{z}_c=-1, \; \tilde{z}_{c1}\simeq -1, \; \tilde{z}_{c2}\simeq 0  \, .
\end{equation}

From the last column of Table~\ref{tab:MainTextTableJump} we now see that, in that limit:
\begin{enumerate}
\item the branches $\Delta_1(z)$ and $\Delta_2(z)$ disappear due to the coalescence of $z_c$ and $z_{c1}$,
\item the next branch recovers the second branch of the KPZ limit, i.e.
\be
\Delta_2(z) - \Delta_1(z) \to \frac{4}{\xi^2} \Delta_{\rm KPZ}(\tilde{z})=\frac{16}{3 \xi^2 }  \log \big(\frac{-1}{\tilde z}\big)^{\frac{3}{2}} \, ,
\ee 
which can be explicitly checked \cite{SM}.
\item When $\tilde{z} $ approaches $\tilde{z}_{c2}$, we obtain in the large $\xi$ limit that the corresponding 
value of $H_{\rm KPZ}$ goes to $+\infty$, hence the branches $\Delta_3(z)-\Delta_1(z)$ and $\Delta_3(z)$ disappear at infinity,
see Fig.~\ref{fig:HeightZTilde}. This corresponds to events where $Z = \mathcal{O}(1)$ which become irrelevant in
that limit \cite{footnoteZ}. 
\end{enumerate}

\paragraph{Fredholm determinant formula.} We can now compare our result \eqref{PsiFinal0} obtained using the inverse scattering method, to a
formula obtained by completely different methods, for a model of sticky Brownian motions
\cite{BarraquandSticky}. That model, which allows for a rigorous formulation, is believed (up to mathematical subtleties) to
be equivalent to the one considered here. The original
formula of \cite{BarraquandSticky} is valid for any time $T$ and any $Y$, and here we 
obtain its limit in the large deviation diffusive scaling regime. Applied to our model this formula reads
\be \label{FDformula} 
\overline{ e^{- u Z(Y,T)} }  = {\rm det} (I - K_u)
\ee 
where the kernel $K_u$ was derived in Ref.~\cite[Theorem 1.11]{BarraquandSticky} 
and is recalled in Eq.~\eqref{eq:KernelGuillaume}. 

We scale $u= \sqrt{T} z$ with $z=\mathcal{O}(1)$ in Eq.~\eqref{FDformula}
so that the l.h.s. of \eqref{FDformula} can be identified to the l.h.s of 
\eqref{eq:gener}. We perform asymptotic analysis on the kernel $K_u$ and 
extract the large $T$ large deviation rate function $\Psi(z)$ by using
the first cumulant method introduced in \cite{KrajLedou2018,ProlhacKrajenbrink,krajenbrink2019beyond}.
The manipulations, sketched in \cite{SM} are quite heuristic, but allow to
recover nicely the algebraic form of formula \eqref{PsiFinal0}. It remains
open how to make it more controled.

\paragraph{Extremal diffusions.}
Consider the rightmost of $N$ independent particles in the same random field, of position
$Y_N(T)=\max_{i=1,\dots,N} Y_i(T)$. Without random field and for $N \gg 1$, 
$Y_N(T)$ has a deterministic part $\simeq 2 \sqrt{T \log N}$ plus
a "thermal" fluctuation part $\simeq G \sqrt{\frac{T}{\log N}}$,
$G$ being a Gumbel random variable. With the random field, 
for $\log N \sim T \gg 1$, there is also a $\mathcal{O}(1)$ sample-to-sample fluctuation part,
with a Tracy-Widom distribution \cite{BarraquandCorwinBeta,BarraquandThesis}. 
In the more accessible regime $\log N \sim \sqrt{T} \gg 1$, as shown in \cite{TTPLD,GBPLDModerate} 
this fluctuating term is distributed as $h(0,T_{\rm KPZ})/\sqrt{\log N}$, the droplet solution of
the KPZ equation. These phenomena go beyond the Gaussian nature of 
Einstein's diffusion. They allow for a detectable fingerprint of the random medium.
Recently, these two regimes have been observed numerically \cite{CorwinPrivate}.
The present results allow to study yet another regime, $\log N \ll \sqrt{T}$,
where diffusive scaling holds and the scaled position $y_N(T)=\frac{Y_N(T)}{\sqrt{T}}$ of the maximum 
converges to
\be 
y_N(T) \simeq 2 \sqrt{\log N} + \frac{ G - c_N + \delta H}{\sqrt{\log N}}
\ee 
where for typical environments $\delta  H=\mathcal{O}(T^{-1/4})$ is an Edwards-Wilkinson
random variable with a computable variance \cite{SM}, and for rare environments 
$\delta H = H-H_{\rm typ}(\xi)=\mathcal{O}(1)$ with the rate function \eqref{largePhi} 
computed here and $\xi=2 \sqrt{\log N}$. We also find that in the regime $N \sim \sqrt{T}$ the
disorder average CDF takes the large deviation form $\overline{\mathbb{P}(y_N(T) < \xi)} 
\sim e^{- \sqrt{T} \Sigma_{\xi}(n)}$, with $n=\frac{N}{\sqrt{T}}= \mathcal{O}(1)$ fixed,
and $\Sigma_{\xi}(n)$ a rate function explicitly obtained
in \cite{SM}. 
\\

\paragraph{Extension to the SSEP.}
Our results are relevant within the class of MFT models with quadratic noise variance $\sigma(\rho)$. These models enjoy a mapping to the $\{R,Q\}$ DNLS system \eqref{DNLS2}, see \cite[Section~\ref{app:ExtensionSSEP}]{SM}. Recently, the exact solution of the MFT of the SSEP was investigated in Ref.~\cite{mallick2022exact} using the well-known gauge transformation of Wadati and Sogo, \cite[Eq.~(4.5)]{wadati1983gauge}, to map the $\{R,Q\}$ DNLS system to the $\{P,Q\}$ NLS system. The remarkable result of \cite{mallick2022exact} is that under this gauge transformation, the annealed initial condition 
of the SSEP is mapped onto the initial condition solved 
by us in \cite{UsWNT2021} (with different coupling constant \cite{footnoteW}). The natural extension of \cite{mallick2022exact} would be to study the statistics of a tracer at arbitrary position in an MFT model with quadratic variance and annealed initial condition. The inverse scattering method we have pursued in this work provides the right tools to answer this question.

\paragraph{Conclusion.} We have elucidated here in great details the crossover upon adding an asymmetry, between the MFT for diffusive systems
and the WNT of the KPZ equation. We have focused on the example of the diffusion of a tracer in a time-dependent random medium 
in an atypical direction and a "droplet" type initial condition. We have obtained the large-deviation functions 
in the context of classical integrability using simple, standard and versatile inverse scattering methods. 
For this model it was based on the integrable crossover between the DNLS and NLS equations.
Obtaining the complete solution of this interpolating system \eqref{interpolating} 
beyond the large-deviation observable requires further efforts involving the use of Fredholm determinants 
similarly to what we have achieved in \cite{UsWNT2021,UsWNTFlat2021} for the complete solution of the WNT. 
This is one open question that we leave to subsequent works \cite{UsInterpolatingNext}, together
with other outstanding questions, such as investigating the MFT-KPZ crossover for more general
models, or within the present model, to study \eqref{FP} for other initial conditions, 
in particular those identified in \cite{TTPLD} to converge in atypical directions to solutions of the
KPZ equation for flat and stationary geometries \cite{UsWNTFlat2021}.

\paragraph{Note added.} After completion, the paper \cite{TsaiDroplet2022} appeared,
where the results of \cite{UsWNTFlat2021} are proved rigorously.

\begin{acknowledgments}
\paragraph{Acknowledgments.}  We thank G.~Barraquand for discussions and collaborations on closely related topics.  AK acknowledges support from ERC under Consolidator grant number 771536 (NEMO)
and PLD from the ANR grant ANR-17-CE30-0027-01 RaMaTraF. This article is based upon work supported by the National Science Foundation under Grant No. DMS-1928930 while the two authors participated in a program hosted by the Mathematical Sciences Research Institute in Berkeley, California, during the Fall 2021 semester.
\end{acknowledgments}

\newpage

\begin{widetext} 

%%%%%%%%%%% Merge with supplemental materials %%%%%%%%%%
%\pagebreak
%%\widetext
%\begin{widetext} 
%\begin{center}
%\textbf{\large Supplemental Materials: Title for main text}
%\end{center}
%%%%%%%%%%% Merge with supplemental materials %%%%%%%%%%
%%%%%%%%%%% Prefix a "S" to all equations, figures, tables and reset the counter %%%%%%%%%%
%\setcounter{equation}{0}
%\setcounter{figure}{0}
%\setcounter{table}{0}
%\setcounter{page}{1}
\makeatletter
\renewcommand{\theequation}{S\arabic{equation}}
\renewcommand{\thefigure}{S\arabic{figure}}
%\renewcommand{\bibnumfmt}[1]{[S#1]}
%\renewcommand{\citenumfont}[1]{S#1}
%%%%%%%%%%% Prefix a "S" to all equations, figures, tables and reset the counter %%%%%%%%%%

\setcounter{section}{0}
\renewcommand{\thesubsection}{\Alph{subsection}}
\renewcommand{\thesubsubsection}{\arabic{subsubsection}}

\setcounter{secnumdepth}{3}

\begin{large}
\begin{center}

Supplementary Material for\\  {\it  
The crossover from the Macroscopic Fluctuation Theory to the Kardar-Parisi-Zhang equation controls the large deviations beyond Einstein's diffusion}

\end{center}
\end{large}

We give the principal details of the calculations described in the main text of the Letter. 
We also give additional information about the results displayed in the text.

\section{Derivation of the interpolating system}

Let us detail the steps performed in the text to obtain $\Psi(z)$ defined from the expectation value in \eqref{eq:gener} via the saddle point method. Introducing
the standard dynamical path integral representation, one has (where overlines represent averages w.r.t. the 
random field $\tilde \eta$)
\bea 
&& \overline{ e^{- z \sqrt{T} Z(Y,T)} } 
= \overline{ e^{- z \sqrt{T} \int_\xi^{+\infty} \rmd x Q_{\tilde \eta}(x,t=1) } }  \\
&& = \overline{  \iint \mathcal{D}\tilde Q \mathcal{D}\tilde P  e^{- \int_0^{1} dt \int_\R dx [ \sqrt{T} \tilde P (\partial_t \tilde Q - \partial_x^2 \tilde Q - \partial_x \sqrt{2} \tilde  \eta(x,t) \tilde Q) ] - z \sqrt{T} \int_\xi^{+\infty} dx \tilde Q(x,t=1) } }  \\
&& = \iint \mathcal{D}\tilde Q \mathcal{D}\tilde P e^{-\sqrt{T} (S[\tilde P,\tilde Q] + z \int_0^{1} \rmd t \delta(t-1) \int_\xi^{+\infty} \rmd x \tilde Q(x,t)  )}  \label{spvalue} 
\eea  
where the equation of motion \eqref{eqmoresc} has been expressed using the response field $\tilde P \sqrt{T}$, and 
the associated dynamical action is
\be
S[\tilde P, \tilde Q] =  \int_0^1 \rmd t \int_\R \rmd x [ \tilde P (\partial_t - \partial_x^2) \tilde Q - \tilde Q^2 (\partial_x \tilde P)^2 ]
\ee 
For $T \to +\infty$ one can use the saddle point method. Here we denote the fields and their saddle point values 
in the original frame as $\{\tilde P,\tilde Q\}$ to distinguish them from the Galilean transformed fields $\{P,Q\}$
introduced below. Taking the functional derivative w.r.t. $\{\tilde P,\tilde Q\}$ we obtain
\bea 
&&  \partial_t \tilde Q = \partial_x^2 \tilde Q+2 \beta \partial_x (\tilde Q^2 (\partial_x \tilde P))  \\
   - && \partial_t \tilde P = \partial_x^2 \tilde P- 2 \beta \tilde Q (\partial_x \tilde P)^2 - z \delta(t-1) \Theta(x-\xi)
\eea 
with $\beta=-1$. We will keep $\beta$ as a parameter but for the application to obtain $\Psi(z)$
it is understood that it is set to $\beta=-1$. Initially the upper boundary in time is $t=+\infty$ but since $\tilde P$ vanishes for $t>1$, 
we can equivalently restrict the
equations for $t \in [0,1]$ and interpret the last term in the second equation (which must be integrated
backward in time) as a boundary condition $P(x,t=1) = - z \, \Theta(x-\xi)$
for $\tilde P(x,t)$, so it drops from the equation. To make the two equations more 
symmetric let us now introduce the derivative field $\tilde R(x,t)=\partial_x \tilde P(x,t)$, leading to
\begin{equation} \label{DNLS2} 
\begin{split}
    \partial_t \tilde Q &= \partial_x^2 \tilde Q+2 \beta \partial_x (\tilde Q^2 \tilde R)  \\
  -  \partial_t \tilde R &= \partial_x^2 \tilde R- 2 \beta \partial_x(\tilde Q \tilde R^2) 
    \end{split}
\end{equation}
with the boundary conditions
\be  \label{inittilde} 
\tilde Q(x,t=0)= \delta(x) \quad , \quad \tilde R(x,t=1) = \Lambda_0 \delta(x-\xi)  \quad , \quad \Lambda_0= - z 
\ee 
This system is the cousin of the DNLS equation (identical to it upon the change $t \to \I t$).\\ 

Now we perform a Galilean transformation $x \to x - \xi t$ 
to bring back $\xi$ to zero. Anticipating a bit, let us introduce the
interpolating system introduced in the text in \eqref{interpolating} which we recall here
\begin{equation}
\label{interpolating2} 
\begin{split}
    \partial_t Q &= \partial_x^2 Q+2 \beta \partial_x (Q^2 R) + 2 g Q^2 R \\
  -  \partial_t R &= \partial_x^2 R- 2 \beta \partial_x(Q R^2) + 2 g Q R^2
    \end{split}
\end{equation}
and notice that if $\tilde Q$, $\tilde R$ satisfies this system with couplings $(\beta,g)$ then
\be
Q(x,t)= \tilde Q(x-v t,t) e^{-\frac{1}{2} x v + \frac{v^2}{4} t} \quad , \quad 
R(x,t)= \tilde R(x-v t,t) e^{\frac{1}{2} x v - \frac{v^2}{4} t} \quad , \quad 
\label{eq:app-boost-dnls}
\ee 
also satisfies the same system with couplings $(\beta,g + \beta \frac{v}{2})$.
Thus consider $\tilde Q$, $\tilde R$ which satisfy the above DNLS system \eqref{DNLS2} 
with boundary conditions $\tilde Q(x,0)=Q_0(x)$ and $\tilde R(x,1)=\Lambda_0 \delta(x-\xi)$. 
We will choose $v=-\xi$ so that

\be
Q(x,t)= \tilde Q(x+\xi t,t) e^{\frac{1}{2} x \xi + \frac{\xi^2}{4} t} \quad , \quad 
R(x,t)= \tilde R(x+ \xi t,t) e^{-\frac{1}{2} x \xi - \frac{\xi^2}{4} t} \quad , \quad 
\ee  
satisfies the interpolating system \eqref{interpolating2} with couplings $(\beta, - \beta \frac{\xi}{2})$
and boundary conditions
\be
Q(x,0)= \tilde Q_0(x) e^{\frac{1}{2} x \xi } \quad , \quad 
R(x,1)= \Lambda_0 \delta(x) e^{ - \frac{\xi^2}{4}} \quad , \quad 
\ee  
which for $\tilde Q_0(x)=\delta(x)$ gives the result \eqref{init} in the text, where we called $\Lambda=\Lambda_0 e^{ - \frac{\xi^2}{4}}$.
\\

{\bf Symmetries}. Note that the DNLS equation \eqref{DNLS2} is invariant by $x \to -x$ and $R \to -R$. The interpolating system $\{R,Q\}$ 
\eqref{interpolating2} is invariant by $x \to -x$, $R \to -R$ and $g \to - g$. 
\\

{\bf Conserved quantities}. Note that the system \eqref{DNLS2} admits a series of conserved (i.e. time independent) quantities, 
the simplest one being $\int_\R \rmd x \,\tilde Q(x,t)=1$ (here its value is fixed to unity by the
initial condition \eqref{inittilde}). This conservation law originates from the conservation of probability in the
Fokker-Planck equation, $\frac{\rmd }{\rmd t} \int \rmd t \,q_\eta(x,t)=0$. Upon a Galilean transformation it becomes 
\be 
\int_\R \rmd x \, Q(x,t) e^{- x \frac{\xi}{2}  - \frac{\xi^2}{4} t} = 1 \label{cons1} 
\ee 
Note that $\tilde R$ also satisfies the conservation law $\int_\R \rmd x \, \tilde R(x,t)=\Lambda$ and after a Galilean transformation
$\int_\R \rmd x \, R(x,t) e^{x \frac{\xi}{2}  + \frac{\xi^2}{4} t} = \Lambda_0= \Lambda e^{\xi^2/4}$. One can check that this is consistent with the symmetry 
\eqref{eq:SymmetrySolution}.
\\

{\bf Coupling constant}. If one compares with Ref.~\cite{UsWNT2021} the true coupling constant 
of the $\{P,Q\}$ (i.e. here $\{R,Q\}$) system used there (called $g$ there) is $\hat g=\Lambda g$. 
Since $\beta \Lambda= z e^{-\xi^2/4}$ and $g=- \beta \frac{\xi}{2}$, this gives 
$\hat g=- z \frac{\xi}{2} e^{-\xi^2/4}$. The special point $z_c$ discussed in the text 
thus corresponds to $\hat g=1$, as for the case of the WNT of the KPZ equation.
\\

{\bf The rate function $\Psi(z)$ from the saddle point}. The value of $\Psi(z)$ defined in \eqref{eq:gener}
is then obtained from the saddle point value in \eqref{spvalue}. One has
\be 
\Psi(z) = \left[ S[\tilde P,\tilde Q] + z  \int_\xi^{+\infty} \rmd x \, \tilde Q(x,1) \right]|_{\rm sp}
\ee 
where $\tilde P,\tilde Q$ must be replaced by the $z$ dependent solutions of the 
system \eqref{DNLS2} with boundary conditions \eqref{inittilde}. Taking a derivative w.r.t. $z$ and using the
saddle point conditions, only the explicit derivation w.r.t. $z$ remains, and one obtains the formula \eqref{Psi} given in the text
\be \label{Psi2} 
\Psi'(z) = \int_{\xi}^{+\infty} \rmd x \, \tilde Q(x,1)  = \int_{0}^{+\infty} \rmd x \, Q(x,1) e^{- \frac{1}{2} x \xi - \frac{\xi^2}{4}} 
\ee 
where $Q(x,1)$ is the $z$-dependent solution of the interpolating system \eqref{interpolating2} with boundary conditions
\eqref{init}. Since by definition $\Psi(0)=0$ this equation is sufficient to
obtain $\Psi(z)$ if the r.h.s. is known as a function of $z$.

\section{Direct scattering solution for the interpolating system}
\label{sec:scatt1} 

In this section we derive the formula \eqref{bbt} and \eqref{aat} for the scattering amplitudes
$\{a(k),\tilde a(k),b(k),\tilde b(k)\}$ given in the text.\\

{\bf Equation for $\bar{\phi}$ at $t=1$ }. This equation allows to obtain the relations involving $\tilde a(k)$ and $\tilde b(k)$.
We call $\bar{\phi}_{1,2}(x,t)$ the two components of $\bar{\phi}$ (the dependence in $k$ is implicit).
Let us recall that at $x \to -\infty$, $\bar \phi \simeq (0,-e^{\I k x/2})^\intercal$. The first equation of
the Lax pair $\partial_x \vec v= U_1 \vec v$ with $\Vec{v}=e^{- k^2 t/2} {\bar{\phi}}$ reads in components at $t=1$,
from \eqref{eq:LaxPairU} and using that $R(x,1)= \Lambda \delta(x)$ 
\begin{equation} \label{eq:barphi12} 
\p_x (e^{\I \frac{k}{2} x} \bar{\phi}_1) =- (g + \I \beta k)  \Lambda \delta(x)  \bar{\phi}_2 e^{\I \frac{k}{2} x}
\quad , \quad \p_x ( e^{- \I \frac{k}{2} x} \bar{\phi}_2) = Q(x,1)\bar{\phi}_1 e^{- \I \frac{k}{2} x} \\
\end{equation}

Let us integrate the first equation from $x=-\infty$ to $x$. Since $\bar{\phi}_1$ vanishes at $x=-\infty$ it gives
\be \label{eq:phi11} 
\bar{\phi}_1(x,1)=- (g + \I \beta k) \Lambda e^{-\I \frac{k}{2} x}\Theta(x) \bar{\phi}_2(0,1)
\ee 
Taking the limit $x \to +\infty$, we thus obtain
\be \label{eq:btilde11} 
\tilde{b}(k,t=1)=- (g + \I \beta k) \Lambda \bar{\phi}_2(0,1) 
\ee
To determine $\bar{\phi}_2(0,1)$ we can integrate the second equation in \eqref{eq:barphi12}, which gives, using 
\eqref{eq:phi11} and \eqref{eq:btilde11}
\begin{equation}
\begin{cases} \label{eq:phibsolu}
e^{-\I \frac{k}{2} x}\bar{\phi}_2(x,1)=\bar{\phi}_2(0,1) + \tilde{b}(k,1)\int_{0}^x \rmd x' Q(x',1)e^{-\I k x'}, \quad x>0\\
\bar{\phi}_2(x,1)=-e^{\I \frac{k}{2} x}, \quad x<0
\end{cases}
\end{equation}
where in the second equation we have used that $\bar{\phi}_2(x,1)\simeq -e^{\I \frac{k}{2} x}$ for $x \to -\infty$.
Assuming continuity of $\bar{\phi}_2(x,1)$ at $x=0$, this leads to $\bar{\phi}_2(0,1)=-1$ and to
\be  \label{eq:tildeb} 
\tilde{b}(k,t=1)= (g + \I \beta k) \Lambda \quad   \Rightarrow \quad \tilde b(k)= (g + \I \beta k) \Lambda e^{-k^2} 
\ee 
since we recall that $\tilde b(k,t)=\tilde b(k) e^{k^2 t}$. Taking the $x\to +\infty$ limit
of \eqref{eq:phibsolu} and using the asymptotics \eqref{eq:plusinfinity} we also obtain the relation 
\begin{equation} \label{eq:tildea} 
\tilde{a}(k,1) = \tilde a(k) = 1- (g + \I \beta k) \Lambda \int_{0}^{+\infty} \rmd x' Q(x',1)e^{- \I k x'}
\end{equation}

{\bf Equation for $\phi$ at $t=0$ }. This equation allows to obtain the relations involving $a(k)$ and $b(k)$.
We call $\phi_{1,2}(x,t)$ the two components of $\phi$ (the dependence in $k$ is implicit).
Let us recall that at $x \to -\infty$, $\phi \simeq (e^{- \I k x/2},0)^\intercal$. The first equation of
the Lax pair $\partial_x \vec v= U_1 \vec v$ with $\Vec{v}=e^{k^2 t/2} \phi$ reads in components at $t=0$,
from \eqref{eq:LaxPairU} and using that $Q(x,1)= \delta(x)$. 
\begin{equation}
\begin{split}
\partial_x (e^{\I \frac{k}{2} x}\phi_1)&=- (g + \I \beta k) R(x,0) \phi_2 e^{\I \frac{k}{2} x}, \qquad \partial_x (e^{-\I \frac{k}{2} x}\phi_2) = \delta(x) \phi_1 e^{-\I \frac{k}{2} x}
\end{split}
\label{eq:SuppMatLaxPairEqGeneral}
\end{equation}
Integrating the second equation of \eqref{eq:SuppMatLaxPairEqGeneral} from $x=-\infty$ to $x$. 
Since $\phi_2$ vanishes at $x=-\infty$ it gives
\be \label{eq:phi12n} 
\phi_2(x,0)= e^{\I \frac{k}{2} x} \Theta(x) \phi_1(0,0)
\ee 
Taking the limit $x \to +\infty$, we thus obtain
\be \label{eq:b11n} 
b(k,t=0)= \phi_1(0,0) 
\ee
To determine $\phi_1(0,0)$ we can integrate the first equation in \eqref{eq:SuppMatLaxPairEqGeneral}, which gives, using 
\eqref{eq:phi12n} and \eqref{eq:b11n}
\begin{equation}
\begin{cases} \label{eq:phibsolu2}
e^{\I \frac{k}{2} x} \phi_1(x,0)=\phi_1(0,0) - (g + \I \beta k) b(k,0) \int_{0}^x \rmd x' R(x',0)e^{ \I k x'}, \quad x>0\\
\phi_1(x,0)= e^{- \I \frac{k}{2} x}, \quad x<0
\end{cases}
\end{equation}
where in the second equation we have used that $\phi_1(x,0)\simeq e^{\I \frac{k}{2} x}$ for $x \to -\infty$.
Assuming continuity of $\phi_1(x,0)$ at $x=0$, this leads to $\phi_1(0,0)=1$ and to
\be  \label{eq:bfinal} 
b(k,t=0)= b(k) = 1 
\ee 
Taking the $x\to +\infty$ limit
of \eqref{eq:phibsolu2} and using the asymptotics \eqref{eq:plusinfinity} we also obtain the relation 
\begin{equation} \label{eq:tildea-R} 
a(k,0) = a(k) = 1- (g + \I \beta k)  \int_{0}^{+\infty} \rmd x' R(x',0)e^{\I k x'}
\end{equation}
At this stage we can use the symmetry \eqref{eq:SymmetrySolution} and obtain 
\begin{equation} \label{eq:tildea2} 
a(k) = 1- (g + \I \beta k) \Lambda \int_{-\infty}^{0} \rmd x' Q(x',1) e^{- \I k x'}
\end{equation}
which completes the derivation of the equations \eqref{aat} and \eqref{bbt} in text. Alternatively one may derive \eqref{eq:tildea2} without using the symmetry \eqref{eq:SymmetrySolution}
by considering the equation for $\phi$ at $t=1$. We now present that derivation.
\\

{\bf Equation for $\phi$ at $t=1$ }. This equation allows to obtain $a(k)$ in \eqref{eq:tildea2}. 
Let us recall that at $x \to -\infty$, $\phi \simeq (e^{-\I k x/2},0)^\intercal$.
The first equation of the Lax pair, $\partial_x \vec v= U_1 \vec v$ with $\Vec{v}=e^{k^2 t/2} {\phi}$ as given
in the text now reads, in components and at $t=1$, using that $R(x,1)= \Lambda \delta(x)$
\begin{equation} \label{2eq} 
\begin{split}
\partial_x ( e^{\I \frac{k}{2} x}\phi_1 ) &=- (g + \I \beta k) \Lambda  \delta(x) \phi_2 e^{\I \frac{k}{2} x}, \qquad \partial_x ( e^{-\I \frac{k}{2} x}\phi_2) =Q(x,1)\phi_1 e^{-\I \frac{k}{2} x}
\end{split}
\end{equation}
Integrating these two equations, and using the asymptotics \eqref{eq:plusinfinity} at $x \to +\infty$ we obtain
\begin{equation}
\begin{split} \label{solusolu} 
\phi_1(x,1)&=e^{-\I \frac{k}{2} x}(\Theta(-x)+a(k)\Theta(x)), \quad a(k)-1= - (g + \I \beta k) \Lambda \phi_2(0,1)\\
\phi_2(x,1)&=e^{\I \frac{k}{2} x }\int_{-\infty}^x \rmd x' Q(x',1)e^{- \I k x'}(\Theta(-x')+a(k)\Theta(x'))
\end{split}
\end{equation}
where we used that $a(k,t)=a(k)$, see the main text. Setting $x=0$ in the second equation we obtain the relation displayed in the text
\begin{equation} \label{eq:a} 
\phi_2(0,1)=\int_{-\infty}^0 \rmd x' Q(x',1)e^{-\I k x'}, \quad 
a(k)=1- (g + \I \beta k) \Lambda \int_{-\infty}^0 \rmd x' Q(x',1)e^{- \I k x'}
\end{equation}

\section{Details of the calculation of the scattering amplitudes}
So far the scattering amplitudes $\{ a, \tilde{a}\}$ have been expressed as half-Fourier transforms in Eqs.~\eqref{eq:tildea} and \eqref{eq:a}. To determine them more explicitly, one wants to solve Eq.~\eqref{conservation}, namely the normalization relation of the scattering amplitudes, which read here
\be  \label{eq1} 
a(k) \tilde a(k) = 1- (g+\I \beta k)\Lambda e^{-k^2} 
\ee
where $a(k)$ and $\tilde a(k)$ satisfy Eq.~\eqref{aat}, which we recall  reads 
\be \label{aat2} 
a(k) = 1- (g + \I \beta k) \Lambda Q_-(k) \, , \quad  \tilde a(k) = 1- (g + \I \beta k) \Lambda Q_+(k) \, , \quad Q_\pm(k)=\int_{\mathbb{R}^\pm} \rmd x \, Q(x,1) e^{-\I k x} \, .
\ee 
Clearly for $k$ complex, $Q_+(k)$, hence $\tilde a(k)$, is analytic in the lower half-plane, and $Q_-(k)$, hence $a(k)$, is analytic in the upper half-plane.  Now we define the parametrization
\bea
&& a(k) = a(\infty)e^{\Phi_+(k)} \; , \quad  a(\infty) = 1+\beta \Lambda Q(0^-,1)\; , \\
&& \tilde{a}(k)= \tilde{a}(\infty) e^{\Phi_-(k)} \; , \quad \tilde{a}(\infty) = 1-\beta \Lambda Q(0^+,1) \; ,  
\eea
where $a(\infty)$ and $\tilde a(\infty)$ where obtained in Eq.~\eqref{largek}, so that $\Phi_\pm(k) \to 0$ as $k \to \pm \infty$. 
One can thus rewrite \eqref{eq1} for real $k$ as 
\begin{equation}
   1-(g+\I \beta k)\Lambda e^{-k^2} = e^{\Phi_+(k)} e^{\Phi_-(k)} 
\end{equation}
where $e^{\Phi_\pm(k)}$ are analytic respectively in the UHP/LHP. This is a typical Riemann-Hilbert \cite{RiemannHilbert1} or Wiener-Hopf problem. 
In some domain, taking the logarithm of this equation, it can be written as
\be  \label{sum} 
\log( 1-(g+\I \beta k)\Lambda e^{-k^2}  ) = \Phi_+(k)+\Phi_-(k) + 2 \I \pi n(k)
\ee 
for some integer-valued function $n(k)$. One can check that for $\hat g=\Lambda g <1$ the l.h.s. of \eqref{sum} is analytic in a strip 
around the real axis in $k$ (see Section~\ref{app:PsiBranch}) and decays fast at infinity along the real axis. One can thus
apply e.g. \cite[Theorem~3.1]{RiemannHilbert2} (see  \cite[Chapter~1.3]{Noble58}), which implies that 
\begin{equation} \label{Mpm2} 
    \Phi_{\pm}(k)= \pm \int_\R \frac{\rmd q}{2\I \pi}\frac{\log(1-(g+\I \beta q)\Lambda e^{-q^2} ) }{q-k\mp \I 0^+}
\end{equation}
with $n(k)=0$ in the strip. Hence for $\hat g=\Lambda g <1$ the multi-valuation occurs only outside this strip.
The formula for $\Phi_+(k)$ is valid only in the UHP and the one for $\Phi_-(k)$ is valid only in the LHP
(at least in a strip around the real axis).
This recovers Eq.~\eqref{Mpm} in the text.
\\

Formally, one can also use as in \cite{NaftaliDNLS} the well-known Sokhotskyi–Plemelj formula 
\begin{equation} \label{f} 
    \int_\R \frac{\rmd q}{2\I \pi}\frac{f(q)}{q-k \pm \I 0^+}=\dashint_\R \frac{\rmd q}{2\I \pi}\frac{f(q)}{q-k}\mp \frac{1}{2}f(k)
\end{equation}
which leads to the decomposition of a general function $f(k)$ 
\be \label{identity} 
f(k) = \int_\R \frac{\rmd k'}{2\I \pi}\frac{f(k')}{k'-k - \I 0^+} - \int_\R \frac{\rmd k'}{2\I \pi}\frac{f(k')}{k'-k + \I 0^+}
\ee 
in parts which are analytic in the UHP and LHP respectively.  
\\

{\bf Validity}. The results above are valid for $\Lambda g<1$. For their continuation beyond that
domain see Section~\ref{app:PsiBranch} below.
\\

{\bf Recovering the case $\beta=0$}. If we set $\beta=0$ in the interpolating system \eqref{interpolating}
we obtain the $\{P,Q\}$ system extensively discussed in Refs.~\cite{UsWNT2021,UsWNTFlat2021}. Let us show that one then recovers the solution
obtained in our previous work \cite{UsWNT2021}. There, we studied a more general initial condition $Q(x,t=0)=Q_0(x)$,
and to identify we must set $g \to g \Lambda$ (since there $R(x,t=1)=\delta(x)$ while here $R(x,t=1)=\Lambda \delta(x)$
while there $\Lambda$ is set to unity). Taking this into account there we obtained $\tilde b(k)=g e^{-k^2}$ which agrees with \eqref{bbt}, and
\be 
a(k) = \sqrt{1 - g b(k) \Lambda e^{-k^2}} e^{- \I \varphi(k) } \quad , \quad 
\varphi(k) = \dashint_\R \frac{\rmd q}{2 \pi} \frac{1}{q- k} \log(1 - g b(q) \Lambda e^{-q^2} )
\ee 
together with $\tilde a(k)=a(-k)$ for real $k$, and $\varphi(-k)=-\varphi(k)$. 
It is easy to see that it agrees with \eqref{Mpm2} for $\beta=0$ using
\eqref{f} with $f(k)= \log( 1 - g b(k) \Lambda e^{-k^2})$, that is, in that case for $k\in \R$
\be 
\Phi_\pm(k) = - \I \varphi(\pm k) + \frac{1}{2} \log( 1 - g b(k) \Lambda e^{-k^2}) 
\ee 
Of course here we restricted to the special case of the droplet initial condition $Q_0(x)=\delta(x)$, 
where $b(k)=1$ as found in \cite{UsWNT2021} and recovered here in Eq.~\eqref{bbt}.\\

{\bf General initial condition for $\beta \neq 0$}. Extending the previous discussion we see that the
solution of the interpolating system for a more general initial condition $Q(x,0)=Q_0(x)$ reads
\be 
\Phi_{\pm}(k)= \pm \int_\R \frac{\rmd q}{2\I \pi}\frac{\log(1-  (g+\I \beta q)\Lambda b(q) e^{-q^2} ) }{q-k\mp \I 0^+}
\ee 
where there is a map between $Q_0(x)$ and $b(q)$ which can be obtained by solving the scattering problem.  

\section{Bounds and symmetries for $\Psi(z)$}

We give here some properties of the function $\Psi(z)$. Since we start from its definition \eqref{eq:gener}
we are dealing here with what we call in the text the "optimal" $\Psi(z)$, also given by the minimization
\eqref{Legendre0}. From its definition \eqref{eq:gener} the expansion of $\Psi(z)$ in powers of $z$ around $z=0$ gives the cumulants of $Z=Z(Y,T)$
\be \label{Psicum} 
\Psi(z) = - \frac{1}{\sqrt{T}} \log \overline{\exp(- z \sqrt{T} Z)} =
- \sum_{p \geq 1} \frac{(-z)^p}{p!} T^{\frac{p-1}{2}} \overline{Z^p}^c 
\ee 
hence the leading behavior of each cumulants at large time is given by
\be 
\overline{Z^p}^c \simeq (-1)^{p+1} T^{\frac{1-p}{2}} \Psi^{(p)}(0) 
\ee 

On the other hand, taking derivatives of $\Psi(z)$ w.r.t $z$ for any $z$ lead to 
\be 
\Psi'(z) = \langle Z \rangle_z \quad , \quad \Psi''(z) = - \sqrt{T} \langle Z^2 \rangle^c_z \quad , \quad 
\langle {\cal O}  \rangle_z = \frac{ \overline{{\cal O} \exp(- z \sqrt{T} Z)} }{ \overline{\exp(- z \sqrt{T} Z)} }
\ee 
where the expectation values are w.r.t. the tilted measure also defined in the text. 
Since the random variable $Z$ obeys $0< Z < 1$ it implies 
\be 
0< \Psi'(z) < 1 \, , \quad \Psi''(z) < 0 \, .
\ee 
The function $\Psi(z)$ must thus be concave. Note that some of the branches obtained in the text 
are not concave hence they do not appear in the optimal $\Psi(z)$. In such cases there is
instead a jump of $\Psi'(z)$ from one branch to another one, for $z=z^*$. As discussed in the text, at this point 
the tilted measure has two degenerate maxima hence the fluctuations are
anomalously large, $\sqrt{T} \langle Z^2 \rangle^c_{z=z^*}=+\infty$.
\\

Let us now make the dependence in $\xi$ apparent and denote $Z_\xi = Z(Y,T)$. By definition one has 
\be 
Z_\xi = \int_\xi^{+\infty} \rmd y \, q_\eta(y,T) \quad , \quad 1- Z_{-\xi} = \int_{-\infty}^{-\xi}  \rmd y \, q_\eta(y,T) 
\ee 
where $0< Z_\xi < 1$. The equation \eqref{FP} is invariant by $y \to -y$ and $\eta(y,\tau) \to - \eta(-y,-\tau)$,
which leaves the PDF of the noise invariant, hence $Z_\xi$ and $1- Z_{-\xi}$ have the same PDF. 
This observation inserted into \eqref{eq:gener} gives 
\be \label{eq:gener22} 
\overline{ \exp( - z \sqrt{T} Z_\xi )  } \sim \exp(- \sqrt{T} \Psi_\xi(z)) 
= \overline{ \exp( - z \sqrt{T} (1- Z_{-\xi}) )  } 
= \exp(- \sqrt{T} (z + \Psi_{-\xi}(-z))) 
\ee 
hence it
implies the symmetry given in the text
\be 
\Psi_{-\xi}(z)= \Psi_{\xi}(-z) + z 
\ee 

{\bf Remark}. To measure $Z_\xi$ and $-Z_{-\xi}$ one can use the DNLS equation with boundary condition $P(x,1)=\Theta(x-\xi)$ i.e. $R(x,1)=\delta(x-\xi)$
for the first, and $P(x,1)=\Theta(-x-\xi)$ i.e. $R(x,1)=- \delta(-x-\xi)$ for the second. Using the symmetry $x \to -x$
and $R \to -R$ one arrives at the same conclusion.

\section{Derivation of the rate function $\Psi(z)$ -- main branch}

Let us give some more details on how \eqref{PsiFinal0} in the text is obtained. 
Taking a derivative w.r.t. $k$ of \eqref{rel2} at $k=\frac{\I g}{\beta}$ and
using \eqref{rel3} one obtains
\be \label{der1} 
\I \beta  \Lambda Q_\mp\left(\frac{\I g}{\beta}\right) = - \Phi_\pm'\left(\frac{\I g}{\beta}\right) 
\ee
Let us verify that this is consistent with the second equality in \eqref{psi2} (which comes from
the conservation of probability). 
For that let us first recall that (from \eqref{sum} with $n(k)=0$) 
\bea
\log(1-(g+\I \beta k)\Lambda e^{-k^2}) = \Phi_+(k) + \Phi_-(k) 
\eea 
where $\Phi_{\pm}(k)$ are given in \eqref{Mpm2} for $k$ in the complex upper/lower half planes (at least in a strip around the real axis).\\

Taken at $k=\frac{\I g}{\beta}$ (which always belongs to the strip) it again shows that $\Phi_\pm(\frac{\I g}{\beta})$ are 
opposite quantities. Taking a derivative w.r.t. $k$ at $k=\frac{\I g}{\beta}= - \I \frac{\xi}{2}$ one obtains
\bea \label{sumsum} 
- \I \beta \Lambda e^{\xi^2/4} = \Phi'_+\left(\frac{\I g}{\beta}\right) + \Phi'_-\left(\frac{\I g}{\beta}\right) 
\eea  
which is exactly equivalent to the second equality in \eqref{psi2}. \\

One must be careful in computing $\Phi_\pm'(k)$ for $k=\frac{\I g}{\beta}= - \I \xi/2$ since the formula \eqref{Mpm} 
given in the text and above in \eqref{Mpm2} for $\Phi_\pm(k)$ are valid only for $k$ in the UHP/LHP respectively. There are thus two cases:

\begin{enumerate}
    \item If $g/\beta>0$, i.e. $\xi<0$, we can use \eqref{Mpm} for $\Phi_+$ for ${\Im}(k) \geq 0$ and one obtains
\be
\Phi_+'\left(\frac{\I g}{\beta}\right) = \int_\R \frac{\rmd q}{2\I \pi}\frac{\log(1-(g+\I \beta q)\Lambda e^{-q^2} )}{(q-\frac{\I g}{\beta}  )^2}
\ee 
while, using \eqref{sumsum} one has
\be 
 \Phi_-'\left(\frac{\I g}{\beta}\right) = - \int_\R \frac{\rmd q}{2\I \pi}\frac{\log(1-(g+\I \beta q)\Lambda e^{-q^2} )}{(q-\frac{\I g}{\beta}  )^2} - \I \beta \Lambda e^{\xi^2/4} \Theta(-\xi) 
\ee 
\item  If $g/\beta<0$, i.e. $\xi>0$, we can use \eqref{Mpm} for $\Phi_-$ for ${\Im}(k) \leq 0$ and one obtains
\be
\Phi_-'\left(\frac{\I g}{\beta}\right) = - \int_\R \frac{\rmd q}{2\I \pi}\frac{\log(1-(g+\I \beta q)\Lambda e^{-q^2} )}{(q-\frac{\I g}{\beta}  )^2}
\ee 
while, using \eqref{sumsum} one has
\be
 \Phi_+'\left(\frac{\I g}{\beta}\right) =  
 \int_\R \frac{\rmd q}{2\I \pi}\frac{\log(1-(g+\I \beta q)\Lambda e^{-q^2} )}{(q-\frac{\I g}{\beta}  )^2} - \I \beta \Lambda e^{\xi^2/4} \Theta(\xi) 
\ee 
\end{enumerate}

Putting the two cases together, we obtain from the equality \eqref{der1}  
\bea \label{appa} 
 \I \beta  \Lambda Q_\mp\left(\frac{\I g}{\beta}\right)  = - \Phi_\pm'\left(\frac{\I g}{\beta}\right)  = \mp \dashint_\R \frac{\rmd q}{2\I \pi}\frac{\log(1-(g+\I \beta q)\Lambda e^{-q^2} )}{(q - \frac{\I g}{\beta})^2}
+  \I \beta \Lambda e^{\xi^2/4} \Theta(\pm \xi) 
  \eea
a result which remains true for $g=\xi=0$ provided the integrals are then interpreted as principal values and that 
we use the convention $\Theta(0)=1/2$. \\

Now using \eqref{psi2} and inserting $\beta=-1$, $g=-\beta \frac{\xi}{2}$ and $\beta \Lambda = z e^{- \xi^2/4}$
we obtain the result \eqref{zPsiFinal0} in the text, from which \eqref{PsiFinal0} is obtained upon
integration over $z$. 
\\

{\bf Remark on the Heaviside function}. The appearance of the term $\Theta(\pm \xi)$ in  \eqref{appa} can also be seen as follows. 
Let us expand \eqref{Mpm} in series of $\Lambda$
\be 
\Phi_\pm'(k) = \mp \sum_{n \geq 1}  \frac{(\I \beta \Lambda)^n}{n} 
\int_\R \frac{\rmd q}{2\I \pi} \frac{(q- \frac{\I g}{\beta})^n  e^{-n q^2} }{(q-k \mp \I 0^+)^2} 
\ee 
Taking $k=\frac{\I g}{\beta}$ we can neglect the term $\pm \I 0^+$ except for $n=1$. The term $n=1$ is
\be
  \mp (\I \beta \Lambda)
\int_\R \frac{\rmd q}{2\I \pi} \frac{  e^{-q^2} }{q + \frac{\I \xi}{2} \mp \I 0^+} =\mp (\I \beta \Lambda)
\dashint_\R \frac{\rmd q}{2\I \pi} \frac{  e^{-q^2} }{q + \frac{\I \xi}{2} } -\I \beta \Lambda e^{\xi^2/4} \Theta(\pm \xi)
\ee 
\\

{\bf Recovering the typical probability}. Similarly the expansion of $z \Psi'(z)$
in powers of $z$ contains a term with a pole at $q=- \I \xi/2$ only for $n=1$. 
As shown in the text, this term recovers the typical probability $\overline{Z}=Z_{\rm typ}$.
In deriving the equation \eqref{Ztyp} we have used the identity, for real $a$
\be 
\dashint_\R \frac{\rmd q}{2 \pi}\frac{1}{\I q + a} e^{-q^2} = \frac{1}{2} e^{a^2} ({\rm Erfc}(a)-2 \Theta(-a)\ee

{\bf Range of validity}
As discussed in the text the formula for $\Psi(z)$ presented in this section is what we call the "main branch"
valid only for $z/z_c<1$ with $z_c=- \frac{2}{\xi} e^{\xi^2/4}$. As a result it only allows to determine
$\Phi(H)$ for $H<H_c$. To obtain the full solution to the problem we need to consider 
analytical continuations, to which we now turn.

\section{Analytic continuation in the case of the WNT for the KPZ equation} 

Before discussing the intricacies of the analytic continuations for the present problem, we
recall here how it works in the case of the WNT for the KPZ equation. It is necessary to do so since
we show below that at large $\xi$ the results for KPZ equation are recovered. We present further details 
than given in Ref.~\cite{UsWNT2021} as they will be very useful below. Indeed the situation in the present paper is already 
quite similar to the one for the KPZ equation
where $\Psi_{\rm KPZ}(\tilde z)$ admits a second branch for $-1 \leq \tilde z <0$. \\

For the KPZ equation one first obtains for $\tilde z \in [0,+\infty)$ 
\be 
\Psi_{\rm KPZ}(\tilde z) = \Psi_{\rm KPZ,0}(\tilde z) := - \frac{1}{\sqrt{4 \pi}} {\rm Li}_{5/2}(- \tilde z) 
\ee 
which can be continued for $\tilde z \in [-1,+\infty)$. 
The polylogarithm function $\mathrm{Li}_{5/2}(-\tilde z )$ is analytic in the complex $\tilde z$ plane except on 
a branch cut for $\tilde z \in (-\infty,-1]$. Across this branch cut it has a jump, which leads to 
\be 
\Psi_{\rm KPZ,0}(\tilde z + \I 0^+)  - \Psi_{\rm KPZ,0}(\tilde z - \I 0^+) = \Delta_{\rm KPZ}(\tilde z) \quad , \quad
\Delta_{\rm KPZ}(\tilde z) = \frac{4 }{3}\I (\log(-\tilde z))^{3/2} 
\ee
for $z \in (-\infty,-1]$.\\ 

\textbf{Analogy with the logarithm.} This situation is analogous to the study of the logarithm which admits different determination in the complex plane. Indeed, along the negative real axis, the logarithm has a jump of value $2\I \pi$. A better understanding of the logarithm is done by considering its domain of definition not in the complex plane but rather on a Riemann surface, see Fig.~\ref{fig:RiemannSheet}, where it does not have any jump. In the case of the logarithm, the Riemann surface is composed of different sheets joined by winding around the origin and the correct definition of the logarithm on the $n$-th sheet is $\log z + 2 \I \pi n$ where $\log z$ is the principal determination or main branch.

\begin{figure}[h!]
    \centering
    \includegraphics[scale=0.4]{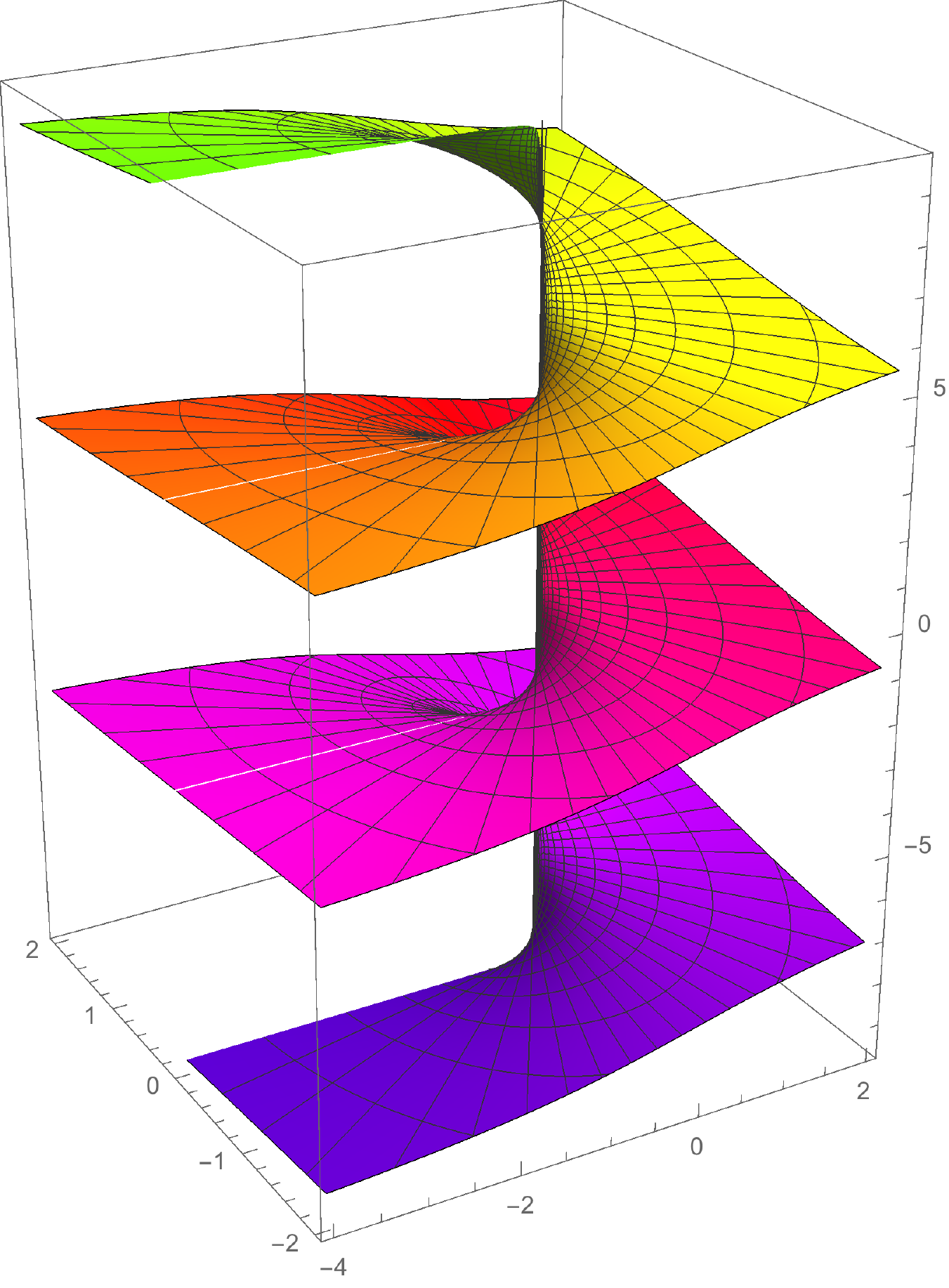}
    \caption{Plot of the Riemann surface of the logarithm $z \mapsto \log z$. This surface is composed of different sheets continuously connected is a staircase manner.}
    \label{fig:RiemannSheet}
\end{figure}

Pursuing the construction of $\Psi_{\rm KPZ}(\tilde z)$ on a Riemann surface rather than on the complex plane, we extend continuously the definition $\Psi_{\rm KPZ,0}(\tilde z)$ to the first Riemann sheet along the branch cut as follows

\begin{equation}
    \Psi_{\rm KPZ}(\tilde{z})=
    \begin{cases}
    \Psi_{\rm KPZ,0}(\tilde{z}), &\Im(\tilde{z})>0\\
    \Psi_{\rm KPZ,0}(\tilde{z})+\Delta_{\rm KPZ}(\tilde{z}), &\Im(\tilde{z})<0\\
    \end{cases}
\end{equation}
On the real axis it is multi-valued, i.e. there is a first branch for 
$\Psi_{\rm KPZ}(\tilde{z})$ given by the first line, and
a second branch given by the second line. One can now continue these two branches for
$\tilde z \in ]-1,0]$ and one finds that the second branch is  
\be 
\Psi_{\rm KPZ,0}(\tilde{z})+\Delta_{\rm KPZ}(\tilde{z}) = - \frac{1}{\sqrt{4 \pi}} {\rm Li}_{5/2}(- \tilde z) 
+ \frac{4}{3} (- \log(-\tilde z))^{3/2} 
\ee 
for $\tilde z \in ]-1,0]$.
\\

\textbf{Another route to find the analytic continuation.} One can arrive at the same result from the integral representation. Indeed, 
for $\tilde z \in ]-1,+\infty[$ one has 
\be 
\tilde z \Psi'_{\rm KPZ}(\tilde z) = 
\int_\R \frac{\rmd q}{2 \pi}\log(1+ \tilde z e^{-q^2 } ) = \tilde z \Psi^{\prime}_{0,\rm KPZ}(\tilde z) 
= - \frac{ 1}{\sqrt{4 \pi}} \mathrm{Li}_{3/2}(-\tilde z )
\label{eq:KPZ-IntegralRepPsip}
\ee

\begin{figure}[t!]
    \centering
    \includegraphics[angle=-90,scale=0.26]{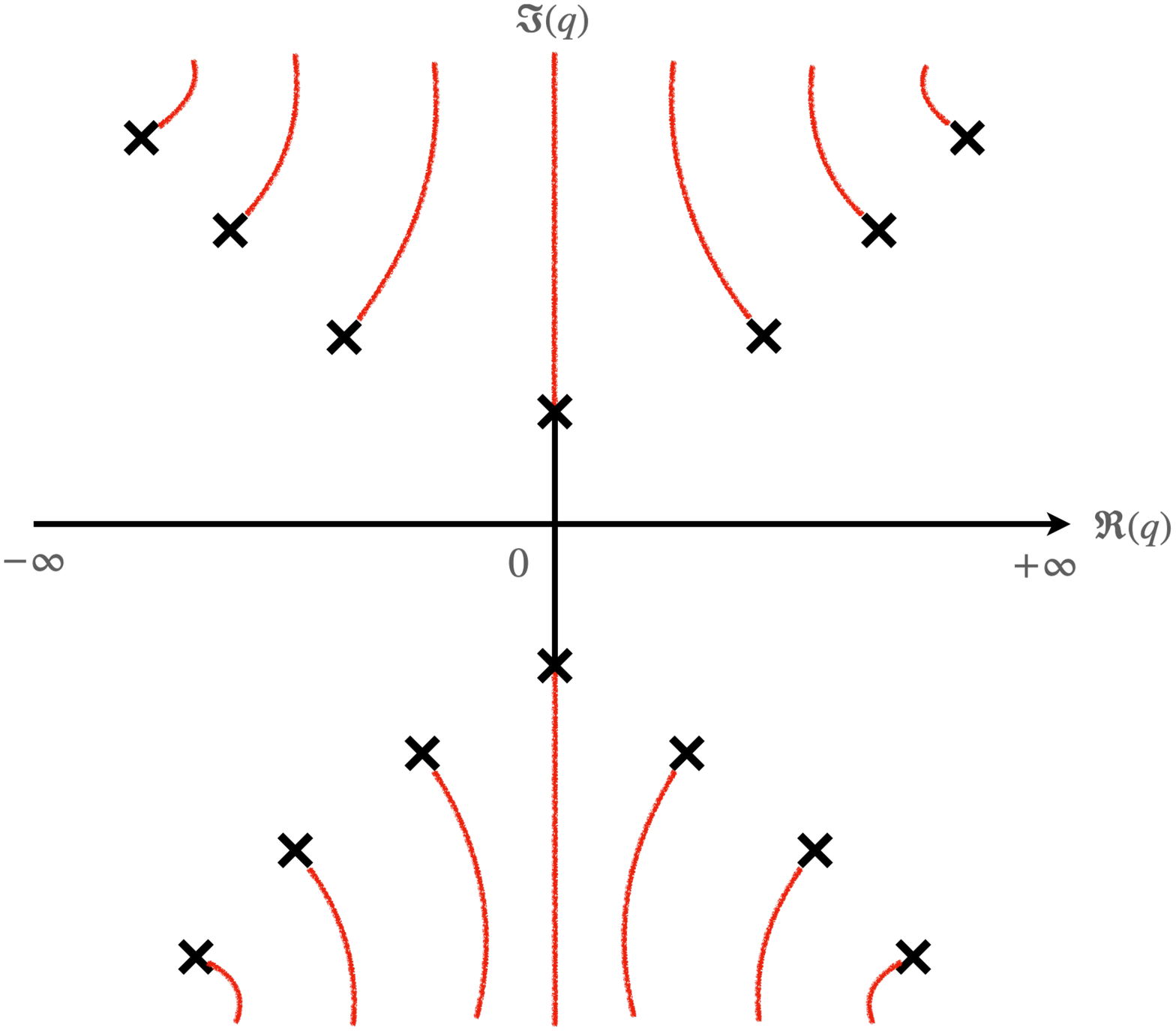}
    \includegraphics[angle=-90,scale=0.26]{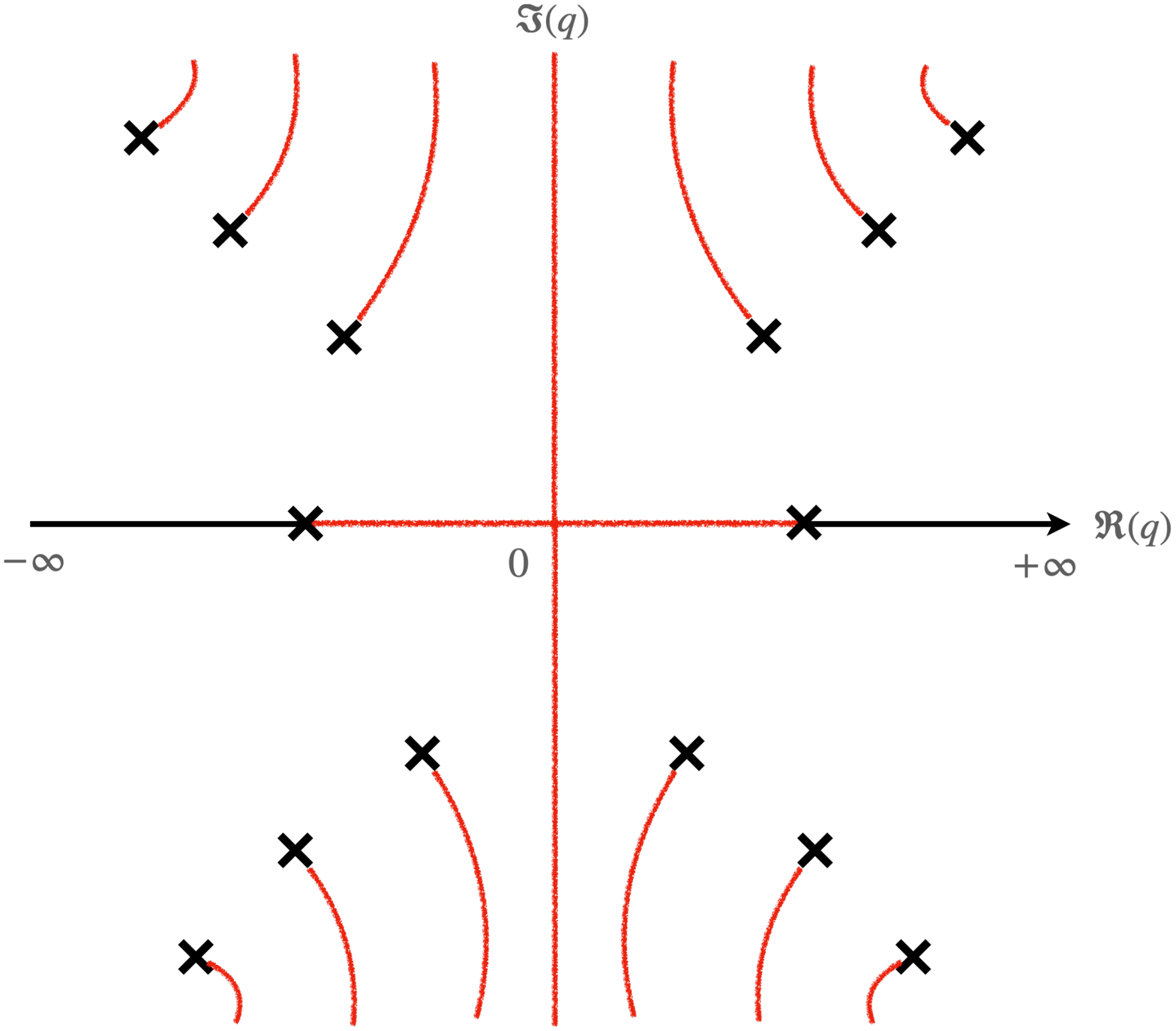}\\
    \includegraphics[angle=-90,scale=0.26]{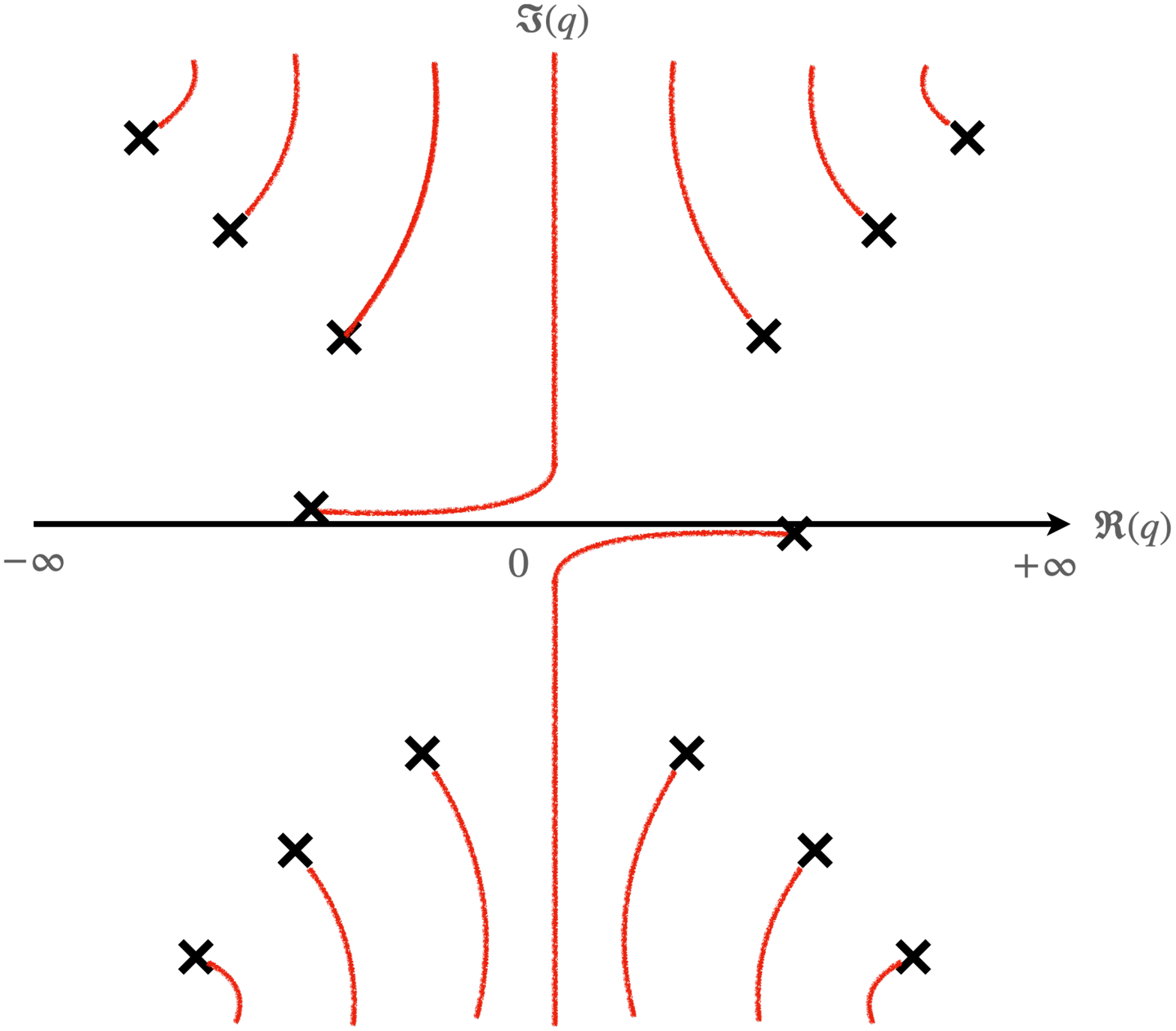}
    \caption{Schematic plot of the argument of the logarithm in Eq.~\eqref{eq:KPZ-IntegralRepPsip} in the complex $q$ plane for various values of $\tilde{z}$. \textbf{Top left.} $\tilde{z}\geq -1$, $\tilde{z}\in \R$. \textbf{Top right.} $\tilde{z}\leq -1$, $\tilde{z}\in \R$. \textbf{Bottom}  $\Re(\tilde{z})<-1$, $\Im(\tilde{z})=0^+$. The black crosses correspond to the locations where the argument $A_{\rm KPZ}(q)$ is zero and the red curves correspond to the branch cuts of $\log A_{\rm KPZ}(q)$. }
    \label{fig:ArgumentLogarithmKPZ}
\end{figure}

Let us plot the argument of the logarithm $A_{\rm KPZ}(q) = 1+ \tilde z e^{-q^2}$ in the complex $q$
plane. This is shown schematically in Fig.~\ref{fig:ArgumentLogarithmKPZ}. For $\tilde z > -1$ no branch cut crosses the real axis (integration axis).
When $\tilde z$ reaches $-1$ the
two symmetric branch cuts along the imaginary axis join. For $\tilde z <-1$ they form a "cross" (see
Fig.~\ref{fig:ArgumentLogarithmKPZ} -- top right) with ends located at $q=\pm \sqrt{\log(-1/\tilde{z})}$. 
It is impossible to integrate over the real axis without crossing them.
However suppose now we consider $\tilde z \to \tilde z \pm \I \epsilon$. We can see that
the two branch cuts then avoid each others and it is possible to deform slightly the integration contour
to avoid crossing them (see
Fig.~\ref{fig:ArgumentLogarithmKPZ} -- bottom). This is consistent 
with the function $\mathrm{Li}_{3/2}(-\tilde z )$ being analytic away from the negative real axis for $\tilde z$.\\

Now one can see that the additional 
contribution $\Delta_{\rm KPZ}(z)$ comes for $\tilde z<-1$ from the jump across the horizontal part 
of the "cross" (see
Fig.~\ref{fig:ArgumentLogarithmKPZ} -- bottom) and is precisely 
\be 
\tilde{z} \Delta_{\rm KPZ}'(\tilde{z}) = 2 \I \pi \int_{-\sqrt{-\log(-1/\tilde{z})}}^{\sqrt{-\log(-1/\tilde{z})}} \frac{\rmd q}{2 \pi} 
= 2 \I [\log(-\tilde z)]^{1/2} 
\ee 
while its continuation for $\tilde z>-1$ - which enters the second branch -- can be obtained as an integral around the complementary of the
branch cut in Fig.~\ref{fig:ArgumentLogarithmKPZ} -- top left)
\be \label{jumpprimeKPZ} 
\tilde{z} \Delta_{\rm KPZ}'(\tilde{z}) = 2 \I \pi \int_{- \I \sqrt{\log(-1/\tilde{z})}}^{\I \sqrt{\log(-1/\tilde{z})}} \frac{\rmd q}{2 \pi} 
= - 2 [-\log(-\tilde z)]^{1/2} 
\ee 

consistent with the previous argument. These considerations will be useful for the next subsection. 

\section{Analytic continuation and additional branches of the rate function $\Psi(z)$}\label{app:PsiBranch}

\begin{figure}[t!]
    \centering
    \includegraphics[angle=-90, scale=0.25]{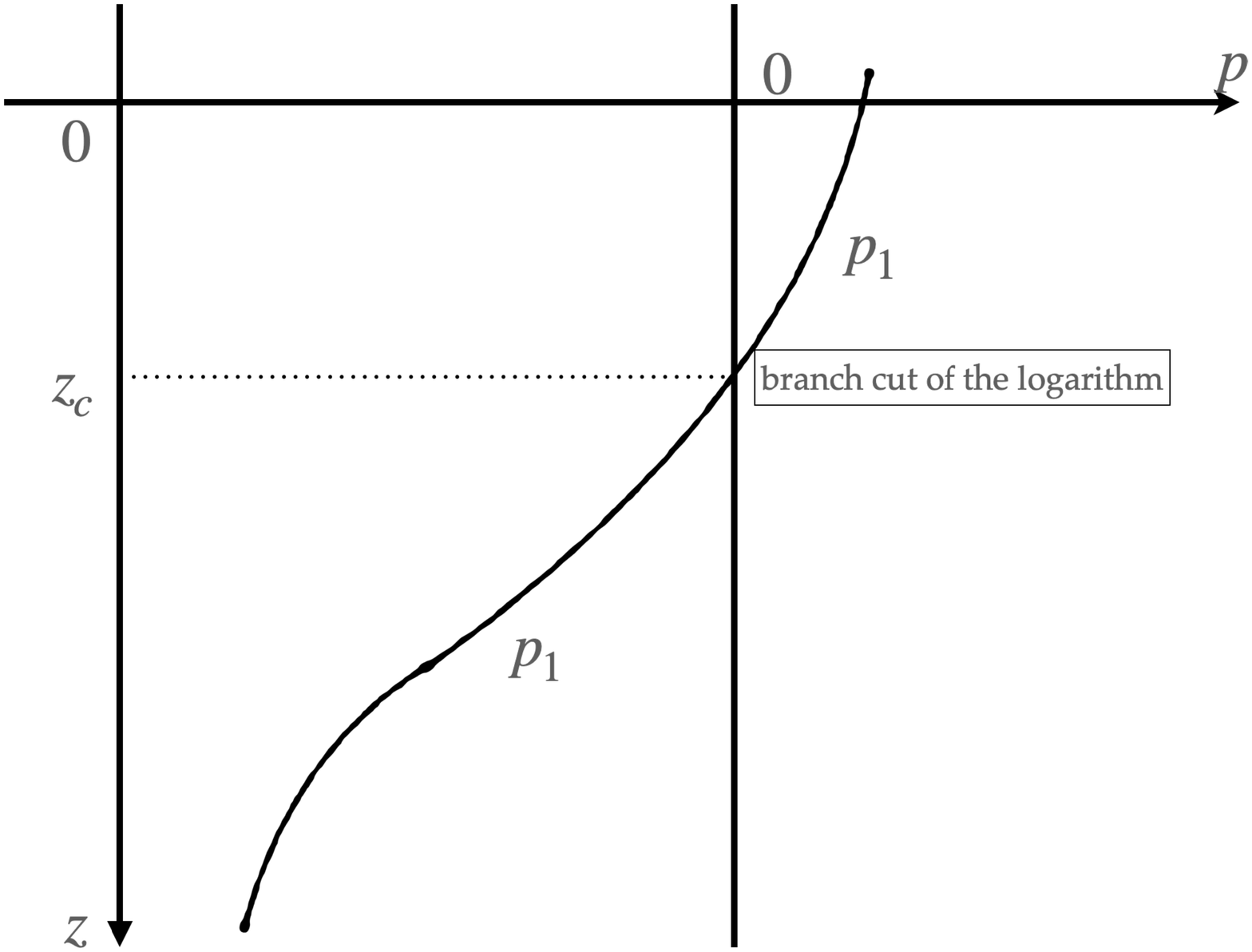}
    \includegraphics[angle=-90, scale=0.25]{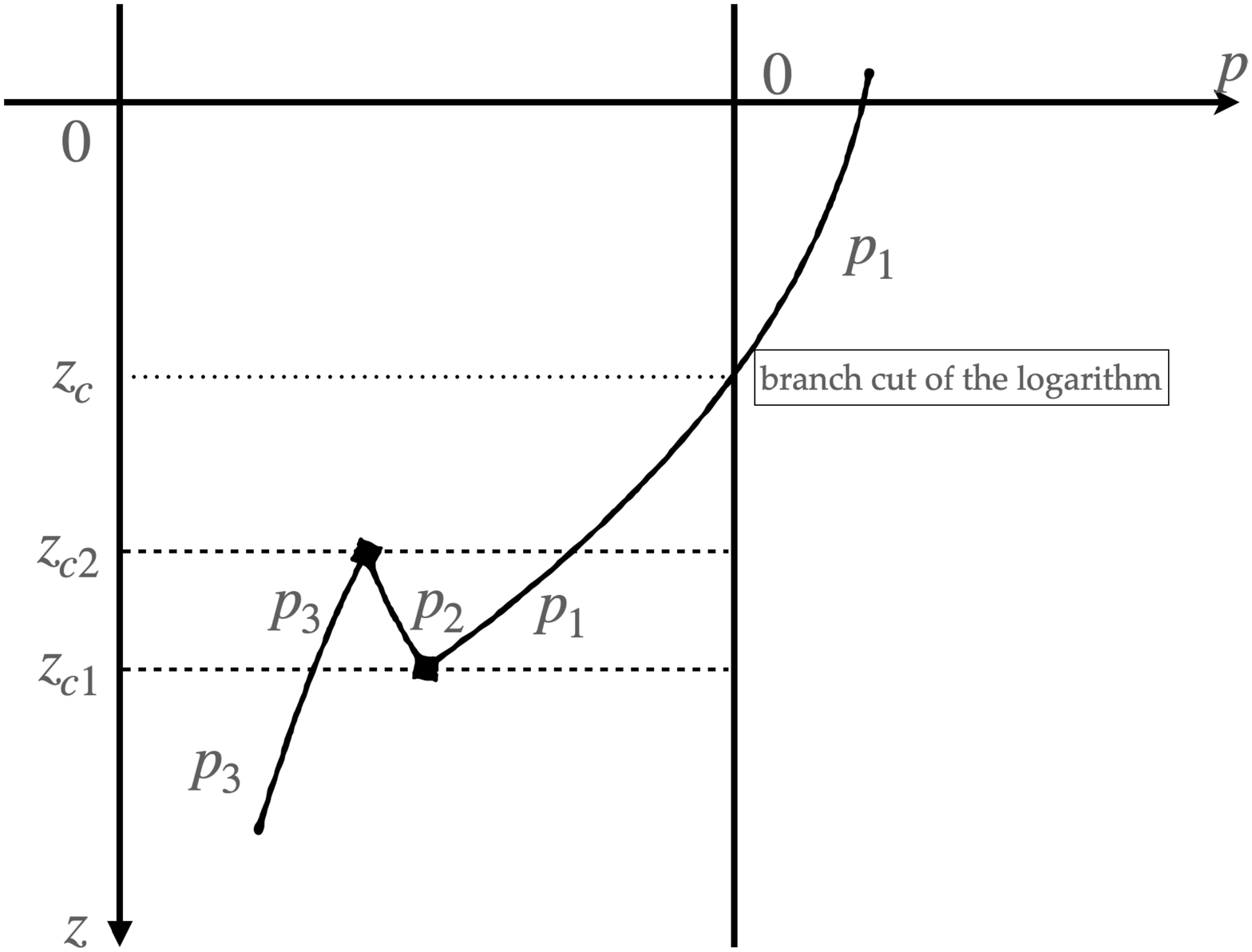}
    \includegraphics[angle=-90, scale=0.25]{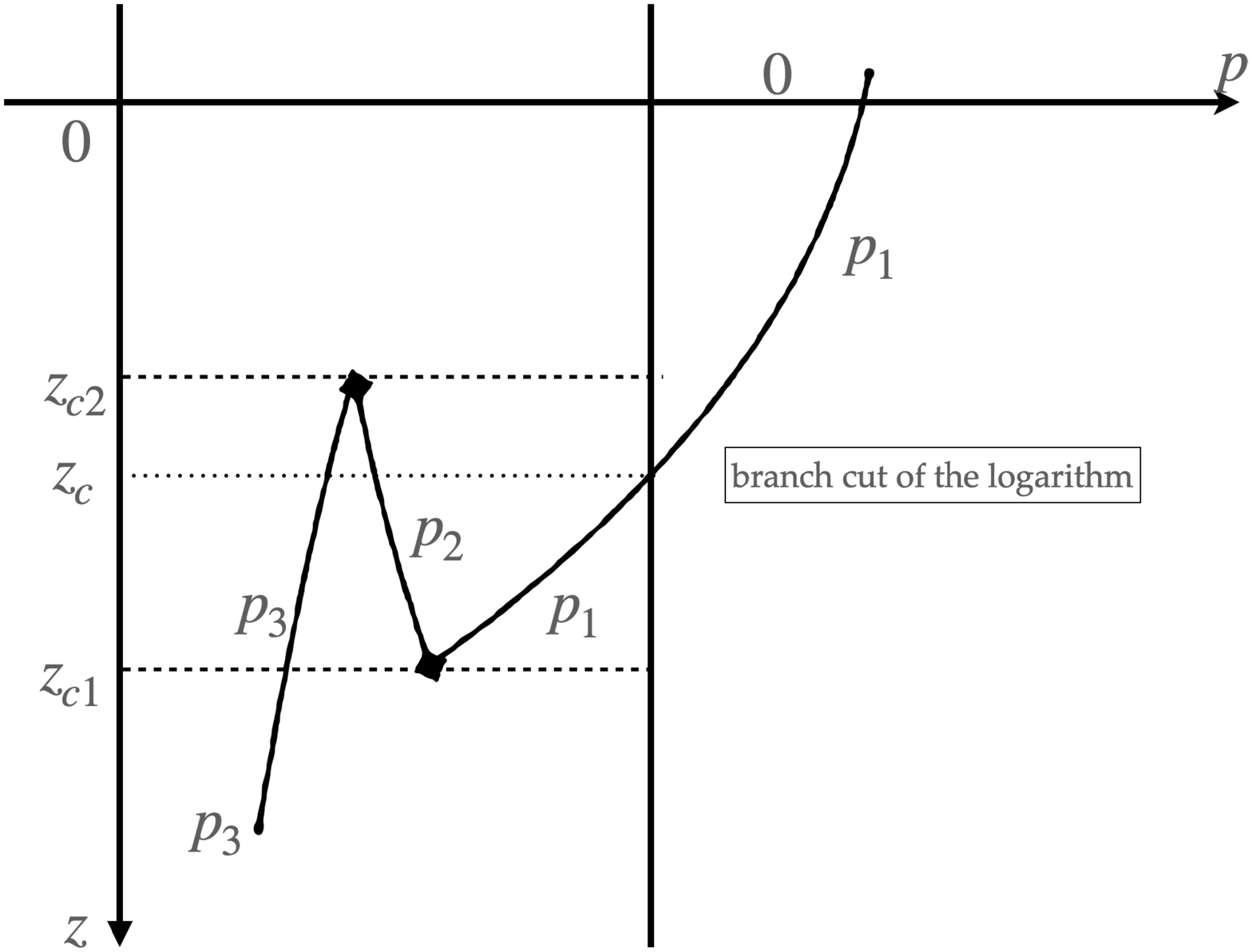}
    \caption{Schematic behavior of the zeroes $p_i$ of the equation $f_{z,\xi}(p_i)=0$ as a function of $z<0$ for $\xi>0$.
    Top Left: the case $\xi<\xi_1=\sqrt{8}$ in which case there is a single zero, which changes sign at $z=z_c$.
    Top Right: the case $\xi_1 < \xi<\xi_2$ in which case there are three zeroes $p_1>p_2>p_3$ in the interval $z \in (z_{c1},z_{c2})$
    and $z_{c2}<z_c$. 
    Bottom: the case $\xi>\xi_2$, same but now $z_{c2}>z_c$. The zeroes $p_2,p_3$ coalesce and annihilate at $z=z_{c2}$
    and $p_2,p_3$ at $z=z_{c1}$. The points $z_{c1}$ and $z_{c2}$ also serve as turning points in the
    definition of the function $\Psi(z)$.}
    \label{fig:zeroes} 
\end{figure}

\subsection{Preliminaries: solutions of Eq.~\eqref{eq:BranchCutEquation} in the text} 

As mentioned in the text, and for the discussion below about the branch cuts in the integration in the formulas \eqref{zPsiFinal0}
and \eqref{PsiFinal0} for $z \Psi'(z)$ and $\Psi(z)$, it is important to study the argument of the logarithm in 
\eqref{zPsiFinal0}, which we denote $A(q)$
\be \label{Aq} 
A(q) := 1- z (\I q - \frac{\xi}{2}) e^{-q^2 - \frac{\xi^2}{4}}
\ee
and in particular to find the points where it vanishes, i.e. the zeroes, solutions of $A(q)=0$. There are many such zeroes but it turns out, see below, that the zeroes 
on the imaginary axis are the one which play an important role. Setting $q= \I p$, it is equivalent to study
$A(\I p)$ or the function $f_{z,\xi}(p)$ defined as
\be 
f_{z,\xi}(p)=  e^{- p^2 + \frac{\xi^2}{4}} A(\I p) = e^{- p^2 + \frac{\xi^2}{4}} + z (p + \frac{\xi}{2}) 
\label{eq:CriticalEquationF}
\ee
and finding its real zeroes, $f_{z,\xi}(p)=0$,
which is Eq.~\eqref{eq:BranchCutEquation} in the text. \\

We will consider $\xi>0$ (the case $\xi<0$ can be studied from the symmetry $(\xi,z) \to (-\xi,-z)$). 
Consider also $z<0$ (see $z>0$ below). Since $f_{z,\xi}(p) \to \mp \infty$ as $p \to \pm \infty$
it has at least one real zero, but in some cases can have three. When there are three zeroes
we will denote them $p_1>p_2>p_3$ in decreasing order. They are functions of $(z,\xi)$ i.e.
$p_i=p_i(z,\xi)$. \\

The evolution of the zeroes when $z<0$ is varied is shown in Fig.~\ref{fig:zeroes}.
There are three cases depending in the values of $\xi$ which we now describe. 
In all three cases the largest zero $p_1$ vanishes for the value of $z=z_c= z_c(\xi) =-\frac{2 e^{\frac{\xi ^2}{4}}}{\xi }<0$. 
One finds that

\begin{enumerate}
    \item for $0<\xi<\xi_1=\sqrt{8}$, and for all $z<0$, there is only one zero, $p_1=p_{1}(z,\xi)$, see
    Fig.~\ref{fig:zeroes} (top left). 
    
    \item for $\xi> \xi_1$ there is an interval of values of $z$, $z \in ]z_{c1},z_{c2}[$, where there 
    are three zeroes. To find this interval one looks for double zeroes $f_{z,\xi}(p)=f_{z,\xi}'(p)=0$, i.e. 
\be \label{equa} 
e^{- p^2 + \frac{\xi^2}{4}} = - z (p + \frac{\xi}{2}) = \frac{z}{2 p}
\ee
For a given $\xi$ one can solve these conditions for the couple $(z,p)$. One finds that there are
no real solutions for $\xi<\sqrt{8}$ but that there are two solutions $(z_{c1},p_{c1})$ and $(z_{c2},p_{c2})$
for $\xi> \xi_1=\sqrt{8}$. These read, with $z_{c1}<z_{c2}$, 
\bea 
&&    z_{c1}=z_{c1}(\xi)=-\frac{1}{2} e^{\frac{1}{8} \left(\xi  \left(\xi+\sqrt{\xi ^2-8} \right)+4\right)} \left(\xi -\sqrt{\xi ^2-8}\right)
\quad , \quad p_{c1} = -\frac{1}{4} \left(\xi -\sqrt{\xi ^2-8}\right) \\
&&   z_{c2}= z_{c2}(\xi)=-\frac{1}{2} e^{\frac{1}{8} \left(\xi \left(\xi-\sqrt{\xi ^2-8}\right)  +4\right)} \left(\xi+\sqrt{\xi ^2-8} \right)
\quad , \quad p_{c2}= -\frac{1}{4}  \left(\xi+\sqrt{\xi ^2-8} \right) \label{zc2} 
\eea 
For any $\xi>\xi_1$, and as can be seen in Fig.~\ref{fig:zeroes}, the two smallest zeroes annihilate at $z=z_{c2}$
where their values are $p_2=p_3=p_{c2}$ and the two largest zeroes annihilate at $z=z_{c1}$
where their values are $p_1=p_2=p_{c1}$. Note that at $\xi=\sqrt{8}$ the interval is a single point 
and one has $z_{c1}=z_{c2}=- \sqrt{2} e^{3/2}$ and $p_{c1}=p_{c2}=-1/\sqrt{2}$.

\item It will turn out to be important below to distinguish the cases where $z_{c2}<z_c$ and $z_{c2}>z_c$, see Fig.~\ref{fig:zeroes}. 
Let us determine the value of $\xi$, denoted $\xi=\xi_2$, at which $z_c(\xi)=z_{c2}(\xi)$. Inserting $z=z_c(\xi)=-\frac{2 e^{\frac{\xi ^2}{4}}}{\xi }$ into \eqref{equa} one gets two
equations
\be 
\xi p  e^{-p^2}= - 1 \quad , \quad p^2 + \frac{p \xi}{2} = -\frac{1}{2} 
\ee 
where here $p$ should be set to $p=p_{c2}(\xi)$ given in \eqref{zc2}. Combining we obtain a closed equation
for $p \xi/2$, i.e.
\be \label{eq4} 
\frac{p \xi}{2} e^{\frac{p \xi}{2}} = - \frac{1}{2} e^{-1/2} \quad \Rightarrow \quad \frac{\xi p_{c2}(\xi)}{2} = W_{-1}\left(-\frac{1}{2 \sqrt{e}}\right)
\ee 
which using $p_{c2}(\xi)$ from \eqref{zc2} and solving for $\xi$ finally leads to $\xi= \xi_2$ with
\be 
   \xi_2 =-2 \sqrt{\frac{2}{-2 W_{-1}\left(-\frac{1}{2 \sqrt{e}}\right)-1}} W_{-1}\left(-\frac{1}{2 \sqrt{e}}\right)\simeq 3.13395\\
\ee
Note that we have discarded the other solution $\xi p_{c2}(\xi)=-1$ of \eqref{eq4} which does
not provide a solution for $\xi_2$.
Hence, we finally find that for $\xi<\xi_2$ one has $z_{c2} < z_c$ and for $\xi > \xi_2$ one has $z_c < z_{c2}$,
see Fig.~\ref{fig:zeroes}. This will be important below.
\end{enumerate}

\subsection{Continuation and branches of $\Psi(z)$ for $0< \xi<\xi_1$} 

We now study the analytical continuations and various branches of $\Psi(z)$.
Let us first recall the expressions of $\Psi(z)$ and $\Psi'(z)$ obtained in the text in \eqref{zPsiFinal0} and \eqref{PsiFinal0} for $\xi>0$
\be \label{zPsiFinal00}
 z \Psi'(z) = 
\dashint_\R \frac{\rmd q}{2 \pi}\frac{\log(1- z (\I q - \frac{\xi}{2}) e^{-q^2 - \frac{\xi^2}{4}} )}{(\I q - \frac{\xi}{2})^2} 
\quad , \quad 
\Psi(z) = -
\dashint_\R \frac{\rmd q}{2 \pi}\frac{\mathrm{Li}_2( z (\I q - \frac{\xi}{2}) e^{-q^2 - \frac{\xi^2}{4}} )}{(\I q - \frac{\xi}{2})^2} 
\ee 

The argument of the logarithm and polylogarithm is $A(q)$ defined in \eqref{Aq}. The integrand has branch cuts 
in the complex plane for $q$ when $A(q) \in \mathbb{R}^-$, i.e. is real negative. In the previous subsection
we found some of the zeroes (those on the imaginary axis) from which the branch cuts originate. There are
additional ones, and the full picture for all $\xi>0$ is shown schematically in Fig.~\ref{fig:ComplexPlane} where the zeroes of $A(q)$ are represented by crosses and the branch cuts by red lines. \\

Here we examine the simplest case $\xi<\xi_1=\sqrt{8}$. Then one finds that for $z_c <z < 0$ (top left in Fig.~\ref{fig:ComplexPlane}) no branch
cut crosses the real axis. This corresponds to the regime with
a single positive zero $p=p_1$ to Eq.~\eqref{eq:BranchCutEquation}. In that
regime the formula in \eqref{zPsiFinal00} are valid. This is the {\it main branch}. \\

For $z \leq z_c(\xi) =-\frac{2 e^{\frac{\xi ^2}{4}}}{\xi }$ the single zero $p=p_1$ becomes negative hence the branch cut along
the positive imaginary axis intersects the real axis at $q=0$. 
This is represented in Fig.~\ref{fig:ComplexPlane} (top right). In the case however, i.e. for $0< \xi < \xi_1$, it is always possible (i.e. for any $z \leq z_c$)
to deform the integration contour of $q$ away from the real axis to pass
below the branch cut (as represented on the Figure). We call this new contour $C$. This provides a natural analytical continuation to all real $z$. This leads to
\be
\label{eq:app-zpsip}
\begin{split}
 z \Psi'(z) &= \int_C \frac{\rmd q}{2 \pi}\frac{\log(1- z (\I q - \frac{\xi}{2}) e^{-q^2 - \frac{\xi^2}{4}} )}{(\I q - \frac{\xi}{2})^2} \\
& =
z \Psi'_0(z)  +  z \Delta_1'(z) \quad , \quad  z \Psi'_0(z) = \int_{\mathbb{R}} \frac{\rmd q}{2 \pi}\frac{\log(1- z (\I q - \frac{\xi}{2}) e^{-q^2 - \frac{\xi^2}{4}} )}{(\I q - \frac{\xi}{2})^2}
\end{split}
\ee
In the second line we have split the integral into an integral over the real axis which passes right through the branch cut, and a contribution denoted $z \Delta_1'(z)$ which represents the contribution of a contour around the branch cut (taking into account the $2 \I \pi$ discontinuity of the logarithm), which reads 
\begin{equation}
   z  \Delta_1'(z) = \int_0^{p_1}  \frac{\rmd p}{(p+ \xi/2)^2}=\frac{4p_1}{\xi(2p_1+\xi)}   \label{zDep1} 
\end{equation}
with $p_1=p_1(z,\xi)$. The first piece, $z \Psi'_0(z)$ is by definition the integral over 
$\mathbb{R}$ computed "naively", that is with a jump of the integrand when the argument of the logarithm crosses the
negative real axis (which occurs for $q=0$) and in such a way that the invariance under the change of variable $\I q \to - \I q$ ensures that the result is real
(in other words one can e.g. replace $\int_{\mathbb{R}} = 2 {\Re} \int_{\mathbb{R}^-}$) and not worry about the branch cut.\\

We can repeat the same procedure for the formula for $\Psi(z)$ itself in 
\eqref{zPsiFinal00}, the branch cut of 
the ${\rm Li}_2$ function being identical to the one of the logarithm with however a different value of the jump 
\be 
{\rm Li}_2(t+\I 0^+) - {\rm Li}_2(t-\I 0^+) = -2 \I \pi \log t 
\ee 
One obtains 
\be
\begin{split}
     \Psi(z) &= -
\int_C \frac{\rmd q}{2 \pi}\frac{\mathrm{Li}_2( z (\I q - \frac{\xi}{2}) e^{-q^2 - \frac{\xi^2}{4}} )}{(\I q - \frac{\xi}{2})^2}  \\
&= \Psi_0(z) + \Delta_1(z) \quad , \quad \Psi_0(z) = 
- \int_{\mathbb{R}} \frac{\rmd q}{2 \pi}\frac{\mathrm{Li}_2( z (\I q - \frac{\xi}{2}) e^{-q^2 - \frac{\xi^2}{4}} )}{(\I q - \frac{\xi}{2})^2}
\label{Psi000} 
\end{split}
\ee
with
\bea \label{De1} 
&& \Delta_1(z) =   -  \int_0^{p_1} \frac{\rmd p}{(p+ \xi/2)^2} \log( -z (p  + \frac{\xi}{2}) e^{p^2 - \frac{\xi^2}{4}} ) \\
&& = \hat \Delta(p_1(z,\xi)) \quad , \quad \hat \Delta(p) = 
\frac{1}{\xi}\left[-(\xi ^2+2) (\log (\xi )-\log (\xi +2 p))+2 p (p-\xi )-\frac{4 p}{\xi +2 p}\right] \label{defhatDe} 
\eea 
where we have defined a new function $\hat \Delta(p)$ which will be useful below.
To obtain this expression for
$\Delta_1(z)$ one can either compute the contribution of the branch cut, as done above, or
integrate the expression \eqref{zDep1} over $z$. In the latter case one uses the following differential relation
for $p_1=p_1(z)$
\begin{equation}
  \frac{\rmd  p_1}{\rmd z} =-\frac{1}{z}\frac{p_1+\frac{\xi}{2}}{1+2p_1(p_1+\frac{\xi}{2})}
\end{equation}
and write
\begin{equation}
\label{eq:FinalFormulaDelta}
\begin{split}
 \Delta_1(z)&=    -\int_z^{z_c}\frac{\rmd z'}{z'}\frac{4p_1(z')}{\xi(2p_1(z')+\xi)} 
 = \int_0^{p_1}\rmd p \frac{2p}{\xi}\frac{1+2p(p+\frac{\xi}{2})}{(p+\frac{\xi}{2})^2}
 %   &=\frac{1}{\xi}\left[-(\xi ^2+2) (\log (\xi )-\log (\xi +2 p_c(z)))+2 % p_c(z) (p_c(z)-\xi )-\frac{4 p_c(z)}{\xi +2 p_c(z)}\right]
    \end{split}
\end{equation}
which also yields \eqref{De1}, showing that the two methods agree. 
\\

The above formula are those used for the plots of $\Psi'(z)$ in the main text for $0<\xi<\xi_1$. We have checked numerically 
that for large negative $z \to -\infty$, $\Psi'(z) \to 1$ since $p_1 \to - \frac{\xi}{2} - \frac{1}{z} + o(1/z)$.
This gives confidence that this is the correct solution. 
\\

{\bf Remark}. We call $\Psi(z)=\Psi_0(z)+\Delta_1(z)$ for $z<z_c$ a new branch different from the main branch,
although in a sense they are the same branch by some choice of integration contour. The important point
here is the identification of the jump function $\Delta_1(z)$ which, as we will see now, plays an important role to
determine the several other branches for $\xi>\xi_1$.
\\

{\bf Remark.} The structure of branch cuts in the complex plane discussed here for $\xi>0$ 
is already present for $\xi=0$, although in that case $z_c=-\infty$ (for $\xi \to 0^+$) so no analytic continuation is needed. 
There is some interpretation of the corresponding zeroes of $a(k)$ and $\tilde a(k)$ in terms of additional solitonic 
solutions of the DNLS equation, as discussed in the main text. For $z>z_c$ these are presumably irrelevant
for the large deviations.

\begin{figure}[t!]
    \centering
   \includegraphics[scale=0.2]{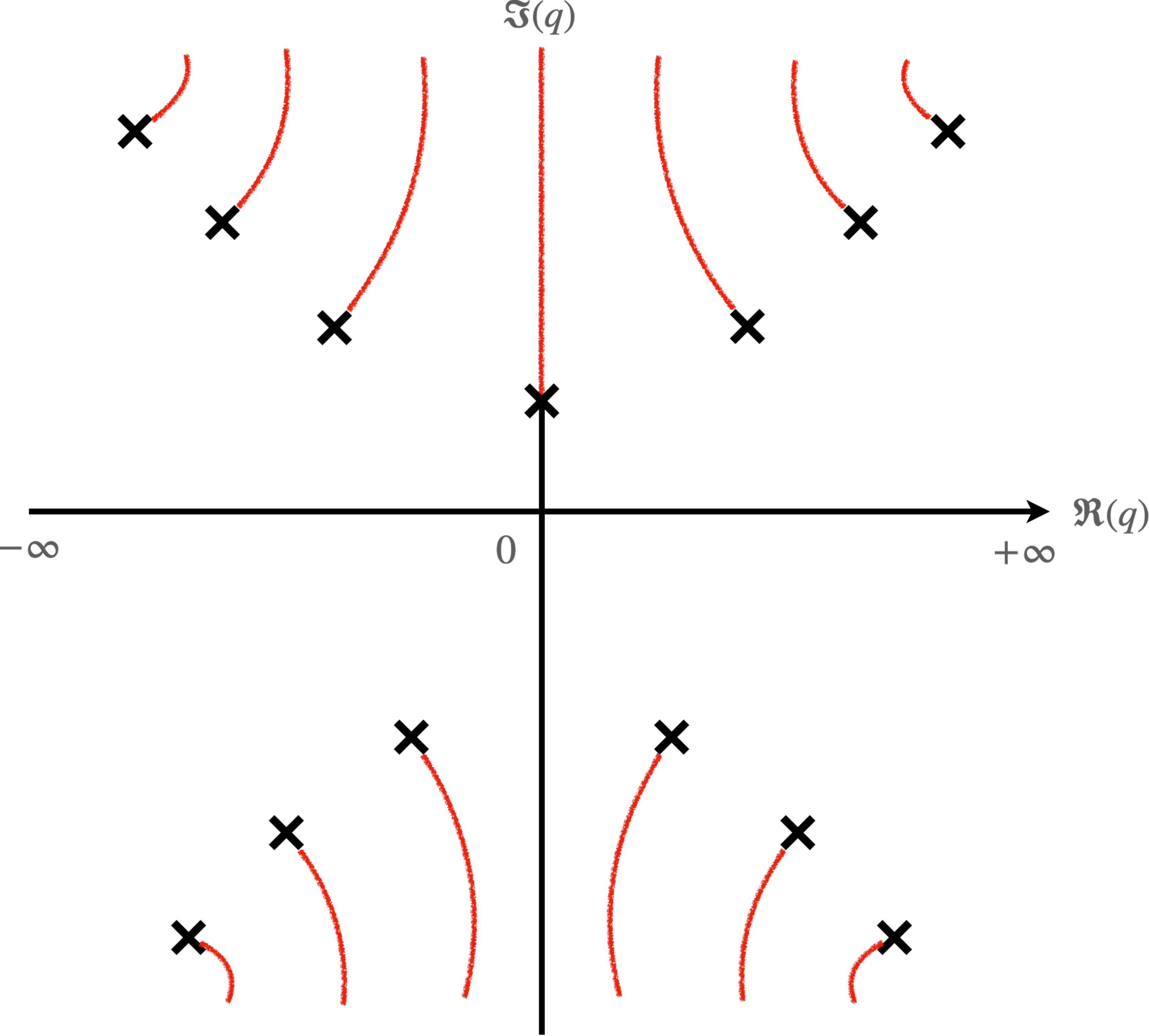}
   \includegraphics[scale=0.2]{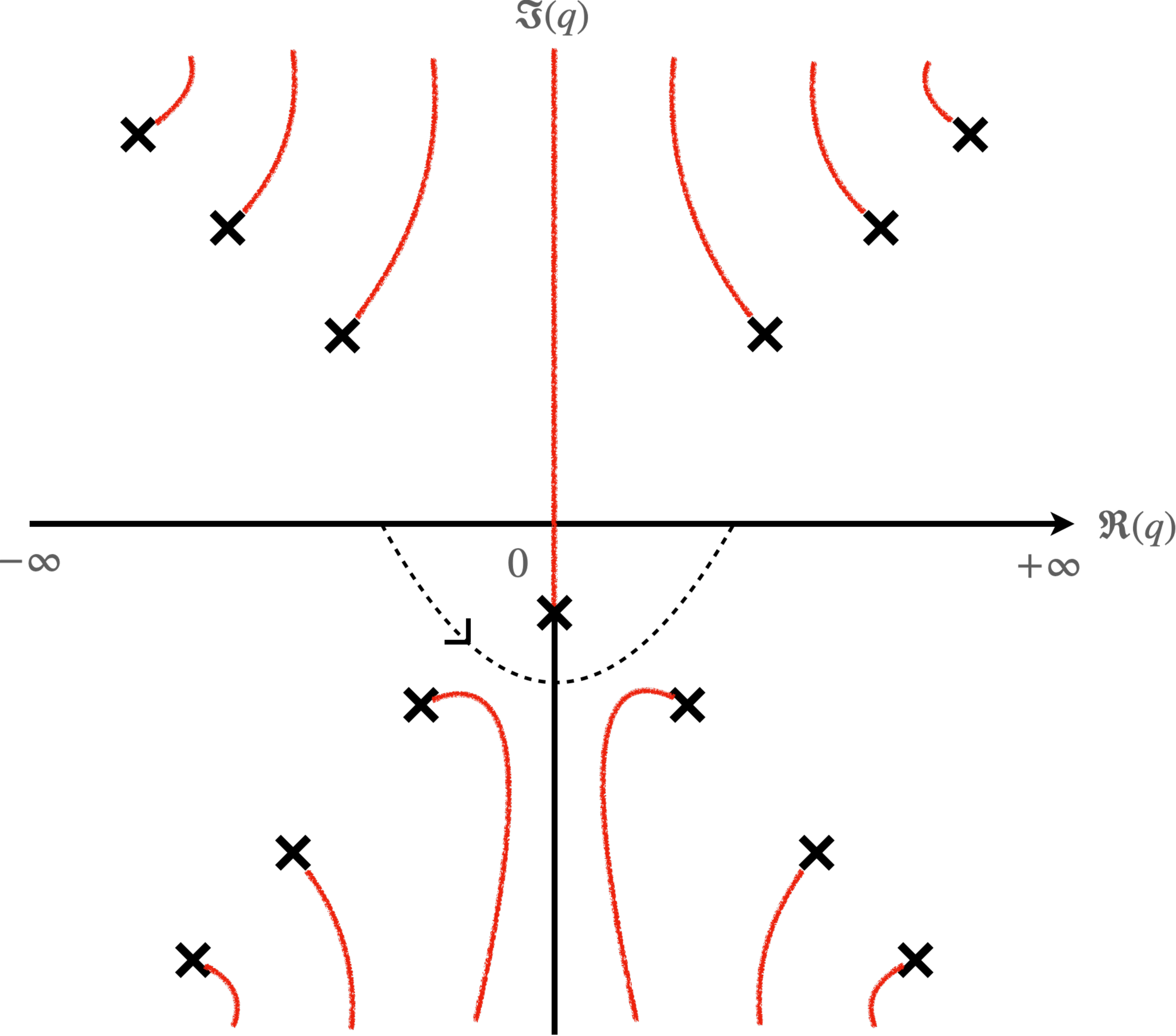}
   \includegraphics[scale=0.292]{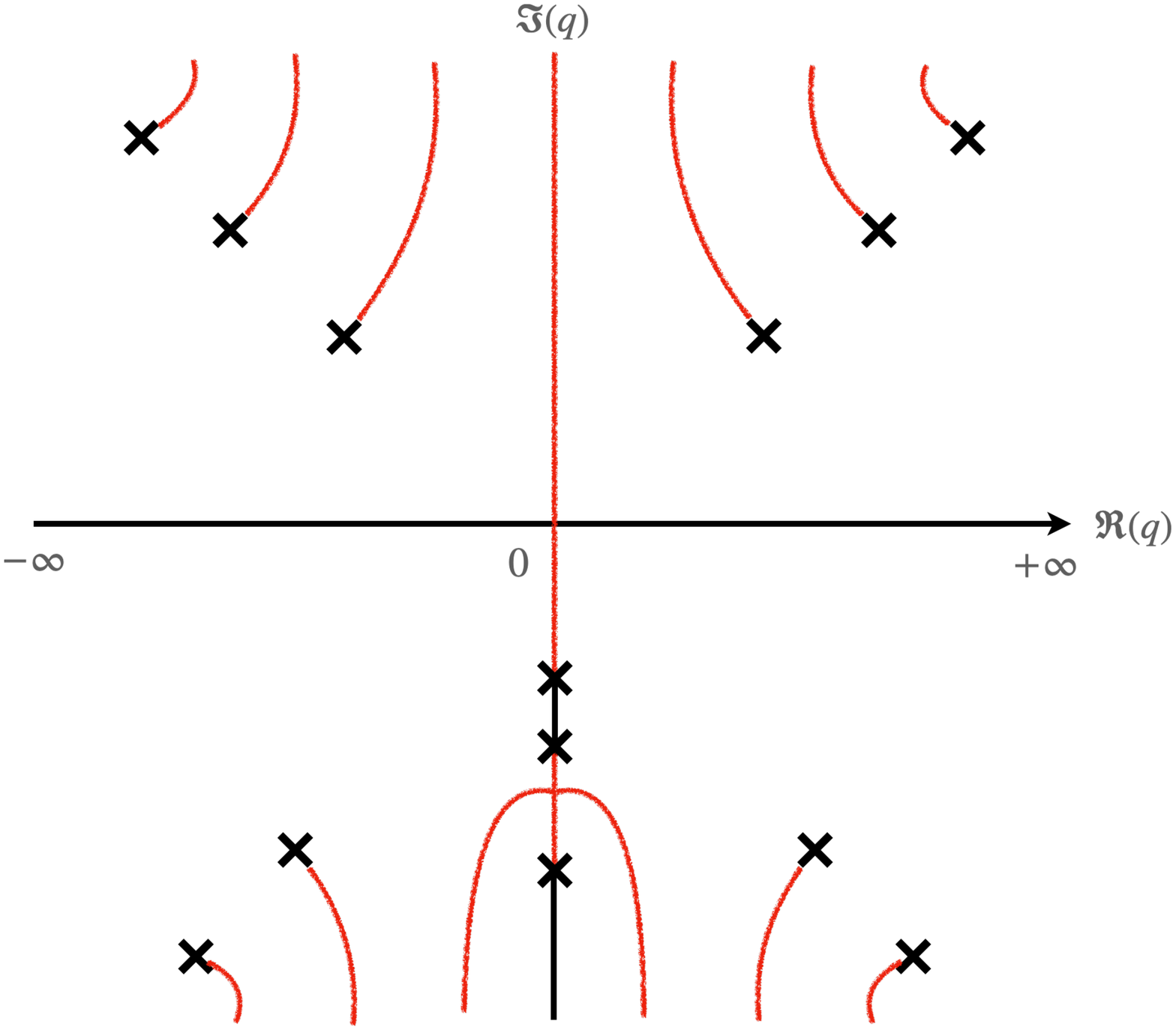}
   \includegraphics[scale=0.2]{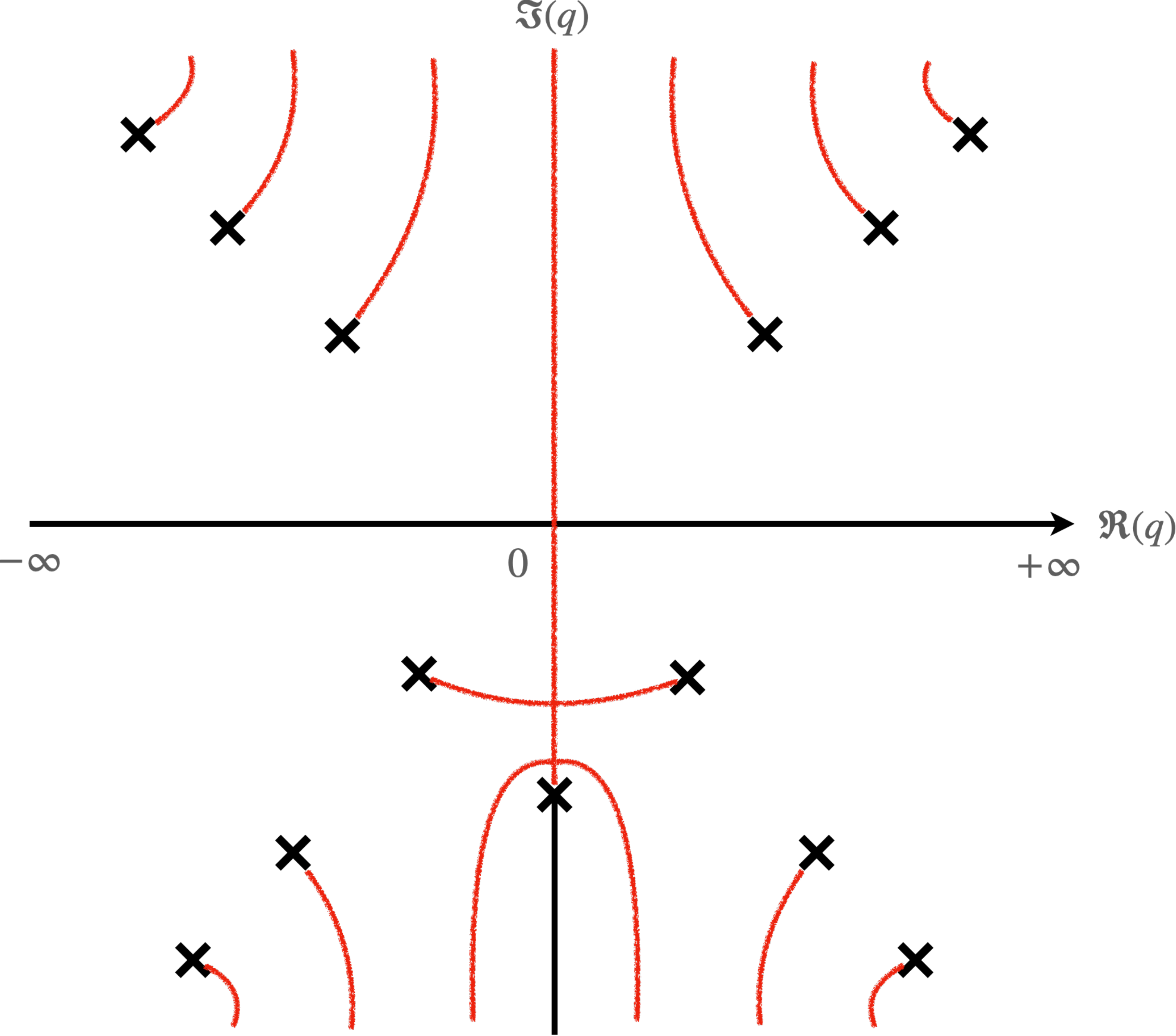}
    \caption{Schematic plots of the argument of the logarithm $A(q)$ given in \eqref{Aq} as a function of $q$ in the complex plane for $\xi>0$ and $z<0$. 
    The crosses indicate the positions of the zeroes
    of $A(q)$ and the red lines are the branch cuts. \textbf{Top Left.} $z_c<z<0$ for all $\xi>0$. No branch cut crosses the real axis. 
    \textbf{Top Right.} $z<z_c$ for $0<\xi<\xi_1=\sqrt{8}$,
    and $z_{c2}< z< z_c$ for $\xi_1<\xi<\xi_2$. In that case one branch cut crosses the real axis. The integration contour
    in $q$ in \eqref{zPsiFinal00} can be deformed (dotted lines) to avoid the branch cut. 
    \textbf{Bottom Left.} $z_{c1} < z < z_{c2}$ for $\xi_1 < \xi $. It is still possible
    to avoid the branch cut. \textbf{Bottom Right.} $\xi>\xi_1$ and $z< z_{c1}$. In that case the branch cuts have met and form a cross, and there is no way to deform the
    integration contour to avoid them. }
    \label{fig:ComplexPlane}
\end{figure} 

\subsection{Continuation and branches of $\Psi(z)$ for $\xi> \xi_1$}

Let us consider now the case $\xi> \xi_1 = \sqrt{8}$. First, for $z_c(\xi)<z<0$ 
it is still true that no branch cut crosses the real axis. This is because the
largest zero $p_1$ is strictly positive hence the branch cut $q \in [\I p_1, +\I \infty  [$
does not cross the real axis. Thus the formula in \eqref{zPsiFinal00} are valid, and 
this is again the {\it main branch}.\\

Next, as discussed in a previous subsection, see Fig.~\ref{fig:zeroes}, there is an interval of values of $z$,
$z \in ]z_{c1},z_{c2}[$, where there are three real zeroes $p_1>p_2>p_3$ to the equation $A(\I  p)=0$. 
Then there are two sub-cases, for $\xi_1 < \xi < \xi_2$ one has $z_{c2}<z_c$, while for 
$\xi_2 < \xi $ one has $z_{c}<z_{c2}$.\\

In terms of branch cuts, as one can see in Fig.~\ref{fig:ComplexPlane}, one finds
that for $z<  z_c$ the branch cut $q \in [\I p_1, +\I \infty [$ crosses the real axis. 
However as long as $z> z_{c1}$ there is a way to deform the contour of integration avoid this branch cut.
For $z< z_{c1}$ something nasty happens, the upper and lower branch cuts meet and form a cross 
see Fig.~\ref{fig:ComplexPlane} (bottom right). The same happened
for the KPZ equation, as discussed in the previous section.
In that case, it is not possible anymore to deform the integration
contour to avoid these branch cuts. We can now use the "jump" function $\Delta_1(z)$ obtained in the previous section
to propose the proper analytical continuations and the ensuing new branches.\\

When $z$ reaches $z_{c1}$ the zeroes $p_1$ and $p_2$ annihilate (corresponding to the merging
of the upper and lower branch cuts) and for $z<z_{c1}$ the only remaining zero is $p_3$. 
Thus one would like to write 
\bea \label{d3} 
\Psi(z)=\Psi_0(z) + \Delta_3(z)  \quad , \quad \Delta_3(z) = \hat \Delta(p_3(z)) 
\eea 
where $\hat \Delta(p)$ was defined in \eqref{defhatDe}. This branch appears indeed
in the Table~\ref{tab:MainTextTableJump}. However since $\Delta_3(z_{c1}) \neq \Delta_1(z_{c1})$ it is not a continuous extension
of $\Psi_0(z) + \Delta_1(z)$. This means that there are other branches that will allow a continuous extension. As we now discuss, they will be constructed
by first decreasing $z$ from $+\infty$ down to a turning point, increasing it up to a second turning point, and finally decreasing it
again down to $-\infty$.\\

We will thus consider the point $z=z_{c1}$ as the first turning point and start to follow the second zero $p_2=p_2(z)$. One then proposes the continuous extension 
defined for $z \in ]z_{c1},z_{c2}[$
 \bea
\Psi(z)=\Psi_0(z) + \Delta_2(z) \quad , \quad \Delta_2(z) = \hat \Delta(p_2(z)) 
\eea 
It is a new branch and since $\Psi_0(z) + \Delta_1(z)$ also exists in the same interval $z \in ]z_{c1},z_{c2}[$ the function $\Psi(z)$ is multi-valued in that interval.\\

When $z$ reaches $z_{c2}$ the zeroes $p_2=p_2(z)$ and $p_3=p_3(z)$ annihilate (corresponding to the disappearance of the lower branch-cut as seen in Fig.~\ref{fig:ComplexPlane} (top right)). We then again consider $z=z_{c2}$ as a second turning point and start following the third zero $p_3$. The candidate for the next continuous extension is therefore
 \bea
\Psi(z)=\Psi_0(z) + \Delta_3(z) \quad , \quad \Delta_3(z) = \hat \Delta(p_3(z)) 
\eea 
which is precisely the one in \eqref{d3}. \\

This procedure is sufficient for $\xi_1 <\xi <\xi_2$ and leads the third column in the Table~\ref{tab:MainTextTableJump}. Using the two turning points
one thus obtains a continuous extension, which is multi-valued in the interval $]z_{c1},z_{c2}[$. \\

In the case $\xi> \xi_2$ this however is insufficient.
Indeed, there is a last feature of the branch cuts to take into account. When $z$ increases from $z_{c1}$ to $z_{c2}$ and then decreases from $z_{c2}$ to $-\infty$ it can cross the value $z=z_c$ depending whether $z_c$ is in the interval $[z_{c1},z_{c2}]$. This is the case when $\xi > \xi_2$. When $z$ crosses the value $z_c$, the branch point $\I p_1$ crosses the real axis, either descending from the upper half plane or ascending from the lower half plane. We have observed for the first branch that crossing $z_c$ from above, i.e. $z=z_c+0^+ \to z_c+0^-$ implies that the function $\Psi_0$ is modified as
\begin{equation}
    \Psi_0 \to \Psi_0+\Delta_1
\end{equation}
Conversely, crossing $z_c$ from below, i.e. $z=z_c+0^- \to z_c+0^+$ implies that the function $\Psi_0$ should be modified as
\begin{equation}
    \Psi_0 \to \Psi_0-\Delta_1
\end{equation}
To obtain the complete solution for $\xi> \xi_2$ one then needs to take into account the coalescence of the zeroes $p_1, p_2, p_3$ shown
in Fig.~\ref{fig:zeroes} and also the different crossing of $z=z_c$ independently. This leads to the fourth column of Table~\ref{tab:MainTextTableJump}.
\\

{\bf Remark.} In all these formula $\Psi_0(z)$ denotes the integral \eqref{Psi000} along the real axis which may or may not have a jump
in the integrand depending on whether $z<z_c$ or $z>z_c$.

\section{Summary of the results: determination of  $\Phi(H)$  }

In this Section we summarize the exact results for the large-deviation rate function $\Phi(H)$ 
of the diffusion in time-dependent random medium for arbitrary position of the tracer $\xi>0$.
The rate function $\hat \Phi(Z)$ defined in the text is simply obtained as $\hat \Phi(Z) = \Phi(\log Z)$.

\subsection{Exact expressions for all rate functions and critical values}
\textbf{Critical values of $\xi$.}
There are two critical values of the position of tracer denoted $\xi_1$ and $\xi_2$. Their value is given as
\begin{equation}
\label{eq:CriticalXi}
    \begin{split}
        \xi_1&=\sqrt{8}\simeq 2.82843\\
        \xi_2&=-2 \sqrt{\frac{2}{-2 W_{-1}\left(-\frac{1}{2 \sqrt{e}}\right)-1}} W_{-1}\left(-\frac{1}{2 \sqrt{e}}\right)\simeq 3.13395\\
    \end{split}
\end{equation} 
where $W_{-1}$ is the second real branch of the Lambert function \cite{corless1996lambertw}.
\begin{enumerate}
    \item For $\xi\geq \xi_1$ there can be three real zeroes to Eq.~\eqref{eq:BranchCutEquation} depending on $z$, 
    whereas for $\xi\leq \xi_1$ there is only one real zero.
    \item The value $\xi_2$ is determined as the solution of $z_c(\xi)=z_{c2}(\xi)$. For $\xi \geq \xi_2$, we have the ordering $z_{c1} < z_c < z_{c2}$.
\end{enumerate}

\textbf{Critical values of $z$.} There are three critical values of the parameter $z$ denoted $z_c$, $z_{c1}$ and $z_{c2}$. Their dependence on the tracer position $\xi$ is given as

\begin{equation}
\label{eq:FormulaZcP}
\begin{split}
    z_c(\xi)&=-\frac{2 e^{\frac{\xi ^2}{4}}}{\xi }\\
    z_{c1}(\xi)&=-\frac{1}{2} e^{\frac{1}{8} \left(\xi  \left(\xi+\sqrt{\xi ^2-8} \right)+4\right)} \left(\xi -\sqrt{\xi ^2-8}\right)\\
    z_{c2}(\xi)&=-\frac{1}{2} e^{\frac{1}{8} \left(\xi \left(\xi-\sqrt{\xi ^2-8}\right)  +4\right)} \left(\xi+\sqrt{\xi ^2-8} \right)\\
    \end{split}
\end{equation}
\begin{enumerate}
    \item The quantities $ z_{c1}$, $ z_{c2}$ are real only for $\xi>\xi_1$. They are determined by the value of $p$ where the function $f_{z,\xi}$ has two degenerate zeroes, i.e. $f_{z,\xi}(p)=f_{z,\xi}'(p)=0$
    \item For $z<z_c$, the largest real zero of $f_{z,\xi}(p)$ is negative whereas for $z>z_c$ it is positive.
\end{enumerate}

\textbf{Critical values of the zeroes $p$.}

\bea 
&& p_{c} = 0 \\
&& p_{c1} = -\frac{1}{4} \left(\xi -\sqrt{\xi ^2-8}\right) \\
&& p_{c2}= -\frac{1}{4}  \left(\xi+\sqrt{\xi ^2-8} \right) 
\eea

\textbf{The rate function $\Phi(H)$}. We obtain the rate function $\Phi(H)$ parametrically. In practice, its numerical determination will be done by parts using all the different branches of $\Psi(z)$. Since $z(H)$ is single-valued, this procedure 
allows to obtain $\Phi(H)$ in the whole range $]-\infty,0]$.
We provide in the following the representations which were used for the numerical plots.
\subsubsection{$\xi=0$}

The rate function, see Table~\ref{tab:app-case1}, reads

\begin{table}[h!]
    \centering
    \begin{tabular}{c| c| c | c }
        interval of $H$ & interval of $z$ & $H=$ & $\Phi(H)=$ \\[2ex]
        \hline 
        \hline &&&\\[-0.5ex]
        $H \in \R^-$ & $z \in \R$ &$\log \Psi_0'(z)$ & $\Psi_0(z)- z \Psi_0'(z)$\\[1ex]
    \end{tabular}
    \caption{Case $\xi=0$}
    \label{tab:app-case1}
\end{table}

\newpage
\subsubsection{$0<\xi \leq \xi_1$} 
Defining the critical height $H_c=\log \Psi_0'(z_c)$, the rate function, see Table~\ref{tab:app-case2}, reads 
\begin{table}[h!]
    \centering
    \begin{tabular}{c| c| c | c }
        interval of $H$ & interval of $z$ & $H=$ & $\Phi(H)=$ \\[2ex]
        \hline 
        \hline &&&\\[-0.5ex]
        $H\leq H_c$ & $z_c\leq z$ &$\log \Psi_0'(z)$ & $\Psi_0(z)- z \Psi_0'(z)$\\[2ex]
        \hline &&&\\[-0.5ex]
        $0 > H> H_c$  & $z_c> z$ &$ \log (\Psi_0'(z)+\Delta_1'(z)) $ & $\Psi_0(z)+\Delta_1(z)- z (\Psi_0'(z)+\Delta_1'(z))$\\[1ex]
    \end{tabular}
    \caption{Case $0<\xi \leq \xi_1$}
    \label{tab:app-case2}
\end{table}

\subsubsection{$\xi_1< \xi \leq \xi_2$}
Defining the three critical heights
\begin{equation}
\begin{split}
    H_c&=\log \Psi_0'(z_c),\\
    H_{c1}&=\log (\Psi_0'(z_{c1})+\Delta_1'(z_{c1}))=\log (\Psi_0'(z_{c1})+\Delta_2'(z_{c1})),\\
    H_{c2}&=\log (\Psi_0'(z_{c2})+\Delta_2'(z_{c2}))=\log (\Psi_0'(z_{c2})+\Delta_3'(z_{c2})),
\end{split}
\end{equation}
the rate function, see Table~\ref{tab:app-case3}, reads 
\begin{table}[h!]
    \centering
    \begin{tabular}{c| c| c | c }
        interval of $H$ & interval of $z$ & $H=$ & $\Phi(H)=$ \\[2ex]
        \hline 
        \hline &&&\\[-0.5ex]
        $H\leq H_c$ & $z_c\leq z$ &$\log \Psi_0'(z)$ & $\Psi_0(z)- z \Psi_0'(z)$\\[2ex]
        \hline &&&\\[-0.5ex]
        $H_c<H\leq H_{c1}$  & $z_{c1}\leq z<z_c$ &$ \log (\Psi_0'(z)+\Delta_1'(z)) $ & $\Psi_0(z)+\Delta_1(z)- z (\Psi_0'(z)+\Delta_1'(z))$\\[2ex]
                \hline &&&\\[-0.5ex]
        $H_{c1}<H\leq H_{c2}$  & $z_{c1}< z\leq z_{c2}$ &$ \log (\Psi_0'(z)+\Delta_2'(z)) $ & $\Psi_0(z)+\Delta_2(z)- z (\Psi_0'(z)+\Delta_2'(z))$\\[2ex]
                \hline &&&\\[-0.5ex]
        $H_{c2}<H<0$  & $z_{c2}> z$ &$ \log (\Psi_0'(z)+\Delta_3'(z)) $ & $\Psi_0(z)+\Delta_3(z)- z (\Psi_0'(z)+\Delta_3'(z))$\\[1ex]
    \end{tabular}
    \caption{Case $\xi_1< \xi \leq \xi_2$}
    \label{tab:app-case3}
\end{table}

It is important to note that the expressions for $H$ and for $\Phi(H)$ as a function of $z$
in the second and third line of the above table merge continuously at $H=H_{c1}$
around the turning point at $z=z_{c1}$. This can be seen from \eqref{zDep1} and \eqref{eq:FinalFormulaDelta} as the jumps
$\Delta_j(z) = \hat \Delta(p_j(z))$, $j=1,2$ are the same function of the 
zeroes $p_j(z)$, hence one has $\Delta_1(z_{c1})=\Delta_2(z_{c1})$ as
well as $\Delta'_1(z_{c1})=\Delta'_2(z_{c1})$, since $p_1(z)=p_2(z)$ at $z=z_{c1}$.
This implies that as $z$ decreases from $+\infty$ down to the turning point $z_{c_1}$ and 
then increases again from $z_{c_1}$, the function $H=H(z)$ smoothly increases, and $\Phi(H)$ is a smooth
function of $H$ around $H_{c1}$. These features can be seen in
Fig.~\ref{fig:MainTextFigs} (top right) in the text. The same holds for
each turning point, and is also valid for the table in the next section. 

\subsubsection{ $ \xi_2< \xi$}
Defining the five critical heights
\begin{equation}
\begin{split}
    H_c&=\log \Psi_0'(z_c)\\
    H_{c10}&=\log (\Psi_0'(z_{c1})+\Delta_1'(z_{c1})),\\
    H_{c11}&=\log (\Psi_0'(z_{c})+\Delta_2'(z_{c})),\\
    H_{c20}&=\log (\Psi_0'(z_{c2})+\Delta_2'(z_{c2})-\Delta_1'(z_{c2})),\\
    H_{c21}&=\log (\Psi_0'(z_{c})+\Delta_3'(z_{c})),
\end{split}
\end{equation}
the rate function, see Table~\ref{app-tab-branches-large-xi}, reads 
\begin{table}[h!]
    \centering
    \begin{tabular}{c| c| c | c }
        interval of $H$ & interval of $z$ & $H=$ & $\Phi(H)=$ \\[2ex]
        \hline 
        \hline &&&\\[-0.5ex]
        $H\leq H_c$ & $z_c\leq z$ &$\log \Psi_0'(z)$ & $\Psi_0(z)- z \Psi_0'(z)$\\[2ex]
        \hline &&&\\[-0.5ex]
        $H_c<H\leq H_{c10}$  & $z_{c1}\leq  z< z_{c}$ &$ \log (\Psi_0'(z)+\Delta_1'(z)) $ & $\Psi_0(z)+\Delta_1(z)- z (\Psi_0'(z)+\Delta_1'(z))$\\[2ex]
                \hline &&&\\[-0.5ex]
        $H_{c10}<H\leq H_{c11}$  & $z_{c1}< z\leq z_{c}$ &$ \log (\Psi_0'(z)+\Delta_2'(z)) $ & $\Psi_0(z)+\Delta_2(z)- z (\Psi_0'(z)+\Delta_2'(z))$\\[2ex]
        \hline &&&\\[-0.5ex]
        $H_{c11}<H\leq H_{c20}$  & $z_{c}< z\leq z_{c2}$ &$  \log (\Psi_0'(z)+\Delta'_2(z)-\Delta_1'(z)) $ & $\Psi_0(z)+\Delta_2(z)-\Delta_1(z)- z (\Psi_0'(z)+\Delta_2'(z)-\Delta_1'(z))$\\[2ex]
        \hline &&&\\[-0.5ex]
        $H_{c20}<H\leq H_{c21}$  & $z_{c}\leq  z< z_{c2}$ &$ \log (\Psi_0'(z)+\Delta'_3(z)-\Delta_1'(z))  $ & $\Psi_0(z)+\Delta_3(z)-\Delta_1(z)- z (\Psi_0'(z)+\Delta_3'(z)-\Delta_1'(z))$\\[2ex]
        \hline &&&\\[-0.5ex]
        $H_{c21}<H<0$  & $z_{c}> z$ &$ \log (\Psi_0'(z)+\Delta_3'(z)) $ & $\Psi_0(z)+\Delta_3(z)- z (\Psi_0'(z)+\Delta_3'(z))$\\[1ex]
    \end{tabular}
    \caption{Case $ \xi_2< \xi$}
    \label{app-tab-branches-large-xi}
\end{table}

\textbf{Optimal rate function $\Psi(z)$} As discussed in the text the "optimal" $\Psi(z)$ follows by definition the minimum of the different branches of $\Psi(z)$ that we have found. 
For $\xi<\xi_1$ there is no multi-valuation of $\Psi(z)$ hence $\Psi(z)$ follows continuously the two branches $z>z_c$ 
and $z<z_c$. For $\xi>\xi_1$ there is multi-valuation of $\Psi(z)$ for $z \in ]z_{c1},z_{c2}[$ leading to a discontinuity, i.e. a jump of $\Psi(z)$ 
The value of $z$ for which $\Psi(z)$ jumps from a branch to the next (see inset of Fig.~\ref{fig:MainTextFigs} ) is given by $z^*$ solution of
\begin{equation} \label{jumpcriterion} 
    \Delta_1(z^*)=\Delta_3(z^*)
\end{equation}
This value is located between $z_{c1}$ and $z_{c2}$. We provide in the next two Tables~\ref{tab:app-optimalFenchel} the value of the optimal Legendre solution.
Note that the jump in the value of $Z$ is always $\Delta'_3(z^*)-\Delta'_1(z^*)$ (jumps between the two maxima of
the tilted measure for $Z$ as discussed in the text). 

\begin{table}[h!]
    \centering
    \begin{tabular}{ c| c  }
         interval of $z$ &  "optimal" $\Psi(z)=$ \\[2ex]
        \hline 
        \hline &\\[-0.5ex]
         $z_c\leq z$ &$\Psi_0(z)$\\[2ex]
        \hline &\\[-0.5ex]
         $z^*\leq z<z_c$  & $\Psi_0(z)+\Delta_1(z)$\\[2ex]
                \hline &\\[-0.5ex]
         $z^*> z$  & $\Psi_0(z)+\Delta_3(z)$\\[2ex]
    \end{tabular}\hspace{2cm}
        \begin{tabular}{ c| c  }
         interval of $z$ &  "optimal" $\Psi(z)=$ \\[2ex]
        \hline 
        \hline &\\[-0.5ex]
         $z^*\leq z$ &$\Psi_0(z)$\\[2ex]
        \hline &\\[-0.5ex]
         $z_{c}\leq z<z^*$  & $\Psi_0(z)+\Delta_3(z)-\Delta_1(z)$\\[2ex]
                \hline &\\[-0.5ex]
         $z_c> z$  & $\Psi_0(z)+\Delta_3(z)$\\[2ex]
    \end{tabular}
    \caption{\textbf{(Left)} Case $\xi_1< \xi \leq \xi_2$. In the inversion of the Legendre-Fenchel transform, $Z$ jumps from $\Psi_0'(z^*)+\Delta'_1(z^*)$ to $\Psi_0'(z^*)+\Delta'_3(z^*)$. \textbf{(Right)} Case $\xi_2< \xi $ (assuming $z_c<z^*$). In the inversion of the Legendre-Fenchel transform, $Z$ jumps from $\Psi_0'(z^*)$ to $\Psi_0'(z^*)+\Delta'_3(z^*)-\Delta'_1(z^*)$. Note that $z_c>z^*$ would lead to a jump
    between the second branch and the last branch with the same jump criterion \eqref{jumpcriterion}.}
    \label{tab:app-optimalFenchel}
\end{table}

\subsection{Additional plots}

In this Section we show the plot of $\Psi(z)$ versus $z$, as well as the plot of 
$\Phi(H(z))$ versus $z$, see Fig.~\ref{fig:PhiHZandPsiZzoomed}. 

\begin{figure}[h!]
    \centering
    \includegraphics[scale=0.55]{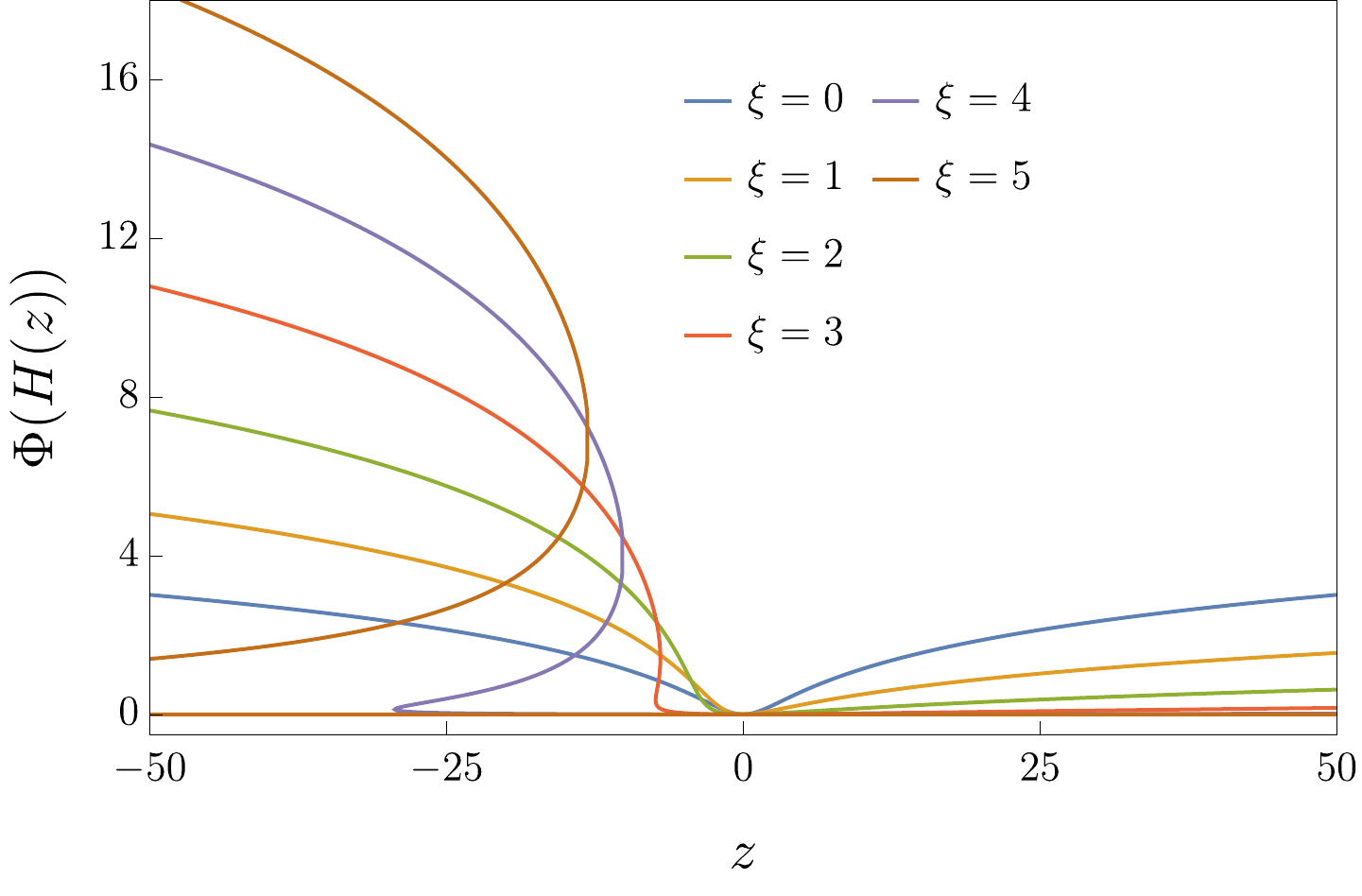}
    \includegraphics[scale=0.55]{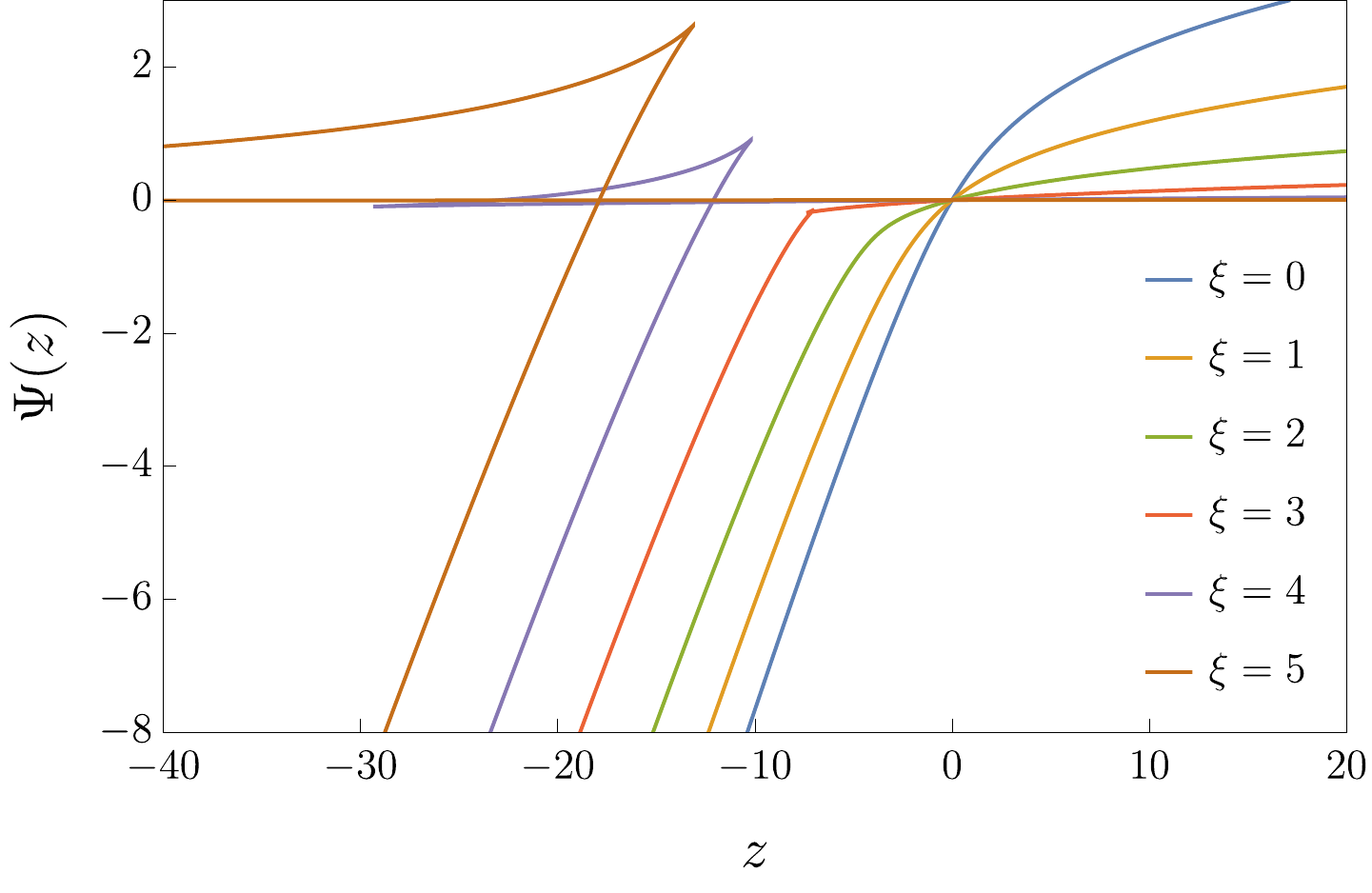}
    \caption{Plots for various values of $\xi=0,1,2,3,4,5$. \textbf{(Left)} Large deviation rate function $\Phi(H(z))$ as a function of $z$ using the definition \eqref{eq:ParametricRepresentation}. The function is symmetric for $\xi=0$ and becomes asymmetric for non-zero values of $\xi$.
    It satisfies the symmetry $\Phi(H(z))|_{-\xi}= \Phi(H(-z))|_{\xi} $. \textbf{(Right)}
    Large deviation rate function $\Psi(z)$. The minimum branch of $\Psi(z)$ defines the "optimal" solution to the Legendre inversion of Eq.~\eqref{Legendre0}. For large negative values of $z$, the function becomes almost linear, i.e. $\Psi(z)\simeq z$. }
    \label{fig:PhiHZandPsiZzoomed}
\end{figure}

\subsection{Result for $\xi=0$ and correspondence with Ref.~\cite{NaftaliDNLS}.}
We provide in this Section a correspondence between the variables and functions studied in this work and in the work \cite{NaftaliDNLS}
in the particular case $\xi=0$. In that case $g=0$. This is summarized in the Table~\ref{tab:CorrespondanceNaftali}.
\begin{table}[h!]
    \centering
    \begin{tabular}{c ||c}
         This present work  & \hspace{1cm} Ref.~\cite{NaftaliDNLS} \hspace{1cm} \\[1ex]
         \hline
         \\[-0.2ex]
        $z  = \beta \Lambda$ & $-\lambda$ \\[1ex]
        \hline
         \\[-0.2ex]
       $\beta R$& $v$\\[1ex]
       \hline
       \\[-0.2ex]
       $Q$& $u$\\[1ex]
       \hline
         \\[-0.2ex]
       $Z=e^H=\Psi'(z)$ & $j+\dfrac{1}{2}$\\[1ex]
       \hline
         \\[-0.2ex]
       $\Phi(H) = \Psi(z) - z e^H = \Psi(z) - z \Psi'(z) $ \hspace{2ex} & $s(j)$\\[1ex]
              \hline
         \\[-0.2ex]
       $- z= e^{- H} \Phi'(H)   $ & $\lambda = \dfrac{\rmd s}{\rmd j}$\\[1ex]
    \end{tabular}
    \caption{Correspondence between the variables of this work and Ref.~\cite{NaftaliDNLS} for $\xi=0$.}
    \label{tab:CorrespondanceNaftali}
\end{table}

Consider the formula (27) and (29) in \cite{NaftaliDNLS}. Taken together
they give
\be 
s(\lambda) - \lambda j(\lambda) = \int_{\mathbb{R}}  \frac{\rmd k}{8 \pi k^2} {\rm Li}_2(- \lambda^2 k^2 e^{-2 k^2}) 
\quad , \quad
\lambda j(\lambda) =  \int_{\mathbb{R}}  \frac{\rmd k}{4 \pi k^2} \log(1+ \lambda^2 k^2 e^{-2 k^2})  
\ee 
From the Table~\ref{tab:CorrespondanceNaftali} we should identify the first result as $\Psi(z)-z/2$ and the second as $z/2- z \Psi'(z)$.
Using the duplication formula for the dilogarithm
\begin{equation}
    \mathrm{Li}_2(z)+\mathrm{Li}_2(-z)=\frac{1}{2}\mathrm{Li}_2(z^2)
\end{equation}
we indeed find agreement with our formula \eqref{PsiFinal0} and \eqref{zPsiFinal0} with $z= - \lambda$ and using $\Theta(\xi=0)=1/2$ as 
discussed in the text. 

\subsection{Cumulant expansion of $Z$ and checks}
\label{sec:cum} 

From its definition \eqref{eq:gener} the function $\Psi(z)$ encodes the cumulant expansion
\be 
\Psi(z) = - \sum_{p \geq 1} \frac{(-z)^p}{p!} T^{\frac{p-1}{2}} \overline{Z(Y,T)^p} 
\ee 
We will now check from perturbation theory that the lowest order matches our exact result.
Since $Z(Y,T)$ is the cumulative probability \eqref{Z}, from \eqref{FP} it satisfies the SDE
\be 
\partial_\tau Z(y,\tau) = \partial^2_{y} Z(y,\tau) - \sqrt{2} \eta(y,\tau) \partial_y Z(y,\tau)
\ee 
which we can call the derivative stochastic heat equation, with initial condition $Z(y,\tau=0)=\Theta(-y)$. 
We rescale the space and time variables as $y=x \sqrt{T}$, $\tau = t T $.
Here we will abuse notations and use the same letter to denote $Z(y,\tau)=Z(x,t)$. 
The original variable is recovered at the end. The rescaling 
yields the dimensionless equation with small noise amplitude
\be 
\partial_t Z(x,t) = \partial^2_{x} Z(x,t) -  \sqrt{2} T^{-1/4} \eta(x,t) \partial_x Z(x,t)
\ee 
with initial condition $Z(x,t=0)=\Theta(-x)$. Denoting $G(x,t)=\frac{1}{\sqrt{4 \pi t}} e^{-x^2/(4 t)}$ the free Green's function, this can also be written as
\be 
Z(x,t) = Z_0(x,t)  -  \frac{\sqrt{2}}{ T^{1/4} } \int_0^t \rmd u \int_\R \rmd x' G(x-x',t-u) 
\eta(x',u) \partial_{x'} Z(x',u) ~ , ~ Z_0(x,t)= \int_\R \rmd x' G(x-x',t) \Theta(- x') 
\ee 
\begin{enumerate}
    \item For the first moment one recovers indeed 
\be 
\overline{ Z(Y,T) } = \overline{ Z(x,1) } = Z_0(x,1)=  \int_x^{+\infty} \rmd y \, G(y,1) =
\frac{1}{2} {\rm Erfc}(\frac{x}{2}) = \Psi'(0)
\ee 
\item To lowest order in $T^{-1/2}$ one finds the second cumulant
\be 
\overline{Z(x,1)^2}^c = 2 T^{-1/2} \int_0^1 \rmd u \int_\R \rmd x' G(x-x',1-u)^2 (\partial_{x'} Z_0(x',u))^2 + \mathcal{O}(T^{-1}) 
\ee 
Using that $\partial_{x'} Z_0(x',u) = - G(x',u)$ one finds the remarkably simple result
\be 
\overline{ Z(Y,T)^2 }^c  = \overline{Z(\xi,1)^2}^c =  T^{-1/2} \frac{e^{-\frac{\xi^2}{2}}}{4 \sqrt{2 \pi }}  + \mathcal{O}(T^{-1}) 
\ee 
On the other hand one must have
\be 
- T^{1/2} \overline{ Z(Y,T)^2 }^c =  \Psi''(0) = - \frac{1}{2} \int_\R \frac{\rmd q}{2 \pi} e^{- 2 q^2 - \frac{\xi^2}{2}} 
= - \frac{1}{4 \sqrt{2 \pi }} e^{- \frac{\xi^2}{2}}
\ee 
since $\Psi''(0)$ is the coefficient of $z^2$ in $z \Psi'(z)$.
\end{enumerate}
This shows that our formula \eqref{PsiFinal0}
for the large-deviation function $\Psi(z)$ yields correctly the two lowest cumulants,
as stated in the text.

\subsection{Cumulants of $H$}

One can obtain the cumulants of $H$ from the derivatives of the rate function $\Phi(H)$ (see e.g. in \cite[Sec.~4.2.5 of the Supp. Mat.]{krajenbrink2017exact}).
Here they scale as $\overline{H^q}^c \sim T^{\frac{1-q}{2}}$. The typical value $H=H_{\rm typ}$ is determined by
$\Phi'(H_{\rm typ})=0$, which leads $e^{H_{\rm typ}}= \Psi'(0)$ (see previous subsection) and the second cumulant reads
\be \label{secondcum} 
\overline{H^2}^c =  \frac{T^{-\frac{1}{2}}}{\Phi''(H_{\rm typ})} = -T^{-\frac{1}{2}} 
\frac{ \Psi''(0)}{\Psi'(0)^2}
= C_2(\xi) \, T^{-\frac{1}{2}} \quad , \quad C_2(\xi) = \frac{1}{\sqrt{2 \pi}} e^{-\frac{\xi^2}{2}} \left({\rm Erfc}(\frac{\xi}{2})\right)^{-2}
\ee 
Indeed, one can relate the derivatives $\Phi^{(q)}(H_{\rm typ})$ to those of $\Psi(z)$ around $z=0$ by differentiating
the relations $\Psi'(z) = e^H$ and $\Phi'(H)=- z e^H$. One obtains $\Phi''(H)-\Phi'(H)=- \frac{\rmd z}{\rmd H} e^H = - \frac{e^{2 H}}{\Psi''(z)}$.
Taken at $z=0$ and $H=H_{\rm typ}$ they lead to \eqref{secondcum}. One has the asymptotics at small and large $\xi$
\be \label{asymptC2} 
C_2(\xi) = \frac{1}{\sqrt{2 \pi}} + \frac{\sqrt{2}}{\pi} \xi + \mathcal{O}(\xi^2) \quad , \quad 
C_2(\xi)  =  \sqrt{\frac{\pi}{2}} ( \frac{\xi^2}{4} + 1) + \mathcal{O}( \frac{1}{\xi^2}) 
\ee

\section{Convergence to the large deviations of the Kardar-Parisi-Zhang equation}
\label{sec:KPZ} 

\subsection{Large $\xi$ limit: matching MFT at large time $T \gg 1$ to WNT at small time $T_{\rm KPZ} \ll 1$}  

We ought to understand in this Section the behavior of our solution in the large $\xi$ limit. In this regime, one first needs to rescale the variable $z$ as
\begin{equation}
    \tilde z = z \frac{\xi}{2} e^{- \frac{\xi^2}{4}} = - \frac{z}{z_c} \, .
\end{equation}

{\bf Values of the main branch of the large-deviation function $\Psi_0(z)$.}
Recalling the definition of $\Psi_0(z)$ in \eqref{PsiFinal0} for $\xi>0$, approximating $\I q-\frac{\xi}{2} \sim -\frac{\xi}{2}$ and using the series expansion of the dilogarithm $\mathrm{Li}_2(y)=\sum_{n>0}y^n/n^2$, we obtain that 
\be 
\begin{split} \label{Psilimit2} 
\Psi_0(z) &\simeq - \frac{1}{(\frac{\xi}{2})^2}\sum_{n=1}^{+\infty}\frac{(-\tilde z )^n}{n^2}
\int_\R \frac{\rmd q}{2 \pi} e^{-n q^2}   \\
&\simeq  \frac{4}{\xi^2} \Psi_{\rm KPZ,0}(\tilde z) 
\end{split}
\ee 

To go from the first line to the second one, we performed the Gaussian integral and used the identity
\be 
\Psi_{\rm KPZ,0}(\tilde z)  = -\frac{ 1}{\sqrt{4 \pi}} \mathrm{Li}_{5/2}(-\tilde z )
\ee

{\bf Critical values of $z$.}
In the same way, we rescale the critical values of $z$ as follows:
\begin{equation}
    \tilde{z}_c=-\frac{z_c}{z_c}, \; \tilde{z}_{c1}=-\frac{z_{c1}}{z_c}, \; \tilde{z}_{c2}=-\frac{z_{c2}}{z_c} \, .
\end{equation}
and take the $\xi \gg 1$ limit to obtain the limiting values
\begin{equation}
    \tilde{z}_c=-1, \; \tilde{z}_{c1}=-1+\mathcal{O}(\frac{1}{\xi^2}), \; \tilde{z}_{c2}=e^{1-\frac{\xi ^2}{4}} \left(1-\frac{\xi ^2}{2}+\mathcal{O}(\frac{1}{\xi^2})\right) \, .
\end{equation}

{\bf Values of the zeroes $p_1, p_2, p_3$.} The equation \eqref{eq:BranchCutEquation} determining the position of the branch cut reads with this variable
\be 
e^{-p^2} + \tilde z (1 + \frac{2 p}{\xi}) = 0
\ee 
At the first order at large $\xi$, the zeroes of this equation read
\begin{equation}
    \begin{split}
        p_1&\simeq\sqrt{\log\left(\frac{z_c}{z}\right)}=\sqrt{-\log(-\tilde{z})}\\
        p_2&\simeq-\sqrt{\log\left(\frac{z_c}{z}\right)}=-\sqrt{-\log(-\tilde{z})}\\
        p_3&\simeq-\frac{\xi}{2}-\frac{1}{z} = -\frac{\xi}{2} - \frac{\xi}{2 \tilde z} e^{- \xi^2/4} \\
    \end{split}
\end{equation}
To study only real zeroes imposes that $\tilde{z}\in [-1,0]$.\\

{\bf Values of the derivative of the jump function.} Recalling that the derivative of the jump function \eqref{eq:JumpPsi} reads
\be 
\tilde z \partial_{\tilde z} \Delta_\ell = \frac{4p_\ell}{\xi(2p_\ell+\xi)}
\ee 
It yields for the different zeroes
\begin{equation}
\begin{split}
\tilde z \partial_{\tilde z} \Delta_1 &\simeq_{\xi \to \infty} \frac{4}{\xi^2}\sqrt{-\log(-\tilde{z})}\, , \\ \tilde z \partial_{\tilde z} \Delta_2 &\simeq_{\xi \to \infty} -\frac{4}{\xi^2}\sqrt{-\log(-\tilde{z})} \, , \\
\tilde z \partial_{\tilde z} \Delta_3 &\simeq_{\xi \to \infty} z = \frac{2 \tilde z}{\xi} e^{\frac{\xi^2}{4} } 
\end{split}
\end{equation}

{\bf Discussion about which branches remain in the large $\xi$ limit.}
Recalling the different branches of the large-deviation function $\Psi(z)$ in Table~\ref{app-tab-branches-large-xi}, we now discuss how the different branches behave in the large $\xi$ limit. Since we have $\tilde{z}_c=\tilde{z}_{c1}$, the branches
\begin{equation}
    \begin{split}
        \Psi(z)&=\Psi_0(z)+\Delta_1(z)\\
        \Psi(z)&=\Psi_0(z)+\Delta_2(z)\\
    \end{split}
\end{equation}
disappear on the $\tilde{z}$ scale. We now explain that the next branch, i.e.
\begin{equation}
        \Psi(z)=\Psi_0(z)+\Delta_2(z)-\Delta_1(z)
\end{equation}
is the only one, besides the main branch, to remain on the $\tilde{z}$ scale. Indeed, looking at the derivative
\begin{equation}
\begin{split}
    \tilde z \partial_{\tilde z} \Psi(z)&=\tilde z \partial_{\tilde z} \Psi_0(z)+\tilde z \partial_{\tilde z}\Delta_2(z)-\tilde z \partial_{\tilde z}\Delta_1(z)\\
    &\simeq \frac{4}{\xi^2}\left(\tilde z \partial_{\tilde z}  \Psi_{\rm KPZ,0}(\tilde z) -2\sqrt{-\log(-\tilde{z})}\right)
    \end{split}
\end{equation}
which has a jump function part identical to \eqref{jumpprimeKPZ}. Hence that branch converges to the
second branch of the KPZ rate function, i.e. to $\Psi_{\rm KPZ}(\tilde z)=\Psi_{\rm KPZ,0}(\tilde z)+\Delta_{\rm KPZ}(\tilde z)$,
as claimed in the text.\\

Furthermore, as explained in the main text, the last two branches with $\Delta=\Delta_3-\Delta_1$ and $\Delta=\Delta_3$ disappear in the region $H\sim 0$ or equivalently $Z \sim 1$, which correspond to $H_{\rm KPZ} \to +\infty$ see discussion below.

\subsection{Matching to the regime $Y \sim T^{4/3}$}

It was predicted in \cite{TTPLD,TTPLDBeta}, and proved in \cite{GBPLDModerate}, that the sample-to-sample fluctuations of the 
probability denoted here as $Z(Y,T)=e^{H(Y,T)}$ -- defined in \eqref{defZ} --  when seen in an atypical
space time direction, are related to those of the random height field 
$h_{\rm KPZ}(x,t)=h(x,t)$ solution of the KPZ equation
\be 
\partial_t h(x,t) = \partial_x^2 h(x,t) + (\partial_x h)^2 + \sqrt{2} \eta(x,t) 
\ee 
with droplet initial condition $e^{h(x,0)}=\delta(x)$, where $\eta$ is a standard space-time white-noise.
The relation to the KPZ solution at finite time, $h_{\rm KPZ}(0,t)$, holds when one scales 
$Y \sim T^{3/4}$. The scaling studied here $Y \sim T^{1/2}$ thus corresponds to short KPZ time,
while the scaling $Y \sim T$ corresponds to the limit of infinite KPZ time, leading to the Tracy-Widom distribution \cite{BarraquandCorwinBeta}. \\

Let us recall the result of \cite[Section~3.2 Eq.~(30)]{GBPLDModerate} established in the scaling regime $Y \sim T^{3/4}$ (we consider here $Y>0$). 
Setting $y=0$ and $t=2  T$ there (to account for the different units) 
it translates into the equality in law in the large $T$ limit (for the diffusion \eqref{FP})
\be 
\log \mathbb{P}\big[y(T) > \tilde x (2 T)^{3/4} \big] + \frac{1}{2} \tilde x^2 (2 T)^{1/2} + \frac{1}{4} \log (2 T) - \log \tilde x
= h_{\rm KPZ}\left(0, \frac{\tilde x^4}{2}\right) 
\ee 
where $\tilde x = \frac{Y}{(2 T)^{3/4}}$.  Hence denoting
\be 
T_{\rm KPZ}= \frac{Y^4}{16 T^3} 
\ee 
we have the equalities in law
\be 
H(Y,T) = h_{\rm KPZ}\left(0, T_{\rm KPZ}\right)- \frac{Y^2}{4 T}    + \log \frac{Y}{2 T} \quad 
\Leftrightarrow \quad Z(Y,T) = \frac{Y}{2 T}  e^{- \frac{Y^2}{4 T} } e^{ h_{\rm KPZ}(0, T_{\rm KPZ}) }
\ee 
valid a priori in the regime $Y \sim T^{3/4}$.  Let us now set $Y=\xi \sqrt{T}$, with $\xi>0$. One gets 
\be \label{relZ} 
Z(Y,T) = \frac{\xi}{2 \sqrt{T}} e^{- \frac{\xi^2}{4} } e^{ h_{\rm KPZ}(0, T_{\rm KPZ}) } \quad , \quad  T_{\rm KPZ}= \frac{\xi^4}{16 T} 
\ee 
valid a priori in the regime $\xi \sim T^{1/4}$. We now show that it holds beyond that, i.e. in the large deviation
regime where $\xi$ is of order one but large, which is also the regime where the KPZ time is small, $T_{\rm KPZ} \ll 1$. 
To compare with the known large deviation results for the KPZ equation at short time, it is
useful to introduce
\be \label{defH} 
H_{\rm KPZ} :=  h_{\rm KPZ}(0, T_{\rm KPZ}) + \log( \sqrt{T_{\rm KPZ}}) 
\ee 
These results read \cite{le2016exact}, given here in the form of \cite[Eqs.~(4) and (22)]{UsWNT2021}
\be \label{LDKPZ} 
\overline{ \exp( - \tilde z e^{ h_{\rm KPZ}(0, T_{\rm KPZ}) } ) } = 
\overline{ \exp( - \frac{\tilde z}{\sqrt{T_{\rm KPZ}}} e^{ H_{\rm KPZ} } ) }
=
\exp\left( - \frac{\Psi_{\rm KPZ}(\tilde z)}{\sqrt{T_{\rm KPZ}}} \right)
\ee 
where $\Psi_{\rm KPZ}(\tilde z) = \Psi_{\rm KPZ,0}(\tilde z)= - \frac{1}{\sqrt{4 \pi}} {\rm Li}_{5/2}(-\tilde z)$. 
While the l.h.s. exists a priori only for $\tilde z>0 $ this formula admits an analytic continuation, called the main
branch, for $\tilde z \in [-1,+\infty]$. Already at this level we can match with the results of the 
present study. Indeed here we obtained for the main branch for $z>z_c$, see Eqs~\eqref{KPZ1} and \eqref{KPZ2} in the text
\be 
\overline{ \exp( - z \sqrt{T} Z ) } = \exp\left( - \sqrt{T} \frac{4}{\xi^2} \Psi_{\rm KPZ,0}(\tilde z) \right) \quad , \quad 
\tilde z = z \frac{\xi}{2} e^{-\xi^2/4}
\ee 
which is in perfect agreement with \eqref{LDKPZ} using the relations in \eqref{relZ}. Hence the large deviations in the
regime $Y \sim T^{3/4}$
and the diffusive regime $Y \sim \sqrt{T}$ match smoothly.\\

\begin{figure}[t!]
    \centering
    \includegraphics[scale=0.55]{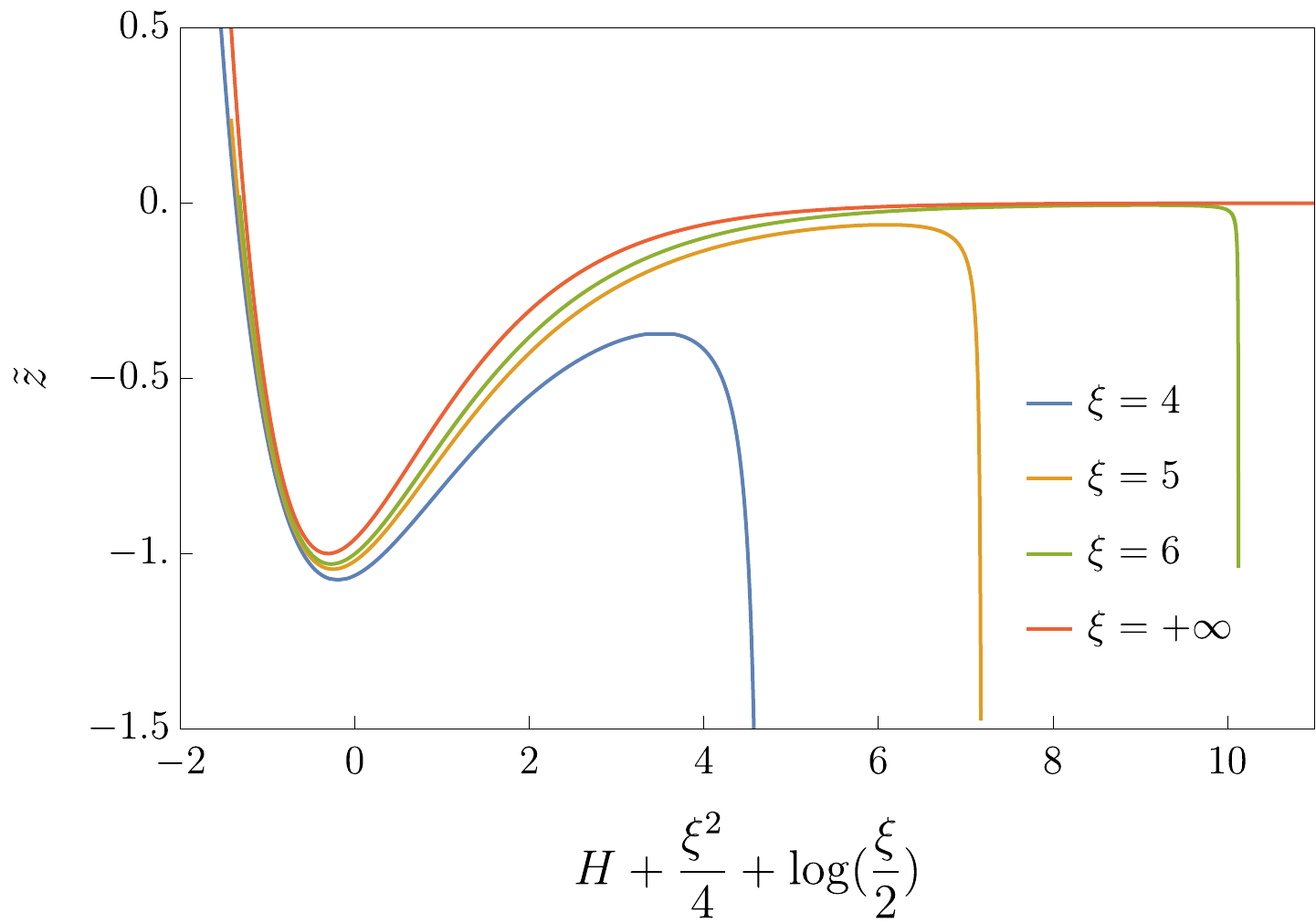}
    \caption{Plot of $\tilde z=-z/z_c$ as a function of $H_{\rm KPZ}=H + \frac{\xi^2}{4} + \log(\frac{\xi}{2})$ for several
    values of $\xi=4,5,6$ as compared with the asymptotic $\xi=\infty$ KPZ expression. All branches are represented. 
    The convergence to KPZ is excellent. The last two branches in Table~\ref{tab:MainTextTableJump} correspond to the sharp decrease to $\tilde{z}\to -\infty$
    and is pushed to $H_{\rm KPZ}=+\infty$ as $\xi=+\infty$. It corresponds to events where $Z \approx 1$ which become
    irrelevant in that limit.}
    \label{fig:HeightZTilde}
\end{figure}

As discussed in \cite{le2016exact,UsWNT2021} one obtains the rate function for the KPZ equation, $\Phi_{\rm KPZ}(H_{\rm KPZ})$,
upon Legendre inversion in the parametric form
\be 
\Phi_{\rm KPZ}(H_{\rm KPZ}) = \Psi_{\rm KPZ}(\tilde z) - \tilde z e^{H_{\rm KPZ}} \quad , \quad 
e^{H_{\rm KPZ}} = \Psi_{\rm KPZ}'(\tilde z) \label{Legkpz} 
\ee 
For the KPZ equation the main branch $\Psi_{\rm KPZ,0}(\tilde z)$ allows to obtain 
$\Phi_{\rm KPZ}(H_{\rm KPZ})$ only for $H_{\rm KPZ} < H_{\rm KPZ,c}= \log \frac{\zeta(3/2)}{4 \pi}$
which corresponds to the field at which $\tilde z = \tilde z(H_{\rm KPZ})$ solution of \eqref{Legkpz}
with $\Psi_{\rm KPZ}\to \Psi_{\rm KPZ,0}$ reaches $\tilde z=-1$. 
For $H_{\rm KPZ} > H_{\rm KPZ,c}$ one needs to use the second branch 
$\Psi_{\rm KPZ}\to \Psi_{\rm KPZ,0} + \Delta_{\rm KPZ}(\tilde z)$ and 
$\tilde z(H_{\rm KPZ})$ increases again from $-1$ to $0$ as $H_{\rm KPZ} \to +\infty$. \\

As we have shown in the previous subsection in the limit $\xi \to +\infty$ one obtains the convergence 
\be 
\Psi(z) \to \frac{4}{\xi^2} \Psi_{\rm KPZ}(\tilde z) 
\ee 
not just for the main branch, but for all the branches which survive in that limit. 
Thus we expect the correspondence between the fields obtained upon Legendre transform
\be \label{corr1} 
 H(z) = \log\left( \frac{\rmd \Psi}{\rmd z} \right) \underset{\xi \gg 1}{\simeq} \log( \frac{4}{\xi^2} \frac{\rmd \tilde z}{\rmd  z} \frac{\rmd  \Psi_{\rm KPZ}}{\rmd  \tilde z})  \underset{\xi \gg 1}{\simeq}  - \frac{\xi^2}{4} - \log(\frac{\xi}{2}) + H_{\rm KPZ}(\tilde z) 
\ee
The prediction is thus that using the results of the present work for $\xi \to +\infty$, one should have that $\tilde z=-z/z_c$
plotted versus $H_{\rm KPZ} = H + \frac{\xi^2}{4} + \log \frac{\xi}{2}$ reaches a
limit curve indentical to $\tilde z(H_{\rm KPZ})$ for the KPZ equation. As one can see
from \eqref{fig:HeightZTilde} this is indeed the case.\\ 

Finally we can check that \eqref{corr1} is indeed consistent with the correspondence 
discussed above from the matching to the $Y \sim T^{3/4}$ regime. Indeed, using \eqref{defH} and \eqref{relZ} one has
\be
\begin{split}
 H_{\rm KPZ}(\tilde z)   &= h_{\rm KPZ}(0, T_{\rm KPZ}) + \log( \sqrt{T_{\rm KPZ}})  \\
& = 
h_{\rm KPZ}(0, T_{\rm KPZ}) - \log( \sqrt{T} ) + \log(\xi^2/4) \\
&= H + \frac{\xi^2}{4} + \log \frac{\xi}{2} 
\end{split}
\ee
which is identical to \eqref{corr1}. \\

In conclusion, inserting into the parametric representation of the Legendre transform one obtains for
$\xi \gg 1$
\be 
\Phi(H) \simeq \frac{4}{\xi^2} \Phi_{\rm KPZ}(H_{\rm KPZ}) \quad , \quad 
H_{\rm KPZ} = H + \frac{\xi^2}{4} + \log \frac{\xi}{2} 
\ee 
as given in the text which means that one can identify the large deviations probabilities
\be 
{\cal P}(H) \sim \exp\left( - \sqrt{T} \Phi(H) \right) \sim \exp\left( - \frac{\Phi(H_{\rm KPZ})}{\sqrt{T_{\rm KPZ}}}  \right)  
\sim {\cal P}_{\rm KPZ}(H_{\rm KPZ}) \, .
\ee

\section{Large-time limit of the Fredholm determinant result for the sticky Brownian motion}
\label{app:stickyBrownian}

In this section we start from the formula of \cite{BarraquandSticky} and study the diffusive limit
where $T$ and $Y$ are large with $\xi= Y/\sqrt{T}$ fixed. This leads to a conjectural form for $\Psi(z)$
which agrees with the one derived in the text using inverse scattering. The manipulations in this
Appendix are quite heuristic but they have the merit to show that the algebraic structure which emerges from 
the Fredholm determinant is similar to the one derived in the text from first principles by the inverse scattering method.
We hope that it will help to obtain in the future a more precise and rigorous derivation.\\

In \cite{BarraquandSticky} the quantity which corresponds to $Z(Y,T)$ was studied. It was denoted
$K_{0,t}(0,[x,+\infty[)$ and called the kernel of the uniform Howitt-Warren flow. The equivalence
between the two objects, mathematically very subtle, was discussed in \cite[Remark 2.4]{BarraquandSticky},
see also \cite{WarrenSticky}.\\

Here, the quantity which we define as $Z(Y,T)$ obeys a backward Fokker-Planck equation.
Indeed, from the definition
\be 
Z(Y,\tau) = \int_Y^{+\infty} \rmd y\,  q_\eta(y,\tau) \quad , \quad q_\eta(y,\tau) = - \partial_y Z(y,\tau) 
\ee 
one easily obtain upon integrating Eq.~\eqref{FP} 
\be 
\partial_\tau Z(Y,\tau) = \partial^2_{Y} Z(Y,\tau) - \sqrt{2} \eta(Y,\tau) \partial_Y Z(Y,\tau)
\ee 
with initial condition $Z(Y,\tau=0)=\Theta(-Y)$. 
This is equivalent to Ref.~\cite[Eq.~(19)]{BarraquandSticky}. The correspondence of notations is that 
the space and time variables there must be replaced by $t \to 2 T$ and $x=Y$. Note that in Ref.~\cite[Eq.~(19)]{BarraquandSticky} the noise term is $- \sqrt{2}  \eta(y,T) =  \frac{\sqrt{2}}{\sqrt{\lambda}} \hat \eta(y,T)$
where $\hat \eta$ is standard space time white noise. Hence below $\lambda$ is set to unity.
\\

The identity proved in \cite[Theorem 1.11]{BarraquandSticky} reads for $u$ non-negative
\be 
\mathbb{E}[ e^{- u  Z(Y,T) } ]
= \Det (I - K_u)\vert_{\mathbb{L}^2(C)} 
\ee 
where $C$ is a positively-oriented circle with radius ${\sf R}$ and centered at ${\sf R}$ and
\bea 
\label{eq:KernelGuillaume}
&& K_u(v,v') = \frac{1}{2 \I \pi}
\int_{1/2 +\I \R} \frac{\pi}{\sin \pi s} u^s \frac{g(v)}{g(v+s)}
\frac{\rmd s}{s+v-v'} 
\eea 
The definition of $g(v)$ is
\be
\label{eq:def_g}
g(v) = \Gamma(v) e^{ a Y \psi_0(v) + b T \psi_1(v) } 
\ee 
where $\psi_{0,1}$ denote polygamma functions and $a=\lambda$ and $b=\lambda^2$. Here $\lambda$ is set to unity.\\ 

We now use $\zeta=v+s$ as a variable, for which the integration contour can be chosen as $1/2+2 {\sf R}+\I \R$ -- see remark in 
\cite[Proposition 2.3]{BarraquandSticky} -- where we recall that $C$ is the integration contour for $v,v'$.
We factorize the kernel \eqref{eq:KernelGuillaume} into the following form
\bea
&& K_u(v,v') = \int_{1/2+2 {\sf R}+\I \R}\frac{\rmd \zeta}{2\I \pi} A(v,\zeta) \tilde A(\zeta,v'), \\
&& A(v,\zeta) = \frac{\pi u^{\zeta-v} }{\sin(\pi (\zeta-v) )}\frac{g(v)}{g(\zeta)}, \qquad \tilde A(\zeta,v') = \frac{1}{\zeta-v'} \, .
\label{eq:KernelFactorization}
\eea 

We now use the identity
\begin{equation}
    \frac{\pi }{\sin(\pi s)}u^s=\int_\R \rmd r \frac{u}{u+e^{-r}}e^{-sr}
\end{equation}
which, inserted into \eqref{eq:KernelFactorization}, allows to factorize the kernel $A$ as 
\begin{equation}
\begin{split}
    A(v,\zeta)&=\int_\R \rmd r \frac{u}{u+e^{-r}}e^{-(\zeta-v)r} \, \frac{g(v)}{g(\zeta)}= \int_\R \rmd r \sigma(r)A_1(v,r)A_2(r,\zeta)=(A_1 \sigma A_2)(v,\zeta)
    \end{split}
\end{equation}
where the kernels $A_1, A_2$ and the function $\sigma$ read
\begin{equation}
A_1(v,r)=g(v) e^{vr}, \quad A_2(r,\zeta)=\frac{e^{-\zeta r}}{g(\zeta)}, \quad \sigma(r)=\frac{u}{u+e^{-r}}.    
\end{equation}
 Hence using Sylvester's identity
\begin{equation}
\begin{split}
    \Det(I-K_u)\vert_{\mathbb{L}^2(C)} &=\Det(I-A\tilde{A})\\
    &= \Det(I-A_1 \sigma A_2 \tilde{A})\\
    &=\Det(I-\sigma A_2 \tilde{A} A_1)\vert_{\mathbb{L}^2(\R)}
    \end{split}
\end{equation}

This last Fredholm determinant has the typical structure for which the first cumulant method, developed in \cite{KrajLedou2018,ProlhacKrajenbrink,krajenbrink2019beyond} to study the relevant asymptotics (here large $T$), applies. Defining a determinantal point process $\{a_\ell\}_{\ell \in \N}$ associated to the kernel $A_2 \tilde{A} A_1$, the following identity holds
\begin{equation}
\begin{split}
    \Det(I-\sigma A_2 \tilde{A} A_1)&=\E\left[ \prod_{\ell=1}^\infty (1-\sigma(a_\ell))\right]=\E\left[ \prod_{\ell=1}^\infty e^{-\varphi(a_\ell)}\right]
    \end{split}
\end{equation}
where $e^{-\varphi}=1-\sigma$. The first cumulant approximation asserts \cite[Section 6]{KrajLedou2018} that as some parameter goes to infinity (here it will be $T$, see below),
we expect the point process to self-average, i.e.
\begin{equation}
    \E\left[ \prod_{\ell=1}^\infty e^{-\varphi(a_\ell)}\right]\sim e^{-\E[\varphi(a)]}=e^{-\Tr(\varphi  A_2 \tilde{A}A_1)}
\end{equation}
If the first cumulant method works, we aim to have under the right scaling
\begin{equation}
    \Det(I-K_u)\vert_{\mathbb{L}^2(C)}\sim \exp\left[-\Tr(\varphi A_2 \tilde{A}A_1)\right]
\end{equation}
The explicit expression of the kernel $A_2 \tilde{A}A_1$ is obtained as 
\be 
(A_2 \tilde{A}A_1)(r,r') = \int_{1/2+2 {\sf R}+\I \R}\frac{\rmd \zeta}{2\I \pi} \int_C \frac{\rmd v'}{2\I \pi} \frac{g(v')}{g(\zeta)} \frac{1}{\zeta-v'} e^{r' v'- r \zeta} 
\ee 
taking into account that the measure on the variables $v$ is $\frac{\rmd v}{2 \I \pi}$. Using that 
\begin{equation}
    \varphi(r)=\log(1+ue^r)=-\mathrm{Li}_1(-ue^{r})
\end{equation}
to apply the first cumulant method we need to calculate the following quantity which only involves the diagonal part of the kernel $A_2 \tilde{A}A_1$
\begin{equation} \label{eq11} 
    \Tr(\varphi A_2 \tilde{A}A_1)=-\int_\R \rmd r \int_{1/2+2 {\sf R}+\I \R}\frac{\rmd \zeta}{2\I \pi} \int_C \frac{\rmd v'}{2\I  \pi} \mathrm{Li}_1(-ue^{r}) \frac{g(v')}{g(\zeta)} \frac{1}{\zeta-v'} e^{r (v'-  \zeta)} 
\end{equation}
We recall that $\Re(\zeta-v') >0$ by construction. We further proceed to an integration by part with respect to $r$ to obtain
\begin{equation}
    \Tr(\varphi A_2 \tilde{A}A_1)=-\int_\R \rmd r \int_{1/2+2 {\sf R}+\I \R}\frac{\rmd \zeta}{2\I \pi} \int_C \frac{\rmd v'}{2 \I \pi} \mathrm{Li}_2(- u e^{r}) \frac{g(v')e^{rv'}}{g(\zeta)e^{r\zeta}} 
\end{equation}
The boundary terms of the integration by part are zero since $e^{r (1+v' -\zeta)} \to 0$ for $r \to -\infty$ and the polylogarithms behave as ${\rm Li}_s(e^r) \sim r^s$ at $r \to +\infty$.\\

At this stage we proceed to the large-time rescaling to the diffusive regime using the rescaled variables
\be \label{scaling} 
\{ Y = \xi \sqrt{T} \quad , \quad v = w \sqrt{T} \quad , \quad \zeta=\omega \sqrt{T} \quad, \quad u = z \sqrt{T}  \}
\ee 

We then rewrite \eqref{eq11} as 
\be \label{S135} 
\Tr(\varphi A_2 \tilde{A}A_1)=-\int_\R \rmd r \mathrm{Li}_2(-ue^{r}) I(r) = -\int_\R \rmd r \mathrm{Li}_2(- z e^{r}) I(r- \log \sqrt{T} ) 
\ee 
where we have shifted the variable $r$ by $- \log \sqrt{T}$, and defined
\be 
\label{eq:densityControlled}
\begin{split}
I(r- \log \sqrt{T} ) =&
\int_{1/2+2 {\sf R}+\I \R}\frac{\rmd \zeta}{2\I \pi} \int_C \frac{\rmd v'}{2\I  \pi}\frac{g(v')e^{(r-\log(\sqrt{T})))v'}}{g(\zeta)e^{(r-\log(\sqrt{T}))\zeta}}  \\
&= \int_{1/2+2 {\sf R}+\I \R} \frac{\rmd \zeta}{2 \I  \pi} e^{ - (\log g(\zeta) + (r - \log \sqrt{T}) \zeta) } \int_C \frac{\rmd v}{2\I  \pi} e^{ (\log g(v) + (r - \log \sqrt{T}) v) }
\end{split}
\ee 

The large-time expansion of the function $g(v)$ given in Eq.~\eqref{eq:def_g} reads 
\bea \label{logg} 
&& \log g(v) = a Y \psi_0(v) + b T \psi_1(v) + \log \Gamma(v) \\
&& =
\sqrt{T} \left(\phi(w) + (w + a \xi) \log \sqrt{T}\right)  + \chi(w) - \frac{1}{2} \log(\sqrt{T}) + o(T) \nn 
\eea
where we defined 
\be
\phi(w) =   \frac{b}{w} - w + (w + a \xi) \log(w)  \quad , \quad \chi(w)= \frac{b}{2 w^2} - \frac{a \xi}{2 w} + \frac{1}{2} \log(2 \pi w)
\ee 

At this stage we will choose the radius of the circle $C$ conveniently to be equal to ${\sf R}=T/Y$ so that its mapping under the large-$T$ limit is a circle $C'$ of radius $1/\xi$ centered at $1/\xi$ (we assume here and below that $\xi>0$). 
Upon the change of variable \eqref{scaling} in the large $T$ limit, inserting \eqref{logg} into \eqref{eq:densityControlled}
and noting that constant terms cancel from the two integrals we obtain
\be 
\begin{split}
\label{eq:densityControlled2}
I(r-\log \sqrt{T}) & \simeq  T \int_{2/\xi + 0^+ +  \I \R} \frac{\rmd \omega}{2 \I \pi} e^{ - \sqrt{T}  (\phi(\omega) + r \omega) - \chi(\omega) } \int_{C'} \frac{\rmd w}{2\I  \pi} e^{ \sqrt{T} (\phi(w) + r w) + \chi(w)}
\end{split}
\ee  
In the large-$T$ limit these integrals are dominated by saddle points. The saddle point equations read
\be 
\phi'(w) = - \frac{1}{w^2} + \frac{ \xi}{w} + \log( w) = - r  
\ee 
and the same for $\omega$. Since $w$ is on the circle $C'$ we can parameterize it in the following way 
\begin{equation} \label{ww} 
    \frac{1}{w}=-\I q+\frac{\xi}{2} \quad , \quad q \in \mathbb{R}
\end{equation}
The saddle point equation becomes 
\begin{equation} \label{er} 
    e^r=(-\I q+\frac{\xi}{2})e^{-q^2-\frac{\xi^2}{4}}
\end{equation}
which is very reminiscent of Eq.~\eqref{eq:BranchCutEquation}. To make this saddle point easily attainable, one way is to deform the integration contour of $r$ 
which is not $\R$ anymore but the image of \eqref{er} as $q$ varies on the real axis, which we call $\gamma$. We will assume that this is possible.
This is a closed curve for $e^r$, touching the real axis at values $e^r=0$ and $e^r=\frac{\xi}{2}e^{-\frac{\xi^2}{4}}$. The solution of \eqref{ww} and \eqref{er} defines a function $w(r)$ so that the saddle point evaluation of \eqref{eq:densityControlled2} gives
\be
I(r-\log \sqrt{T})  
\simeq - \frac{\sqrt{T}}{2\I\pi}\frac{1}{ \phi''(w(r))}
\ee  
where we have also assumed that the integration contour of $\omega$ could be deformed to be folded around $C'$.
This ensures that the dominant exponential at the saddle point cancel.\\

To summarize, the first cumulant \eqref{S135} of the Fredholm determinant reads in the large $T$-limit
\begin{equation} \label{eqla} 
    \Tr(\varphi A_2 \tilde{A}A_1) = \frac{\sqrt{T}}{2\I \pi} \int_\gamma \rmd r \, 
    \mathrm{Li}_2(-ze^{r})  \frac{1}{\phi''(w(r))}
\end{equation}
We will now perform the change of variable \eqref{er}. Using the saddle point equation $\phi'(w(r))=-r $ we obtain upon derivation
the Jacobian of this change of variable
\begin{equation}
     \phi''(w(r)) \frac{\rmd  w(r)}{\rmd q}\frac{\rmd  q}{\rmd r}= - 1, \qquad \frac{1}{\phi''(w(r)) }\rmd  r= - \I\frac{\rmd q}{(\I q-\frac{\xi}{2})^2}
\end{equation}
Inserting into \eqref{eqla} we finally obtain
\be \label{PsiFinal} 
\Psi(z) = \frac{1}{\sqrt{T}}  \Tr(\varphi A_2 \tilde{A}A_1) =  -
\int_\R \frac{\rmd q}{2 \pi}\frac{\mathrm{Li}_2( z (\I q - \frac{\xi}{2}) e^{-q^2 - \frac{\xi^2}{4}} )}{(\I q - \frac{\xi}{2})^2} \\
\ee 
which is in agreement with Eq.~\eqref{PsiFinal0} in the text.

\section{Extension to the extremal diffusion beyond Einstein's diffusion theory}

In this section we study the position of the maximum of $N$ walkers in the same random field
(by sample below we mean one given environment, i.e. random field). 
Previous works started with Ref. \cite{BarraquandCorwinBeta,BarraquandThesis} which studied the Beta random walk
and pointed out that for $N \gg 1$, and in the regime $\log N \sim T$, the position
of the maximum has sample to sample fluctuations given by the Tracy-Widom distribution.
Another regime, $\log N \sim \sqrt{T}$, was obtained in \cite{TTPLD} and \cite{GBPLDModerate} where these
fluctuations are described by the solution of the KPZ equation at finite time. Numerical simulations 
which confirm these regimes have been performed recently \cite{CorwinPrivate}. 
Extending these arguments, our present work allows to study another regime, $\log N \ll \sqrt{T}$,
not studied previously.\\

Consider $N$ independent particles in the same environment. One denotes $Y_N(T)=\max_i Y_i(T)$ with $i=1,\dots,N$
and $Z_N(Y,T)=\mathbb{P}(Y_N(T) > Y)$. One has the exact relation
\be  \label{identity1} 
1- Z_N(Y,T)= \mathbb{P}(Y_N(T) < Y) = \mathbb{P}(Y(T) < Y)^N = (1- Z(Y,T))^N 
\ee 

We focus below on the diffusive scaling $Y \sim \sqrt{T}$ at large $T$, not considered previously in the
discussion of the extremal diffusion. We will thus denote $y_N(T)=\frac{1}{\sqrt{T}} Y_N(T)$.
There are several observables of interest.\\ 

{\bf Large deviations of the CDF of the maximum}.
The first observable is $Z_N=\mathbb{P}(y_N(T) > \xi)$, which is simply the analog of $Z$ for the maximum  position of $N$ particles. 
One can ask, for any finite $N$, what are the large deviations of the PDF of $Z_N$ for $T \gg 1$. From the above simple
relation \eqref{identity1} one finds for 
\be 
{\cal P}(Z_N) \sim \exp\left( - \sqrt{T} \hat \Phi_\xi \left( 1 - (1- Z_N)^{1/N} \right) \right)
\ee 
where the rate function $\hat \Phi_\xi(Z)$ is the one obtained in the present work (for $N=1$). Here and below we indicate explicitly the dependence in $\xi$ of the rate functions. 
\\

{\bf Averaged CDF of the maximum}. Another observable is the following average over the environment
\be 
\overline{\mathbb{P}(y_N(T) < \xi)} = \int_0^1 \rmd Z \overline{e^{N \log(1-Z)}} \sim \int_0^1 \rmd Z
e^{N \log(1-Z) - \sqrt{T} \hat \Phi_\xi(Z) } 
\ee
where in the last equation we have substituted the large deviation form. Note that considering instead of the average moments of order $q$ is equivalent to
substitute $n \to n q$. \\

There are several regimes depending on $N$. If $N \ll \sqrt{T}$ then the
second term dominates and implies that $Z \approx Z_{\rm typ}(\xi)$ so that 
\be \label{res1} 
\overline{\mathbb{P}(y_N(T) < \xi)} \simeq e^{N \log(1-Z_{\rm typ}(\xi)) } 
\ee
and the result is identical as the CDF of the maximum position for $N$ particles in the absence of random field. \\

If $N = n \sqrt{T} \gg 1$ with $n=\mathcal{O}(1)$ fixed, the two terms can balance each others and one finds
that this observable takes the large deviation form 
\be 
\overline{\mathbb{P}(y_N(T) < \xi)} 
\sim e^{- \sqrt{T} \Sigma_{\xi}(n) }  \quad , \quad
\Sigma_{\xi}(n) = \min_Z ( \hat \Phi_\xi(Z) - n \log(1-Z))  
\ee 
with a rate function obtained from a non trivial variational formula.
Here for a given $\xi$ the value of $Z$ which realizes the optimum is different from $Z_{\rm typ}(\xi)$
and thus involves rare environments. Upon some simple manipulations, recalling that 
$Z=\Psi'(z)$ and $\hat \Phi'(Z)=-z$ we obtain 
$z=\frac{n}{1-Z}$ leading to the parametric representation 
\be
\Sigma_{\xi}(n) = \Psi_\xi (z)-z + n- n \log(\frac{n}{z}) \quad , \quad z (1 - \Psi_\xi'(z)) = n
\ee
Note that the approximation $N \log(1-Z) \simeq - N Z$ valid for $Z= \Psi_\xi'(z) \ll 1$ 
would instead lead to $z \simeq n$ and $\Sigma_{\xi}(n) \simeq \Psi_\xi (z)$. Although we leave this study to the future, it is 
quite likely that a phase transition similar to the one of $\Psi(z)$ for $\xi>\xi_1$ for
and for some values of $z$ should also occur here. 
For $n \to 0$ one has $n \simeq z(1-\Psi_\xi'(0))=z(1-Z_{\rm typ}(\xi))$ and one recovers \eqref{res1}. 
More precisely one has the expansion
\be 
\Sigma_{\xi}(n) = - n \log(1- \Psi_\xi'(0) ) + \frac{\Psi_\xi''(0) n^2}{2 (1- \Psi_\xi'(0))^2 } + \mathcal{O}(n^3) \quad , \quad \Psi_\xi'(0)=Z_{\rm typ}(\xi)
\ee 
\\

{\bf Position of the maximum: typical behavior}. One can ask about the position of the maximum and its fluctuations. Let us
introduce $N$ i.i.d exponential random variables $g_i$ of PDF $P(g)=e^{-g} \Theta(g)$, and
call $G_N= \max_i g_i - \log N$. At large $N$, $G_N \to G$ a Gumbel random variable
with $\mathbb{P}(G<g)=e^{- e^{- g}}$. For any $N$ one has $\mathbb{P}(G<g)=(1- \frac{1}{N} e^{- g})^N$. 
In a given environment one can write
\be \label{theta} 
\mathbb{P}(Y_N(T) < Y) = e^{N \log(1-Z(Y,T))} = \overline{\Theta(G_N + \log N + H(Y,T) <0) }^{G_N}
\ee
This formula is valid for any $N$ and for large $N$ one obtains the same formula with $G_N \to G$
by approximating $e^{N \log(1-Z)} \simeq e^{-N Z}$. Note that $G_N$ and $G$ in this formula
are independent of $H(Y,T)$. As discussed below, the approximation $Z \ll 1$
is also realized for any $N$ with large probability when $\xi=Y/\sqrt{T}$ is large. 
The random position of the maximum $Y_N(T)$, in a given environment is then given by 
\be \label{condsample1} 
G_N + \log N + H(Y,T) < 0 \quad \Leftrightarrow \quad Y_N(T) < Y 
\ee 
Note that $G_N + \log N$ is a positive random
variable. Since $Z(Y,T)$ and thus $H(Y,T)$ is a positive decreasing function of $Y$ in any sample,
one may argue (by taking a derivative w.r.t. $Y$ in \eqref{theta}) that \eqref{condsample1} is equivalent to
\be \label{condsample2} 
G_N + \log N + H(Y_N(T),T) = 0 
\ee 
This formula generalizes \cite[Eq.~(50)]{GBPLDModerate} to any $N$. 
\\

Until now this is exact. Let us again consider the diffusive scaling regime $Y \sim \sqrt{T}$ at large $T$. 
In a typical environment, one has $H(Y,T) \simeq H_{\rm typ}(\xi)$ where 
$\xi=Y/\sqrt{T}$ and $H_{\rm typ}(\xi)= \log\left( \int_\xi^{+\infty} \frac{e^{-x^2/4}}{\sqrt{4 \pi}} \right) = 
\log( \frac{1}{2} {\rm Erfc}(\frac{\xi}{2})) = - \frac{\xi^2}{4} - \log(\sqrt{\pi} \xi) + \mathcal{O}(\xi^{-1})$.
Note that $H_{\rm typ}(\xi)$ varies from $0$ for $\xi \to -\infty$ to
$-\infty$ for $\xi \to +\infty$. Let us denote $y_N^{\rm typ}$ the scaled position of the maximum in a typical environment. 
At large $T$ it reaches a finite limit in distribution such that 
\be 
G_N + \log N + H_{\rm typ}( y_N^{\rm typ} ) = 0  \quad \Leftrightarrow \quad y_N^{\rm typ} = 
H_{\rm typ}^{-1}( - G_N - \log N )
\ee 
where $H_{\rm typ}^{-1}(h)=\xi$ is the reciprocal function of $H_{\rm typ}(\xi)=h$. This is correct
for any $N$. 
The distribution of $y_N^{\rm typ}$ is exactly the same as the one for the maximum of $N$ Brownian motions
at time $t=1$, performing each diffusion $\rmd B_i(t)^2 = 2 \rmd t$, started at $B_i(0)=0$ at $t=0$,
i.e. for the problem without the quenched random field. For $N \gg 1$, using the asymptotics
of $H_{\rm typ}(\xi)$ one finds the standard result
\be 
y_N^{\rm typ} \simeq 
 2 \sqrt{\log N} + \frac{ G - \frac{1}{2} \log(4 \pi \log N)}{\sqrt{\log N}} + \dots 
\ee 

We can now study the typical fluctuations from sample to sample. To lowest order one should take into
account the typical fluctuations of $H(Y,T)$, which are $\delta H = \mathcal{O}(T^{1/4})$. The variance was obtained in \eqref{secondcum} 
as $\overline{H^2}^c = C_2(\xi) T^{-1/2}$, where the function $C_2(\xi)$ was given there. The position of the maximum is now determined by
\be 
G_N + \log N + H_{\rm typ}( y_N ) + \sqrt{ C_2(y_N) } T^{-1/4} \omega = 0 
\ee 
where $\omega$ is a Gaussian random variable of unit variance. Inverting to leading 
order at large time we find (an equation valid for any $N$)
\bea \label{yN} 
 y_N &=& H_{\rm typ}^{-1}\left( - G_N - \log N - \sqrt{ C_2(y_N^{\rm typ}) } T^{-1/4} \omega \right) + o(T^{-1/4}) \\
& = & y_N^{\rm typ} - \frac{\sqrt{ C_2(y_N^{\rm typ}) }}{H_{\rm typ}'(y_N^{\rm typ})} T^{-1/4} \omega + o(T^{-1/4}) 
\eea 
If $N \gg 1$ one finds 
\be \label{yNN} 
y_N = 2 \sqrt{\log N} + \frac{ G - \frac{1}{2} \log(4 \pi \log N) + \sqrt{ C_2(y_N^{\rm typ}) } T^{-1/4} \omega }{\sqrt{\log N}  } + \dots  
\ee  
where we recall that the $\omega$ term represents the sample-to-sample fluctuations and the Gumbel variable $G$ the "thermal" fluctuations, the
two random variables being uncorrelated. 
\\

We can compare this result with Ref.~\cite[Eqs.~(57-58)]{GBPLDModerate} setting $D=1$ and $r_0=2$ there, which were obtained
when $N$ and $T$ are large with the parameter $g=\frac{\log N}{\sqrt{T}}$ kept fixed. The KPZ time there is 
$T_{\rm KPZ} = g^2 = \frac{(\log N)^2}{T}$. This agrees perfectly with the 
KPZ time in the present work $T_{\rm KPZ} = \xi^4/(16 T)$ where $\xi \sim y_N \sim 2 \sqrt{\log N}$
from \eqref{yNN}. For the matching to \cite[Eqs.~(57-58)]{GBPLDModerate} to be perfect we need 
the variance of the KPZ height field at very short time (i.e. in the Edward-Wilkinson regime for droplet initial condition)
which is given by \cite{le2016exact} 
\be 
\overline{ h(0,T_{\rm KPZ})^2}^c \simeq C_2^{\rm KPZ} T_{\rm KPZ}^{1/2} \quad , \quad C_2^{\rm KPZ} = \sqrt{\frac{2}{\pi}}
\ee 
One then easily checks that it exactly matches the amplitude of the fluctuating term $\sim \omega$ in \eqref{yNN} 
using the large $\xi$ behavior \eqref{asymptC2}, $C_2(\xi) \simeq \frac{\xi^2}{4} \sqrt{\frac{2}{\pi}}$. \\

To summarize, \eqref{yN} and \eqref{yNN} extend the results 
of \cite{GBPLDModerate} about "typical" extremal diffusion to the diffusive regime $Y \sim \sqrt{T}$.
In that new regime $\log N \ll \sqrt{T}$, i.e. $T \gg (\log N)^2$ and the fluctuations are of the Edwards-Wilkinson type. If $N$ is 
large, $\log N$ does not need to be very large. As $(\log N)^2/T$ is increased
there is a perfect match to the predictions of \cite{GBPLDModerate} in the 
regime $Y \sim T^{3/4}$ where the sample-to-sample fluctuations are governed by the finite-time KPZ equation. 
\\

{\bf Remark.} The two independent random contributions in \eqref{yNN} can be separated by considering  
simultaneously the "quantile" as done in numerical simulations \cite{CorwinPrivate}, that is, instead of $y_N(T)$, $x_N(T)=\frac{X_N(T)}{\sqrt{T}}$ 
defined by $\int_{X_N(T)}^{+\infty} \rmd y q_\eta(y,T) = \frac{1}{N}$ in a given sample, or in other words
\be 
\log H(X_N(T),T) = - \log N \quad , \quad Z(X_N(T),T)= \frac{1}{N} 
\ee

{\bf Position of the maximum: large deviations}.
Finally, our results yield additional information about the large deviations of extremal diffusion, i.e. for rare environments such that 
$H - H_{\rm typ} = \mathcal{O}(1)$. In that case if one heuristically replaces in \eqref{condsample2}, 
$H(Y_N(T),T) \to H_{\rm typ}(y_N(T)) + (H - H_{\rm typ}(y_N^{\rm typ}))$ one obtains
\bea \label{yN2} 
 y_N &\simeq & H_{\rm typ}^{-1}\left( - G_N - \log N - (H - H_{\rm typ}) \right) 
\eea 
and for $N \gg 1$ 
\be \label{yNN2} 
y_N \simeq 2 \sqrt{\log N} + \frac{ G - \frac{1}{2} \log(4 \pi \log N) + (H - H_{\rm typ}(\xi)) }{\sqrt{\log N}  } + \dots  
\ee  
with $\xi=2 \sqrt{\log N}$, for rare environments which occur with probability $\sim \exp(- \sqrt{T} \Phi_\xi(H))$. 
Since $\xi$ is large, rewriting $H=- \frac{\xi^2}{4} - \log \frac{\xi}{2} + H_{\rm KPZ}$, 
this is equivalent to extend the estimate of of \cite{GBPLDModerate} for the fluctuations of the position of the 
maximum to the large deviations regime of the KPZ equation (with rare environments occuring with probability $\sim \exp(- \frac{1}{\sqrt{T_{\rm KPZ}}}
\Phi_{\rm KPZ}(H_{\rm KPZ}))$ and with $T_{\rm KPZ}=\frac{\xi^4}{16 T}= \frac{(\log N)^2}{T} \ll 1$.

\section{Extension to general quadratic models in the MFT: diffusion in random medium and the
symmetric simple exclusion process}\label{app:ExtensionSSEP}

One definition of the MFT is as the Langevin equation of a diffusive gas with particle density $q(x,t)$
\cite{SpohnFluct}
\begin{equation} \label{stochMFT} 
    \partial_t q=\partial_x[D(q)\partial_x q-\sqrt{\sigma(q)}\xi(x,t)]
\end{equation}
where $\xi(x,t)$ is a standard space-time white noise. The model solved in this present paper corresponds to $\sigma(q)=2q^2$ and $D(q)=1$. 
Averages of solutions of \eqref{stochMFT} over the noise can be obtained from the dynamical action
$S[q,p] = \iint\rmd x \rmd t \, [ p \p_t q - \mathcal{H}(q,p)]$ with Hamiltonian 
$\mathcal{H}(q,p)=-D(q)\partial_x q \partial_x p + \frac{1}{2}\sigma(q)(\partial_x p)^2$, 
and where $p(x,t)$ is the response field. At large time these averages can be 
obtained from the solutions to the saddle point equations $\p_t q = \frac{\delta \mathcal{H}}{\delta p}$
and $\p_t p = - \frac{\delta \mathcal{H}}{\delta q}$, which admit the conservation law 
$\frac{\rmd }{\rmd t} \mathcal{H}(p,q)=0$.\\

We will focus below on a subclass of models within the MFT called quadratic models and show how the work of this present paper is relevant to solve them.

\subsection{Mapping of quadratic models in the MFT to the coupled DNLS system}

Consider here the quadratic MFT models which have a noise variance parameterized as
\begin{equation}
    \sigma(q)=2Aq(B-q) \label{param} 
\end{equation}
and a diffusion constant $D(q)=1$. This class contains both the SSEP and the present model of diffusion in random medium. 
The MFT hydrodynamic equations (i.e. the saddle point equations) read
\begin{equation}
\begin{split}
    \partial_t q &= \partial_x [\partial_x q-2Aq(B-q)\partial_x p] \label{spq} \\
    \partial_t p &= - \partial_x^2 p-A(B-2q)(\partial_x p)^2
    \end{split}
\end{equation}
We introduce the generalized derivative Cole-Hopf transform
\begin{equation}
    R(x,t)=A\p_x p(x,t) e^{AB p(x,t)}, \quad Q(x,t)=q(x,t)e^{-AB p(x,t)} \, .
    \label{eq:derivativeColeHopfMFT}
\end{equation}
The variables $\{R,Q \}$ then verify the coupled DNLS system  \eqref{DNLS2} with $\beta=1$ 
\begin{equation} 
\begin{split}
    \partial_t  Q &= \partial_x^2 Q+2  \partial_x ( Q^2  R)  \\
  -  \partial_t  R &= \partial_x^2  R- 2  \partial_x( Q  R^2) 
    \end{split}
\end{equation}

\subsection{Gauge transformation between NLS and DNLS and relation with the non-local transformation of \cite{mallick2022exact}}

\textbf{Change of variable of Wadati and Sogo.} Wadati and Sogo proved in 1982 \cite{wadati1983gauge} that the non-linear Schrodinger equation and the derivative non-linear Schrodinger equation were gauge equivalent. Indeed, consider the following systems in the conventions of \cite{wadati1983gauge}, firstly the coupled NLS
\be 
\begin{split}
& \I  q_{1t} + q_{1xx} - 2 r_1 q_1^2 = 0 \\
& \I  r_{1t} - r_{1xx} + 2 r_1^2 q_1 = 0
\end{split}
\label{eq:Wadati1}
\ee 
and secondly the coupled DNLS
\be
\begin{split}
& q_{2t} - \I  q_{2xx} - (r_2 q_2^2)_x = 0 \\
& r_{2t} + \I  r_{2xx} - (r_2^2 q_2)_x = 0
\end{split}
\label{eq:Wadati2}
\ee 
Wadati and Sogo showed that the following change of variables allows to map the coupled DNLS system to the coupled NLS system.
\be
\begin{split}
& q_1= \frac{q_2}{2} \exp\big( - \I \int_{-\infty}^x r_2 q_2\big) \\
& r_1 = (- \I r_{2 x} + r_2^2 q_2/2) \exp\big(  \I \int_{-\infty}^x r_2 q_2\big) 
\end{split}
\label{eq:WadatiGauge1}
\ee
To show the relation with the non-local transformation of \cite{mallick2022exact}, one needs to relate the conventions of Wadati to the ones of this present work and of \cite{mallick2022exact}. We first transform the time in \eqref{eq:Wadati1} and \eqref{eq:Wadati2} as $t\to \I t$, and choose
\begin{equation}
\begin{split}
&q_1=v, \quad r_1=u\\
&q_2=-2R, \quad r_2=\I Q\\
\end{split}
\end{equation}
We obtain that \eqref{eq:Wadati2} is the $\{ R,Q\}$ system with $\beta=1$ (that is e.g. \eqref{interpolating} with $g=0$ 
or \eqref{DNLS2})
and that \eqref{eq:Wadati1} is the $\{P,Q\}$ system with $g=-1$. This $\{P,Q\}$ system is precisely the equations verified by the functions $\{ v, u\}$ of \cite{mallick2022exact} (with $v=P$ and $u=Q$). \\

Now, considering the MFT for the SSEP, we have shown in \eqref{eq:derivativeColeHopfMFT} that the derivative Cole-Hopf transform of the MFT variables verify the DNLS $\{ R,Q\}$ system. Performing the gauge transformation \eqref{eq:WadatiGauge1} with our new variables thus leads to
\be
\begin{split} 
& u =  ( Q^2 R+\p_x Q ) \exp\big(2 \int_{-\infty}^x\rmd y \,  Q R\big) \\
& v = - R  \exp\big(-2 \int_{-\infty}^x \rmd y \,  Q R\big)
\end{split}
\ee 
We can now go back to the variables $q$ and $p$ of the MFT using the
generalized derivative Cole-Hopf transform \eqref{eq:derivativeColeHopfMFT}, and we obtain 
\bea 
&& u =  \left(-A q(B-q) \partial_x p + \partial_x q \right) \exp\left(- \int_{-\infty}^x \rmd y \,  A(B-2q)  \partial_y p\right) \\
&& v = - A \partial_x p  \exp\left( \int_{-\infty}^x \rmd y\,  A(B-2q) \partial_y p\right)
\label{eq:gauge-transfo-wadati-sasamoto}
\eea 
which is valid for any quadratic theory. This recovers the "generalized Cole-Hopf equations" obtained very recently 
in \cite[Eqs.~(10)--(11)]{mallick2022exact} (which use the notations $H=p$ and $\rho=q$).
Note however the missing the factor $A$ in the second equation in that work.

\subsection{Stationary measure}

The stochastic equation \eqref{stochMFT} admits generically a family of stationary measures. For instance if one fixes the boundary conditions as $q(0)=q(L)=\rho$, and if the problem is taken on a finite-size interval,
the stationary measure is \cite{DerridaGershenfeld,BertiniMFT2009,SpohnFluct,DerridaMFTReview2007}
\be 
\mathcal{P}_{\rm eq}(\{ q(x) \}) \sim e^{- \int_0^L \rmd x \left( f(q(x)) - f(\rho) - (q(x)-\rho) f'(\rho) \right) } 
\ee 
where $f''(q) = \frac{2 D(q)}{\sigma(q)}$. The linear term is determined so that the maximum probability is for $q=\rho$.\\

Consider the model of diffusion in a random environment studied here in Eq.~\eqref{FP}, with a more general amplitude
for the noise. In that case one has $D(q)=1$ and $\sigma(q)= 2 \alpha q^2$, hence $f''(q)=1/(\alpha q^2)$.
This leads to $f(q)= - \frac{1}{\alpha} \log q + k q + c$, and to the stationary measure
\be \label{statdiff} 
\mathcal{P}_{\rm eq}(\{ q(x) \}) \propto e^{- \frac{1}{\alpha}  \int_0^L \rmd x \left( - \log(q(x)/\rho)   + \frac{q-\rho}{\rho}   ) \right) } 
\ee 

{\bf Remark}. The stationary measure \eqref{statdiff} is the analog in the continuum of 
a discrete measure on a lattice defined as a product of independent Gamma variables at each site, i.e. $\prod_x w_x$,
with PDF $p(w) \propto w^{\gamma-1} e^{-w}$. Indeed that measure appeared as a stationary measure
in the Beta polymer problem, in a (long time) one point version in 
\cite{TTPLDBeta}, and for a more general discussion see \cite{BarraquandBetaHalfSpace}. 
For the more general quadratic model parametrized as \eqref{param}, in particular for the SSEP, the corresponding discrete
stationary measures are instead
factorized Bernoulli.\\

{\bf Remark}. For the diffusion model, one has $B=0$ in \eqref{param}. Hence $R= A \partial_x p$
and $Q=q$ satisfy the DNLS system with $\beta=1$. By choosing here $A=-\alpha$ one can vary
the exponent $\gamma$ of the local Gamma distribution to any value in the stationary measure.\\

\subsection{Extension of \cite{mallick2022exact} to quadratic MFT models with annealed initial condition and tracer away from the origin}

Let us consider a model within the MFT where the noise variance is parametrized as \eqref{param}. We study here the annealed case where the initial condition of the hydrodynamic equations \eqref{spq} is fluctuating according to the stationary measure of the MFT \cite{DerridaMFTReview2007,grabsch2021closing,poncet2021generalized}. We choose the initial condition as a local equilibrium configuration with two different densities on the positive and negative axis
\begin{equation}
    \mathcal{P}(q(x,0))\sim e^{-\sqrt{T}\mathcal{F}(q(x,0))}, \qquad     \mathcal{F}(q(x,0))=\int_\R \rmd x \int_{\bar{q}(x)}^{q(x,0)}\rmd z \, \frac{2D(z)}{\sigma(z)}(q(x,0)-z) 
\end{equation}
with $\tilde{q}(x)=q_- \Theta(-x)+q_+ \Theta(x)$ is the step density profile.\\ 

We will be interested in the position $X_t$ of a tracer initially located at position $X_0=0$ and at final position $X_1=\xi$. Its position at any time is defined as
\begin{equation}
    \int_0^{X_t} \rmd x \, q(x,t)=\int_0^{\infty} \rmd x \, (q(x,t)-q(x,0))
\end{equation}
If we focus on the generating function of $X_1$ or the current at the right of $X_1$, i.e. $Z(\xi)=\int_{\xi}^\infty \rmd x \, (q(x,1)-q(x,0))$, then it was shown \cite{DerridaGershenfeld,grabsch2021closing,poncet2021generalized} that the mixed-time boundary conditions of the hydrodynamic system \eqref{param} read
\bea 
&& p(x,1) = \lambda \Theta(x-\xi) \\
&& p(x,0)= \lambda \Theta(x) +\int_{\bar{q}(x)}^{q(x,0)}\rmd r \, \frac{2D(r)}{\sigma(r)}
\eea 
for some constant $\lambda$. Using the gauge transformation \eqref{eq:gauge-transfo-wadati-sasamoto} along with the same manipulations as the ones in \cite[ below Eqs.~(14)-(15)]{mallick2022exact} allows to transform these boundary conditions for $\{p,q \}$ into simple boundary conditions for $\{u,v \}$
\be 
u(x,0)= \frac{\omega}{K}  \delta(x) \; , \quad v(x,1) = K \delta(x-\xi) \; .
\ee 
for some constant $K$ to be determined as in \cite{mallick2022exact}. These boundary conditions have an asymmetry due to the presence of $\xi$ that we can cancel using the same boost transformation as in \eqref{eq:app-boost-dnls}
\be
U(x,t)= u(x-v t,t) e^{-\frac{1}{2} x v + \frac{v^2}{4} t} \quad , \quad 
V(x,t)= v(x-v t,t) e^{\frac{1}{2} x v - \frac{v^2}{4} t} \quad , \quad 
\ee 
Note that this boost leaves the coupled NLS system \eqref{eq:Wadati1} invariant. We choose $v=-\xi$ so that
\be
U(x,t)= u(x+\xi t,t) e^{\frac{1}{2} x \xi + \frac{\xi^2}{4} t} \quad , \quad 
V(x,t)= v(x+ \xi t,t) e^{-\frac{1}{2} x \xi - \frac{\xi^2}{4} t} \quad , \quad 
\ee  
which yields for boundary conditions
\be
U(x,0)=\frac{\omega}{K} \delta(x)   \, , \quad 
V(x,1)= Ke^{ - \frac{\xi^2}{4}} \delta(x)  \; .
\ee  
One can then proceed as in this work to complete the scattering analysis and solve the large-deviation problem. 

\subsection{Discussion on the quench and annealed initial conditions}
The quadratic models of MFT have been investigated through the spectrum of classical integrability in three works and two contexts of initial conditions:

\begin{itemize}
    \item Reference \cite{mallick2022exact} considered the SSEP with an initial condition in the annealed class and solved the problem through the mapping to the coupled NLS $\{P,Q \}$ system and the use of its scattering theory. The remarkable feature of that work is that the annealed initial condition for the SSEP admits a simple quenched $\delta, \delta$  mixed-time boundary conditions interpretation in the coupled NLS $\{P,Q \}$ system.
    \item The present work as well as Ref.~\cite{NaftaliDNLS} considered the diffusion in random media, equivalent to the KMP model, with a quenched initial condition and solved the problem using the scattering theory of the coupled DNLS $\{R,Q \}$ system. 
\end{itemize}
At this stage, the observation is that depending on whether the quench or annealed initial condition is considered, a specific integrable model might be more suited to obtain the exact solution of the problem. Since other gauge transformations between integrable models have been proposed in \cite{wadati1983gauge}, it would be interesting to investigate whether mappings to other integrable models would allow to answer new questions.

\newpage

\end{widetext}


\begin{thebibliography}{200}
\bibitem{footnote1} We will keep $\beta$ as a parameter but for the application to obtain $\Psi(z)$
it is understood that it is set to $\beta=-1$.
\bibitem{footnotePsi0} When the integrand has a jump at $q=0$, one
integrates respecting the symmetry $\I q \to - \I q$, equivalently
one integrates on $\mathbb{R}^-$ and take twice the real part.
\bibitem{footnoteZ} For the KPZ equation the partition sum $Z_{\rm KPZ}=e^{H_{\rm KPZ}}$ is an 
unbounded random variable. As a result the 
rate function $\Psi(\tilde z)$ is undefined for $\tilde z<-1$.
The present convergence results show precisely how, in
the large $\xi$ limit, one goes from a bounded random 
variable $Z$ to an unbounded one $Z_{\rm KPZ}$.
\bibitem{footnoteW} Although it maps to the $\{P,Q\}$ system with $\delta-\delta$ initial conditions, as the WNT of the
KPZ equation with droplet initial condition, because of the highly non-local nature of the mapping it does not map SSEP observables to KPZ ones in any obvious way.
\end{thebibliography}
\end{document}